\documentclass[aps,twocolumn,pre,amsmath,amssymb,notitlepage,superscriptaddress,longbibliography,nofootinbib,longbibliography]{revtex4-2}

\usepackage{graphicx} 
\usepackage{amsmath,amssymb,amsfonts}
\usepackage{ragged2e}
\usepackage{tikz}
\usepackage{enumitem}

\usepackage{booktabs}   
\usepackage{multirow}   
\usepackage{makecell}
\usepackage{adjustbox}

\usepackage{graphicx,color,xcolor}
\usepackage{bbm, bm}
\usepackage{framed}

\usepackage{hyperref}
\hypersetup{
    colorlinks=true,
    linkcolor=blue,
    urlcolor=magenta,
    citecolor=red,
    pdfpagemode=FullScreen,
    }
\colorlet{shadecolor}{gray!20}

\begin{document}

\def\ito{It\^{o} }
\newcommand{\dt}{\partial_t}
\newcommand{\id}{\mathbbm{1}}
\newcommand{\dx}{\partial_x}
\newcommand{\Dt}{\frac{d}{dt}}

\newcommand{\n}{\nonumber}

\newcommand{\rmd}{{\rm d}}

\newcommand{\avg}[1]{\left\langle #1 \right\rangle}
\renewcommand{\vec}[1]{\mathbf{#1}}
\renewcommand{\tensor}[1]{\vec{\vec{#1}}}
\newcommand{\tensorr}[1]{\vec{\tensor{#1}}}
\newcommand{\alphaa}{\vec{\tensorr{\alpha}}}
\newcommand{\nabl}{\vec{\nabla}}
\newcommand{\da}{\partial^{\alpha}}
\newcommand{\db}{\partial_{\beta}}
\newcommand{\dg}{\partial_{\gamma}}
\newcommand{\dab}{\partial^{\alpha\beta}}
\newcommand{\emphasize}[1]{\textbf{\textit{#1}}}
\newcommand{\approximation}[1]{$\implies$ \textbf{#1}}
\newcommand{\rf}{r_{\rm F}}
\newcommand{\bfJ}{\mathbf{J}}
\renewcommand{\rq}{r_{\rm QS}}
\newcommand{\pIK}{p^{\rm IK}}
\newcommand{\mueff}{\mu^{\rm eff}}
\newcommand{\avgline}[1]{ \langle #1\rangle}

\newcommand{\cA}{\mathcal{A}}
\newcommand{\cF}{\mathcal{F}}
\newcommand{\cD}{\mathcal{D}}
\newcommand{\dD}{\mathscr{D}}
\newcommand{\dF}{\mathscr{F}}
\newcommand{\cG}{\mathcal{G}}
\newcommand{\cH}{\mathcal{H}}
\newcommand{\cL}{\mathcal{L}}
\newcommand{\cK}{\mathcal{K}}
\newcommand{\cV}{\mathcal{V}}
\newcommand{\cP}{\mathcal{P}}
\newcommand{\dP}{\mathscr{P}}
\newcommand{\cS}{\mathcal{S}}
\newcommand{\cO}{\mathcal{O}}
\newcommand{\cT}{\mathcal{T}}
\newcommand{\cZ}{\mathcal{Z}}

\newcommand{\bet}{\boldsymbol{\eta}}
\newcommand{\dd}{\text{d}}
\newcommand{\ee}{\text{e}}
\newcommand{\bff}{\mathbf{f}}
\newcommand{\bg}{\mathbf{g}}
\newcommand{\bnabla}{\boldsymbol{\nabla}}
\newcommand{\ii}{\text{i}}
\newcommand{\p}{\partial}
\newcommand{\bfp}{\mathbf{p}}
\newcommand{\bfa}{\mathbf{a}}
\newcommand{\bfb}{\mathbf{b}}
\newcommand{\bfw}{\mathbf{w}}
\newcommand{\bfq}{\mathbf{q}}
\newcommand{\bfr}{\mathbf{r}}
\newcommand{\bfe}{\mathbf{e}}
\newcommand{\bfv}{\mathbf{v}}
\newcommand{\bfu}{\mathbf{u}}
\newcommand{\bfx}{\mathbf{x}}
\newcommand{\bfy}{\mathbf{y}}
\newcommand{\bfR}{\mathbf{R}}
\newcommand{\bfV}{\mathbf{V}}
\newcommand{\bfX}{\mathbf{X}}
\newcommand{\bfI}{\mathbf{I}}
\newcommand{\bfC}{\mathbf{C}}
\newcommand{\bxi}{\boldsymbol{\xi}}
\newcommand{\bchi}{\boldsymbol{\chi}}
\newcommand{\bomega}{\boldsymbol{\omega}}
\newcommand{\bzeta}{\boldsymbol{\zeta}}
\newcommand{\bvarphi}{\boldsymbol{\varphi}}

\newcommand{\plm}[1]{{\color{red}[#1]}}
\newcommand{\ad}[1]{{\color{magenta}[#1]}}
\newcommand{\rev}[1]{#1}

\newcommand{\rrangle}{\right\rangle}
\newcommand{\llangle}{\left\langle}

\newcommand{\resume}{\medskip
\begin{center}
\noindent\textcolor{red}{\texttt{\%\%\%\%\%\%\%\%\%\%\%\%\%\%\%\%\%\%\%\%\%\%\%\%\%\%\%\%\%\%\%\%\%\%\%\%\%\%\%\%\%\%\%\%}}
\\
\textcolor{red}{\texttt{\%\%\%\%\%\%\%\%\%\%\%\%\% RESUME FROM HERE \%\%\%\%\%\%\%\%\%\%\%\%\%}}\\
\textcolor{red}{\texttt{\%\%\%\%\%\%\%\%\%\%\%\%\%\%\%\%\%\%\%\%\%\%\%\%\%\%\%\%\%\%\%\%\%\%\%\%\%\%\%\%\%\%\%\%}}
\end{center}}

\newtheorem{hypx}{Hypothesis}
\newenvironment{hypotheses}{
  \renewcommand{\thehypx}{H.\arabic{hypx}}
  \begin{enumerate}[label=\textbf{H.\arabic*\>}, ref=H.\arabic*]
}{
  \end{enumerate}
}

\newcommand{\hypitem}[2]{%
  \item \label{#1} #2
}

\title{Multiscale Perturbative Approach to Active Matter with Motility Regulation}
\author{Alberto Dinelli}
\altaffiliation[Corresponding authors.]{ Email: {alberto.dinelli@unige.ch and pierluigi.muzzeddu@unige.ch}}
\affiliation{Department of Biochemistry, University of Geneva, 1211 Geneva, Switzerland}
\affiliation{Department of Theoretical Physics, University of Geneva, 1211 Geneva, Switzerland}
\author{Pietro Luigi Muzzeddu}
\altaffiliation[Corresponding authors.]{ Email: {alberto.dinelli@unige.ch and pierluigi.muzzeddu@unige.ch}}
\affiliation{Department of Biochemistry, University of Geneva, 1211 Geneva, Switzerland}
\affiliation{Department of Theoretical Physics, University of Geneva, 1211 Geneva, Switzerland}
\date{\today}

\begin{abstract}
We present a coarse-graining method applicable to dry scalar active matter with motility regulation.
Our approach, based on a multiscale perturbative expansion of the backward Kolmogorov equation, does not rely on any specific microscopic dynamics for the particles' orientations. 
Its generality allows us to address different forms of motility regulation, from space-dependent self-propulsion speed to taxis, and to extend the analysis to a class of non-Markovian orientational dynamics.
Furthermore, we identify general conditions on the microscopic dynamics that ensure the existence of an effective large-scale equilibrium regime.
When the latter are violated, our theoretical framework is able to quantitatively capture the emergence of large-scale particle currents. 
We directly apply our coarse-grained theory to several models of self-propelled agents, ranging from single particles to active polymers, and test our analytical predictions with numerical simulations.
Finally, we show that our theory naturally extends to active matter with density-mediated interactions, such as quorum sensing, with potential applications to self-organizing soft materials.
\end{abstract}
\maketitle

\tableofcontents

\section{Introduction}
Motility regulation is a widespread feature of active systems, by which active agents adapt their motion in response to external cues or interactions with peers. 
On the one hand, the response to external signals is crucial for the dynamics of biological systems, from chemotaxis in bacteria~\cite{budrene1991complex} to phototactic algae~\cite{polin2009chlamydomonas}, but it can also be engineered in the lab via self-phoretic colloids or light-activated particles~\cite{jiang2010active,palacci2013living,bauerle2018self,lavergne2019group}. 
On the other hand, interactions between particles can drive the emergence of a wealth of out-of-equilibrium phases. For instance, quorum-sensing (QS) interactions, whereby particles modulate their motility based on the local density of their peers~\cite{miller2001quorum,hammer2003quorum,daniels2004quorum}, is known to lead to motility-induced phase separation (MIPS)~\cite{tailleur2008statistical,liu2011sequential,cates2015motility,curatolo2020cooperative,zhao2023chemotactic}. Such phase separation relies on a positive feedback between local density and particle speed~\cite{fily2012athermal,cates2015motility}, which destabilizes homogeneous states and leads to the coexistence of dense and dilute phases. 
Beyond the classical MIPS paradigm, the emergent phenomenology can be enriched by the interplay with population dynamics~\cite{cates2010arrested}, non-reciprocal interactions~\cite{dinelli2023nonreciprocity,duan2023dynamical}, or chirality~\cite{pisegna2024emergent,caprini2025bubble}.

To predict the large-scale physics arising from motility-regulation, it is essential to connect microscopic dynamics to macroscopic behavior through a systematic coarse-graining approach. 
Here we focus on dry scalar active matter, for which the total momentum of the system is not conserved (`dry'), and where the emergent phenomenology is solely captured by the density field(s) (`scalar').

In this context, numerous coarse-graining schemes have been developed in the literature~\cite{fox1986uniform,fox1986functional,faetti1988projection,schnitzer1993theory,tailleur2008statistical,cates2013when,wittmann2017effective,o2020lamellar,duan2023dynamical,dinelli2024fluctuating,pisegna2025spinning,burekovic2026active}, which typically rely on explicit descriptions of the microscopic orientational dynamics: paradigmatic examples are run-and-tumble particles (RTPs)~\cite{schnitzer1993theory,berg2004coli,kurzthaler2024characterization}, active Brownian particles (ABPs)~\cite{golestanian2007designing,jiang2010active,theurkauff2012dynamic,palacci2013living}, and active Ornstein-Uhlenbeck particles (AOUPs)~\cite{sepulveda2013collective,szamel2014self,wittmann2017effective,wittmann2017effective2,martin2021statistical}. 
For motility-regulated active systems, macroscopic mappings between different models have been explicitly derived~\cite{cates2013when,solon2015active,o2020lamellar,dinelli2024fluctuating}, suggesting a large-scale equivalence to lowest order in gradients. 
Nonetheless, the extent and generality of such a large-scale equivalence require further investigation. 

Moreover, for self-propelled systems with internal structure, such as active polymers~\cite{ghosh2014dynamics,isele2015self,winkler2017active, bianco2018globulelike, winkler2020physics, pfreundt2023controlled, dedenon2026importance}, the range of possible choices for the microscopic dynamics becomes even broader,
encompassing active particles carrying passive cargoes~\cite{hu2025cargo,vuijk2021chemotaxis, muzzeddu2023taxis}, chiral chains~\cite{muzzeddu2022active, caprini2025spontaneous,valecha2025active}, and polymers with diverse topologies. 
Therefore, changes in the microscopic dynamics often require a fresh coarse-graining derivation.

In a companion Letter~\cite{dinelli2026PRL} we presented a general hydrodynamic description of active polymers with motility regulation, whose derivation does not rely on a specific dynamics for the orientation vectors of the microscopic constituents.
On the one hand, our results lead to the identification of general criteria for the existence of an effective equilibrium regime for the system. 
On the other hand, they constitute the starting point to derive fluctuating hydrodynamic theories for scalar active systems. 
Importantly, within our theoretical framework, we revealed a novel class of (anti) motility-induced phase separation or anti-MIPS, in which, at odds with the standard MIPS paradigm, the denser phase is also the most motile one.

In this work, we provide a detailed account of our coarse-graining procedure and extend it to a broader class of systems which include, for instance, heteropolymers and systems with tactic interactions. We illustrate the consequences for the existence of large-scale equilibrium mapping, and, with the aim of providing a pedagogical presentation, we study applications to a broad range of specific active particle models. 

The paper is organized as follows. In Sec.~\ref{sec:model}, we introduce the microscopic dynamics of an active polymer and detail the assumptions underlying our derivation. After identifying the fast and slow degrees of freedom in the system, in Sec.~\ref{sec:adiabatic_elimination} we perform the coarse-graining of the active dynamics, ultimately obtaining a Fokker–Planck equation for the center of mass of the polymer. 
The method relies on a multiscale perturbative expansion of the backward Kolmogorov equation, following Ref.~\cite{pavliotis2008multiscale}.
We extend it to additional cases, including taxis, non-Markovian processes with Markovian embedding, and heteropolymers, as discussed in Sec.~\ref{sec:generalization}.

The second part of the paper explores the implications of the resulting large-scale theory. In Sec.~\ref{sec:equilibrium}, we determine the conditions under which motility-regulated active matter admits an effective equilibrium description at large scales. Using Schwarz’s theorem~\cite{o2020lamellar,dinelli2023nonreciprocity,duan2025phase,o2024geometric,o2025geometric}, we derive general constraints on the auto-correlation tensor of the orientations, $\mathbb{C}^{\alpha\beta}$, for such an equilibrium regime to exist. When these conditions are met, we obtain simple expressions for the steady-state distribution of the center of mass.

Because the coarse-grained theory explicitly depends on the auto-correlation tensor, Sec.~\ref{sec:correlations} briefly reviews its form for several standard models of orientational dynamics. Building on this, in Secs.~\ref{sec:cg_particle} and~\ref{sec:cg_polymer} we derive explicit hydrodynamic descriptions for representative models of single active particles and active polymers, respectively. In addition to recovering known results, we analyze new examples and determine the corresponding steady-state distributions and macroscopic currents. 
For polymers, this analysis reveals a fundamental macroscopic difference between taxis and space-dependent self-propulsion speed. 

Finally, in Sec.~\ref{sec:interacting_case}, we conclude our work by providing a detailed derivation of the large-scale hydrodynamics of active polymers interacting via quorum-sensing motility regulation. 
Our method can be directly extended to tactic interactions, heteropolymers and active mixtures, thus providing an interacting hydrodynamic description for a broad class of scalar active systems.

\section{Microscopic dynamics}
\label{sec:model}
We consider a minimal model for an active agent composed by $N$ particles in $d$ spatial dimensions.
In the case of $N>1$, we will refer to the system as an active polymer, and to its constituents as monomers. 
The latter are described by a position vector $r_i^\alpha$ and an orientation vector $u_i^\alpha$, where $i$ labels the monomer index and $\alpha$ the spatial component.

The positions of each monomer evolve according to the following overdamped It\^o-Langevin dynamics:
\begin{eqnarray}
    \dot r^\alpha_i &=& -\gamma^{-1} \partial^\alpha_{\bfr_i}\mathcal{H} +{\rm v}^\alpha_i(\bfr_i,\bfu_i) + \sqrt{2 D_{\rm t}} \xi^\alpha_i \;, \label{eq:SDE-monomers}    
\end{eqnarray}
where  ${\rm v}^\alpha_i(\bfr_i,\bfu_i)$ is a local function that introduces an active coupling between positional and orientational variables.
In the following, we adopt the shorthand notation $\Theta \equiv \{u_i^\alpha\}_{i=0}^{N-1}$ to denote the whole set of orientational degrees of freedom. The evolution of the latter is governed by a Markovian stochastic process with no dependence on the positions. 
The vectors $\xi_i^\alpha$ represent zero-mean Gaussian white noises with unit variance, accounting for  fluctuations of the monomers due to the interaction with a thermal bath at temperature $T$.
The friction coefficient $\gamma$ and translational diffusion $D_{\rm t}$ satisfy the Einstein relation $D_{\rm t}=k_BT\gamma^{-1}$.
The topology of the polymer is described by an adjacency matrix ${\rm A}_{ij}$, whose entries are equal to one if two monomers are connected, and to zero otherwise.
We then define the associated connectivity matrix as the Laplacian matrix $M_{ij} = \deg[i] \delta_{ij} - {\rm A}_{ij}$, where $\deg[i]$ is the number of connections of monomer $i$.
The $N$ monomers interact via pairwise-additive interaction forces deriving from the quadratic Hamiltonian:
\begin{equation}
    \cH = \frac{\kappa}{2}\sum_{ij} M_{ij} \> r_i^\alpha  r_j^\alpha \;,
    \label{eq:Hamiltonian}
\end{equation}
with $\kappa$ the stiffness parameter. Note that, in Eq.~\eqref{eq:Hamiltonian} as in the rest of the manuscript, summations over repeated spatial indices (Greek letters) are implied.

In the absence of the active coupling ${\rm v}_i^\alpha$, the stochastic differential equation (SDE) in Eq.~\eqref{eq:SDE-monomers} corresponds to the so-called Rouse model for a passive polymer in an equilibrium thermal bath \cite{doi1988theory}. 
For this reason, we introduce the Rouse modes $\{ \chi_i^\alpha\}_{i=0}^{N-1}$ defined by:
\begin{equation}
    \chi_i^\alpha = \sum_{j=0}^{N-1}\varphi_{ij} r_j^\alpha\,,
    \label{eq:rouse-modes-def}
\end{equation}
with $\varphi_{ij}$ defined such that its rows contain the orthonormal eigenvectors of the connectivity matrix $M_{ij}$:
\begin{equation}
    \sum_{jk}\varphi_{ij} {M}_{jk} \varphi_{lk} = \sigma_i \delta_{il} \;, \quad \text{with} \quad \sum_{j=0}^{N-1} \varphi_{ij} \varphi_{kj} = \delta_{ik} \;.
    \label{eq:phi-def}
\end{equation}
In Eq.~\eqref{eq:phi-def}, $\{\sigma_i\}$ denote the eigenvalues of $M_{ij}$.
In particular, we isolate the center of mass $R^\alpha = \chi_0^\alpha/\sqrt{N}$ from the higher-order Rouse modes $\{\chi_j^\alpha\}_{j=1}^{N-1}\equiv\chi$ related to the internal structure of the polymer, as their evolution occurs on very different time scales. 
Inverting the identity in Eq.~\eqref{eq:rouse-modes-def} and using the fact the $\varphi_{ij}$ is orthogonal, we can express the position of the $i$-th monomer as
\begin{equation}
     r_i^\alpha = R^\alpha + \sum_{j=1}^{N-1} \varphi_{ji} \chi_j^\alpha \;.
    \label{eq:r_i}
\end{equation}
In light of the previous change of variables, the SDE in Eq.~\eqref{eq:SDE-monomers} can be rewritten as:
\begin{align}
    \dot R^\alpha &= \frac{1}{N} \sum_{j=0}^{N-1} {\rm v}_j^\alpha(\bfr_j, \bfu_j) + \sqrt{\frac{2D_{\rm t}}{ N}} \eta_0^\alpha \;, \label{eq:SDE-com}\\
    \dot{\chi}_i^\alpha &= - \lambda_i \chi_i^\alpha + \sum_{j=0}^{N-1} \varphi_{ij} {\rm v}^\alpha_j(\bfr_j, \bfu_j) + \sqrt{2D_{\rm t} } \eta_i^\alpha \label{eq:SDE-chi}\;,
\end{align}
where $i\in \{1,...,N-1\}$. 
Here, we introduced the relaxation rates $\lambda_i=\tau^{-1}_{\rm r} \sigma_i$ of the higher-order Rouse modes, where $\tau_{\rm r} = \gamma/\kappa$. 
For later convenience, we also define the longest relaxation timescale $\tau_{\chi}=\max\{ \lambda_i\}$.

Before proceeding, we introduce a central quantity in our derivation.
%
This is the stationary auto-correlation function of the orientational degrees of freedom, given by:
\begin{equation}
    \mathbb{C}_{ij}^{\alpha\beta}(t) = \llangle u_i^\alpha(t) u_j^\beta(0) \rrangle \;,
\end{equation}
where the bracket notation $\langle \cdot\rangle$ denotes the average over the stochastic realizations of the $\Theta$-dynamics, provided the process has already reached steady state. 
By integrating over time the correlation function $\mathbb{C}_{ij}^{\alpha\beta}(t)$, we can now introduce the characteristic times $\{\tau_{ij}^{\alpha \beta}\}$ defined as:
\begin{equation}
    \tau^{\alpha\beta}_{ij} = d \int \rmd t \> \mathbb{C}_{ij}^{\alpha\beta}(t) < \infty \;,
    \label{eq:tau-ij-alphabeta}
\end{equation}
and the persistence time $\tau_{\rm p} \equiv \max{| \tau^{\alpha\beta}_{ij}|}$.
A few remarks are in order. Based on the above discussion, it follows that, while the $\Theta$-dynamics has thus far been kept general, it must admit a stationary state and its auto-correlation function must decay sufficiently fast to ensure the convergence of the integral in Eq.~\eqref{eq:tau-ij-alphabeta}.
We also require all orientational degrees of freedom to have a vanishing average at steady state.
The persistence time $\tau_{\rm p}$ and the typical relaxation time of the Rouse modes, $\tau_{\chi}$, set the fast timescales of our dynamics, compared to the slower evolution of the center of mass $R^\alpha$. 
For the sake of simplicity, we introduce a single fast timescale given by $ \tau = \max \{\tau_{\rm p} , \tau_{\chi} \}$.
\section{Adiabatic elimination in the presence of space-dependent self-propulsion speed}
\label{sec:adiabatic_elimination}
Starting from Eqs.~\eqref{eq:SDE-com}, \eqref{eq:SDE-chi}, we aim to perform an adiabatic elimination of the fast variables, i.e., the orientations $\Theta$ and the higher-order Rouse modes $\chi$. 
Our goal is thus to derive an effective Fokker-Planck description, valid at times $t \gg \tau$, for the center-of-mass distribution $p(\bfR,t)$:
\begin{equation}
    \partial_t p(\bfR, t) = -\partial^{\alpha} [ V^\alpha p - \partial^\beta( \cD^{\alpha\beta} p) ],
\end{equation}
where $V^\alpha$ and $\cD^{\alpha\beta}$ respectively represent the macroscopic drift and diffusion tensor. The method we propose is based on the multiscale expansion discussed in~\cite{pavliotis2008multiscale}. 
Our starting point is the time evolution of the joint probability distribution $P(\bfR,\chi,\Theta ,t)$, which, being the system Markovian, is given by the forward Kolmogorov equation~\cite{gardiner2004handbook}:
\begin{equation}
    \partial_t P(\bfR,\chi,\Theta ,t) = \left[\cL^{\dagger}_{\bfR}  + \cL^{\dagger}_{\chi} + \cL^{\dagger}_{\Theta}  \right] P \;.
    \label{eq:forward-kolmogorov}
\end{equation}
The action of the forward operators $\cL^{\dagger}_{\bfR}$ and $\cL^{\dagger}_{\chi}$ on a test function $f$ is given by:
\begin{align}
    \cL^{\dagger}_\bfR f &= -\partial^\alpha \Big[ \frac{1}{N} \sum_{j=0}^{N-1} {\rm v}^\alpha_j f - \frac{D_{\rm t}}{ N}\partial^\alpha f \Big]\,,  \\[0.2cm]
    \cL^{\dagger}_\chi f &= -\sum_{i=1}^{N-1} \partial_{\bchi_i}^\alpha \Big[\sum_{j=0}^{N-1} \varphi_{ij} {\rm v}_j^\alpha f -\lambda_i\chi_i^\alpha f - D_{\rm t}\partial_{\bchi_i}^\alpha f \Big]\,,
\end{align}
where we introduced the shorthand notation $\partial^\alpha \equiv \partial_{\bm{R}}^\alpha$.
As shown below, under the working assumptions adopted here, the microscopic details of the orientational degrees of freedom $\Theta$ do not affect the derivation of the effective large-scale dynamics of the center of mass $R^\alpha$.
Accordingly, the forward operator 
$\cL^\dagger_\Theta$
is kept generic throughout the analysis and will be specified only when our theory is applied to particular cases in Sec.~\ref{sec:correlations}.
Furthermore, since the orientational degrees of freedom $\Theta$ are independent of the positional variables, the marginal probability density $P_{\Theta}(\Theta,t)$ satisfies:
\begin{equation}
    \partial_t P_{\Theta}(\Theta,t) = \cL^\dagger_\Theta P_{\Theta}(\Theta,t)
\end{equation}
We denote by $\psi(\Theta)$ the steady state probability density associated to the orientations, namely the one solving the homogeneous problem $\cL^\dagger_\Theta \psi=0$. \par

\subsection{The large-scale diffusive regime}
\label{subsec:diffusive-regime}
In the following, we identify the relevant spatial and temporal scales of the problem. Relying on the scale separation between the evolution of the center of mass $R^\alpha$ and that of the fast variables $(\chi,\Theta)$, we then non-dimensionalize Eq.~\eqref{eq:forward-kolmogorov} and introduce a small parameter $\varepsilon$, which will enable us to coarse-grain the dynamics via a perturbative expansion.
For the sake of concreteness, we focus on the case in which the active coupling ${\rm v}_i^\alpha(\bfr_i, \bfu_i)$ varies on a single macroscopic lengthscale $\delta$, which we assume to be of the order of the system size $L$.

By assuming a diffusive scaling at the macroscopic level, we introduce the macroscopic timescale $\cT$ as the typical time over which the center of mass $R^\alpha$ diffuses a lengthscale of the order of $\delta$, i.e., $\delta \sim \cT^{1/2}$.
In contrast, on shorter timescales comparable with the persistence time $\tau_{\rm p}$, the microscopic dynamics of the active polymer is characterized by a persistence length $\ell_{\rm p} = \tau_{\rm p} v_0$, where $v_0$ sets the typical magnitude of the active coupling, i.e., ${\rm v}_i^\alpha(\bfr_i, \bfu_i) \sim \mathcal{O}(v_0)$, and by the typical size of the polymer, given, e.g., by the gyration radius $R_{\rm g}$.
We recall here that $R_{\rm g}^2$ depends on the power spectrum of the higher-order Rouse modes (see, e.g., \cite{doi1988theory}), and therefore it scales as their typical amplitude $\langle \chi_i^\alpha\chi_i^\alpha\rangle$.
In the following, we assume that the microscopic lengthscales described above are of the same order $\ell_{\rm p} \sim R_{\rm g}\sim\ell$, which is much smaller compared to the length of variation $\delta$ of the active coupling.

To formalize such separation of scales, we introduce the adimensional parameter $\varepsilon \ll 1$, such that:
\begin{equation}
\varepsilon = \ell/\delta \;,   \qquad \varepsilon ^2 =\tau/\cT \;.
\label{eq:scaling}
\end{equation}
Note that the different scaling of spatial and temporal  variables follows from the diffusive relation $\delta \sim \cT^{1/2}$ between their corresponding macroscopic scales.
\subsection{Non-dimensional equations of motion}
\label{subsec:adimensional_eom}
Having identified the relevant spatio-temporal scales of the problem, we now proceed to nondimensionalize the equations of motion given by Eqs.~\eqref{eq:SDE-com} and~\eqref{eq:SDE-chi}.
To this aim, we first introduce the adimensional variables $\tilde{R}^\alpha=R^\alpha/\delta$, $\tilde{\chi}_i^\alpha=\chi_i^\alpha/\ell$ and $\tilde{t} = t/\cT$. 
Furthermore, since the active coupling ${\rm v}_i^\alpha$ is characterized by a single length scale $\delta$, we find convenient to introduce the function $\tilde{{\rm v}}_i^{\alpha}(\bfr_i/\delta, \bfu_i) \equiv (\tau/\ell){\rm v}_i^\alpha(\bfr_i, \bfu_i)$.
Together with the definition of the adimensional variables and Eq.~\eqref{eq:r_i}, this allows us to decompose the active coupling into contributions of different order in $\varepsilon$ as:
\begin{align}
    &\tilde{{\rm v}}_i^{\alpha}(\bfr_i/\delta, \bfu_i) = \tilde{{\rm v}}_i^{\alpha}\left(\tilde{\bfR} + \varepsilon\sum_{j=1}^{N-1} \varphi_{ji} \tilde{\bchi}_j, \bfu_i\right)     \label{eq:expansion-active-coupling}\\ & = \tilde{{\rm v}}_i^{\alpha}(\tilde{\bfR}, \bfu_i) + \varepsilon \sum_{j=1}^{N-1}\varphi_{ji} \tilde{\chi}^\beta_j \tilde{\partial}^\beta \tilde{{\rm v}}_i^{\alpha}(\tilde{\bfR}, \bfu_i) + \cO(\varepsilon^2)\;,\n
\end{align}
where we used the notation $\tilde{\partial}^\alpha \equiv \partial^\alpha_{\tilde{\bfR}}$.
The above decomposition paves the way for the perturbative approach that we discuss in Sec.~\ref{subsec:multiscale-expansion}.
Multiplying Eq.~\eqref{eq:SDE-com} by $\cT/\delta = \varepsilon^{-1} \tau/\ell$ and using Eq.~\eqref{eq:expansion-active-coupling}, we obtain the dimensionless stochastic evolution of the center of mass:
\begin{equation}
\partial_{\tilde{t}}\tilde{R}^\alpha = \frac{1}{\varepsilon}\mathcal{V}_0^{0\alpha}+ \mathcal{V}_0^{1\alpha} + \sqrt{2\tilde{D}_{\rm t}N^{-1}}\tilde{\eta}^\alpha_0 +\cO(\varepsilon)\;,
\label{eq:dimensionless_SDE_com}
\end{equation}
where we introduced the functions:
\begin{align}
\mathcal{V}_0^{0\alpha}(\tilde{\bfR}, \Theta) &= \frac{1}{N} \sum_{i=0}^{N-1}\tilde{{\rm v}}_i^{\alpha}(\tilde{\bfR}, \bfu_i)\;, \label{eq:expansion_cV00}\\
\mathcal{V}_0^{1\alpha}(\tilde{\bfR}, \tilde{\chi},  \Theta) &= \frac{1}{N} \sum_{i=0}^{N-1} \sum_{j=1}^{N-1} \varphi_{ji} \tilde{\chi}^\beta_j \tilde{\partial}^\beta \tilde{{\rm v}}_i^{\alpha}(\tilde{\bfR}, \bfu_i)\;.
\label{eq:expansion_cV0}
\end{align}
Moreover, we defined the dimensionless translational diffusion coefficient $\tilde{D}_{\rm t}=D_{\rm t}\tau/\ell^2$ and the dimensionless noise $\tilde{\eta}^\alpha_0 = \eta^\alpha_0\cT^{1/2}$. 
Analogously, we adimensionalize the equation of motion of the higher-order Rouse modes in Eq.~\eqref{eq:SDE-chi} multiplying it by $\cT/\ell=\varepsilon^{-2} \tau/\ell$. This yields:
\begin{equation}
    \partial_{\tilde{t}}\tilde{\chi}_i^{\alpha} = \frac{1}{\varepsilon^2}\mathcal{V}_i^{0\alpha} + \frac{1}{\varepsilon}\mathcal{V}_i^{1\alpha} + \mathcal{V}_i^{2\alpha} + \frac{1}{\varepsilon}\sqrt{2\tilde{D}_{\rm t}}\tilde{\eta}^\alpha_i + \cO(\varepsilon)
    \label{eq:dimensionless-SDE-chi}
\end{equation}
with:
\begin{align}
    \mathcal{V}_i^{0\alpha}(\tilde{\bfR},\tilde{\chi}, \Theta) &= -\tilde{\lambda}_i\tilde{\chi}_i^{\alpha} + \sum_{j=0}^{N-1}\varphi_{ij}\tilde{{\rm v}}_j^{\alpha}(\tilde{\bfR}, \bfu_j)\;, \\
    \mathcal{V}_i^{1\alpha}(\tilde{\bfR},\tilde{\chi}, \Theta) &= \sum_{k=0}^{N-1} \sum_{j=1}^{N-1} \varphi_{ik} \varphi_{jk} \tilde{\chi}^\beta_j \tilde{\partial}^\beta \tilde{{\rm v}}_k^{\alpha}(\tilde{\bfR}, \bfu_k)\;,
    \label{eq:expansion_cVi}
\end{align}
while the expression of the term $\mathcal{V}_i^{2\alpha}(\tilde{\bfR},\tilde{\chi}, \Theta)$, which can be readily obtained by considering the next order in the expansion of Eq.~\eqref{eq:expansion-active-coupling}, is not reported here since, as will be clarified in the Sec.~\ref{subsec:multiscale-expansion}, it is irrelevant for the derivation of the large-scale dynamics of the center of mass at leading order in $\varepsilon$.\par
Concerning the orientational degrees of freedom $\Theta$, we assume that the associated forward Kolmogorov operator $\cL^\dagger_{\Theta}$ introduced in Eq.~\eqref{eq:forward-kolmogorov} is of order $\cO(\tau^{-1})$.
As a consequence, upon rescaling time by the macroscopic scale $\cT$, the following identity will hold:
\begin{equation}
    \partial_{\tilde{t}} P_{\Theta}(\Theta,\tilde{t}) = \varepsilon^{-2}\tilde{\cL}^\dagger_\Theta P_{\Theta}(\Theta,\tilde{t})\;,
    \label{eq:tilde-LTheta}
\end{equation}
with $\tilde{\cL}^\dagger_\Theta = \tau \cL^\dagger_\Theta$.
To simplify the notation, all superscript symbols $\sim$ will be omitted throughout the derivation presented in Sec.~\ref{subsec:multiscale-expansion}. Thus, unless otherwise stated, all parameters, degrees of freedom and operators therein are to be interpreted as dimensionless.
\subsection{Multiscale expansion}
\label{subsec:multiscale-expansion}
Similarly to the adimensional equations of motion reported in Eqs.~\eqref{eq:dimensionless_SDE_com} and~\eqref{eq:dimensionless-SDE-chi}, we regroup all the contributions to the forward Kolmogorov operator based on their order in the small parameter $\varepsilon$, namely:
\begin{equation}
    \partial_{t} P(\bfR, \chi,\Theta, t) = [\varepsilon^{-2}
    \cL_0^{\dagger} + \varepsilon^{-1} \cL_1^{\dagger} +  \cL_2^{\dagger}] P + \cO(\varepsilon) \;.
    \label{eq:tilde-forward-Kolmogorov}
\end{equation}
In the following, we will omit the corrections of order $\cO(\varepsilon)$ and truncate the expansion up to order $\cO(\varepsilon^0)$.
For analytical reasons that will be clearer in the coming derivation, we find convenient to work within the framework of the backward Kolmogorov equation.
Therefore, by computing the adjoint of each forward operator, we get:
\begin{equation}
\partial_{t} W(\bfR, \chi,\Theta, t) = [\varepsilon^{-2}
    \cL_0 + \varepsilon^{-1} \cL_1+  \cL_2] W\;,
    \label{eq:backward-kolmogorov}
\end{equation}
with:
\begin{align}
    \label{eq:backward-operatorsL0}
    \cL_0 &= \cL_{\Theta} + \sum_{j=1}^{N-1}\cV_j^{0\alpha}\partial^\alpha_{\bchi_j} +\sum_{j=1}^{N-1}D_{\rm t}\partial^\alpha_{\tilde\bchi_j}\partial^\alpha_{\tilde\bchi_j}\;,
    \\
    \label{eq:backward-operatorsL1}
    \cL_1 & = \cV_0^{0\alpha}\partial^\alpha + \sum_{j=1}^{N-1}\cV_j^{1\alpha}\partial^\alpha_{\bchi_j}\;, \\
    \cL_2 & = \cV_0^{1\alpha}\partial^\alpha + \sum_{j=1}^{N-1}\cV_j^{2\alpha}\partial^\alpha_{\bchi_j} + \frac{D_{\rm t}}{N}\partial^\alpha\partial^\alpha\;.
    \label{eq:backward-operators}
\end{align}
The function $W(\bfR, \chi,\Theta, t)$ introduced in Eq.~\eqref{eq:backward-kolmogorov} can be interpreted as the conditional expected value at time $t$ of a generic observable $O(\bfR(t), \chi(t),\Theta(t))$, given the initial condition $(\bfR(0)=\bfR, \chi(0)=\chi,\Theta(0)=\Theta)$.
Thus, while the forward Kolmogorov equation in Eq.~\eqref{eq:tilde-forward-Kolmogorov} evolves the probability measure $P(\bfR, \chi, \Theta,t)$, its associated backward equation governs the time-evolution of the observables.
In the spirit of multiscale expansion, we rewrite $W$ as:
\begin{equation}    
    W = W_0 + \varepsilon W_1 + \varepsilon^2W_2 + \cO(\varepsilon^3)\;.
    \label{eq:W-expansion}
\end{equation}
By adopting a perturbative approach, our ultimate goal is to perform an adiabatic elimination of the fast variables $(\chi,\Theta)$ and to derive an approximate backward equation for $W_0$ valid at leading order in $\varepsilon$. 
By combining Eqs.~\eqref{eq:backward-kolmogorov} and~\eqref{eq:W-expansion} we get the following set of coupled differential equations:
\begin{align}
    \mathcal{L}_0 W_0 &= 0 \;, \label{eq:DE-W0}\\[0.2cm]
    \mathcal{L}_0 W_1 &= -\mathcal{L}_1 W_0 \;, \label{eq:DE-W1}\\[0.2cm]
    \mathcal{L}_0 W_2 &= \partial_{t} W_0 - \mathcal{L}_1 W_1  - \mathcal{L}_2 W_0 \;.\label{eq:DE-W2}
\end{align}
Before proceeding, a few comments are in order. 
The backward operator $\cL_0$ is to be interpreted as a differential operator in the fast variables $(\chi,\Theta)$, with a parametric dependence on the center of mass $R^\alpha$ introduced by the active coupling.
Following Ref.~\cite{pavliotis2008multiscale}, we assume that the stochastic process associated to $\cL_0$ is ergodic, and thus $\cL_0$ is characterized by a one-dimensional null space.
As a result, the homogeneous differential equation in Eq.~\eqref{eq:DE-W0} is uniquely solved by constant functions parametrized by $R^\alpha$.
The ergodicity assumption also implies that the stochastic process generated by $\cL_0^\dagger$ admits a unique steady-state probability distribution, that we denote with $\Psi(\chi,\Theta;\bfR)$.
Thus, we have:
\begin{align}
    \cL_0f(\bfR) &= 0\;, \\
    \cL_0^\dagger \Psi(\chi,\Theta;\bfR)&=0\;,
    \label{eq:ergodic-L0}
\end{align}
with $f$ being a generic function of the center of mass $R^\alpha$.
Hence, from Eq.~\eqref{eq:DE-W0}, we obtain that:
\begin{equation}
    \cL_0 W_0 = 0 \quad \Leftrightarrow \quad W_0 = W_0(\bfR,t) \;,
\end{equation}
i.e. the lowest-order function $W_0$ does not depend on the fast variables.
In the following, we additionally assume the \emph{centering condition} on $\cV_0^{0\alpha}$, namely:
\begin{equation}
    \llangle \cV_0^{0\alpha}(\bfR,\Theta) \rrangle_\Psi = \llangle \cV_0^{0\alpha}(\bfR,\Theta) \rrangle_\psi = 0\;.
    \label{eq:centering-condition}
\end{equation}
In our problem, this amounts to requiring that all the active forces $\langle \rm v_i^\alpha(\bfR, \bfu_i) \rangle_{\psi} = 0$. Note that this is satisfied for all the cases studied in this work, under the assumption that $\langle u_i^\alpha\rangle_\psi=0$.
\if{
Similarly to~\cite{pavliotis2008multiscale}, we also assume that $\cV_0^{0\alpha}$ satisfies the so-called \emph{centering condition}, i.e., we have:
\begin{equation}
    \llangle \cV_0^{0\alpha}(\bfR,\Theta) \rrangle_\Psi = \llangle \cV_0^{0\alpha}(\bfR,\Theta) \rrangle_\psi = 0\;.
    \label{eq:centering-condition}
\end{equation}
In our problem, this amounts to the requirement that all active couplings $\rm v_i^\alpha(\bfR, \bfu_i)$ vanish when averaged over the stationary distribution $\psi(\Theta)$ of the orientational degrees of freedom.
We emphasize here that one of the most commonly adopted active couplings in the context of self-propelled particles, namely ${\rm v}_i^\alpha(\bfR,\bfu_i)=v(\bfR)u_i^\alpha$, trivially satisfies the centering condition of Eq.~\eqref{eq:centering-condition}.
Indeed, as anticipated in Sec.~\ref{sec:model}, our derivation relies on the assumption that $\langle u_i^\alpha\rangle_\psi=0$.\par
}\fi

We now shift our attention to the differential equation of order $\cO(\varepsilon^{-1})$, given by Eq.~\eqref{eq:DE-W1}.
By using the definition of the backward operator $\cL_1$ and the fact that $W_0$ does not depend on the higher-order Rouse modes $\chi$, we can rewrite Eq.~\eqref{eq:DE-W1} as:
\begin{equation}
    \cL_0W_1 = -\cV_0^{0\alpha} \partial^\alpha W_0\,.
    \label{eq:DE-W1-v2}
\end{equation}
Also in this case, the center of mass $R^\alpha$ appears only parametrically in the differential operator $\cL_0$.
According to the Fredholm alternative (see, e.g.,~\cite{pavliotis2008multiscale} or Appendix~\ref{app:fredholm}), a necessary condition for Eq.~\eqref{eq:DE-W1-v2} to admit a solution is that its right hand side is orthogonal to the null space of $\cL_0^\dagger$.
In light of Eq.~\eqref{eq:ergodic-L0}, such null space is spanned by the stationary density $\Psi(\chi,\Theta)$.
Hence, we can easily show that Eq.~\eqref{eq:DE-W1-v2} satisfies the Fredholm alternative as:
\begin{equation}
    \langle -\cV_0^{0\alpha} \partial^\alpha W_0 \rangle_\Psi=-\partial^\alpha W_0\langle \cV_0^{0\alpha}  \rangle_\Psi=0\;,
    \label{eq:fredholm-alterntaive}
\end{equation}
where the last equality follows from the centering condition of Eq.~\eqref{eq:centering-condition}.
It can be easily verified that the formal solution of Eq.~\eqref{eq:DE-W1-v2} is given by:
\begin{equation}
        W_1 = \Phi^\alpha(\bfR, \chi, \Theta)  \partial^\alpha W_0\;,
    \label{eq:formal-solution-W1}
\end{equation}
where the newly introduced vector field $\Phi^\alpha$ satisfies:
\begin{equation}
    - \cL_0 \Phi^\alpha(\bfR, \chi, \Theta) = \cV_0^{0\alpha}(\bfR,\Theta) \;, \quad \text{with}  \quad \llangle \Phi^\alpha \rrangle_\Psi = 0.
    \label{eq:cell-problem}
\end{equation}
Within the theory of periodic homogenization for PDEs, the above equation goes under the name of \emph{cell problem}.
For the moment, we keep $\Phi^\alpha$ unspecified and we defer the solution of the cell problem to Sec.~\ref{subsec:cell-problem}.
The formal solution given by Eq.~\eqref{eq:formal-solution-W1} expresses $W_1$ as a differential operator acting on $W_0$, with $\Phi^\alpha$ to be determined according to Eq.~\eqref{eq:cell-problem}.
We proceed by combining Eq.~\eqref{eq:formal-solution-W1} with the differential equation of order $\cO(\varepsilon^0)$ reported in Eq.~\eqref{eq:DE-W2}. This yields:
\begin{equation}
    \mathcal{L}_0 W_2 = \partial_{t} W_0 - \mathcal{L}_1 \Phi^\alpha  \partial^\alpha W_0  - \mathcal{L}_2 W_0 \,.
    \label{eq:DE-W2-v2}
\end{equation}
Once again, a necessary condition for the existence of a solution to Eq.~\eqref{eq:DE-W2-v2} is that its right hand side is orthogonal to the null space of $\cL_0^\dagger$. In other words, we require:
\begin{equation}
\llangle \partial_t W_0 - \cL_1 [ \Phi^\beta \partial^\beta W_0 ] - \cL_2 W_0\rrangle_\Psi = 0\;.
\label{eq:fredholm-alternative-2}
\end{equation}
This means that the solvability condition~\eqref{eq:fredholm-alternative-2} provides us a backward reduced dynamics for $W_0$ that does not depend on the fast variables $\chi$ and $\Theta$.
Indeed, the latter are integrated out due to the averaging with the stationary density $\Psi(\chi,\Theta)$.

We now separately deal with each of the three contributions of Eq.~\eqref{eq:fredholm-alternative-2}.
Since $W_0$ does not depend on the fast variables, it is straightforward to see that $\langle \partial_tW_0 \rangle_\Psi = \partial_tW_0$. 
The second term reads:
\begin{align}
&\llangle \cL_1 [ \Phi^\beta \partial^\beta W_0]\rrangle_\Psi  \label{eq:reduced-W0-term2}
\\[0.1cm]&=  \sum_{j=1}^{N-1}\big\langle \cV_j^{1\alpha} \partial^\alpha_{\bchi_j} \Phi^\beta \big\rangle_\Psi \partial^\beta W_0 + \llangle \cV_0^{0\alpha} \partial^\alpha  [\Phi^\beta \partial^\beta W_0 ] \rrangle_\Psi \n\\[0.1cm]
    &=  \left[\sum_{j=1}^{N-1}\big\langle \cV_j^{1\alpha} \partial^\alpha_{\bchi_j} \Phi^\beta \big\rangle_\Psi  +  \big\langle \cV_0^{0\alpha} \partial^\alpha  \Phi^\beta  \big\rangle_\Psi\right]\partial^\beta W_0 \n \\[0.1cm]&+\llangle \cV_0^{0\alpha}  \Phi^\beta   \rrangle_\Psi  \partial^\alpha\partial^\beta W_0 \;, \n
\end{align}
providing both a drift and a diffusive contribution to the reduced dynamics. Finally, the last term gives:
\begin{align}
\langle \cL_2 W_0\rangle_\Psi = \langle \cV_0^{1\alpha} \rangle_\Psi  \partial^\alpha W_0 + \frac{D_{\rm t}\delta^{\alpha \beta}}{N} \partial^\alpha \partial^\beta W_0 \;.
    \label{eq:reduced-W0-term3}
\end{align}
All in all, we can write an effective dynamics for $W_0$ as a backward Fokker-Planck equation:
\begin{equation}
    \partial_t W_0(\bfR,t) = \cL_{\rm eff} W_0(\bfR,t) \;,
    \label{eq:backward-FP}
\end{equation}
with $\cL_{\rm eff} = V^\alpha \partial^\alpha + \cD^{\beta\alpha} \partial^{\alpha} \partial^{\beta}$. Here, the effective drift $V^\alpha$ and the diffusion tensor $\cD^{\alpha\beta}$ are given by:
\begin{align}
    V^\alpha &=\sum_{j=1}^{N-1}\big\langle \cV_j^{1\beta} \partial^\beta_{\bchi_j} \Phi^\alpha \big\rangle_\Psi  +  \big\langle \cV_0^{0\beta} \partial^\beta  \Phi^\alpha  \big\rangle_\Psi + \big\langle \cV_0^{1\alpha} \big\rangle_\Psi
    \label{eq:drift-Phi}\;,\\
    \cD^{\alpha \beta} &= \frac{D_{\rm t}}{N}\delta^{\alpha \beta} + \big\langle \Phi^\alpha \cV_0^{0\beta}    \big\rangle_\Psi \;.\label{eq:diffusion-Phi}
\end{align}
Finally, by taking the adjoint of Eq.~\eqref{eq:backward-FP}, we obtain the Fokker-Planck equation for the probability density $p(\bfR,t)$ of the center of mass $R^\alpha$:
\begin{equation}
    \partial_t p = -\partial^\alpha \left[V^\alpha p -\partial^\beta(\cD^{\alpha \beta}p)\right]\;.
    \label{eq:forward-FP-Phi}
\end{equation}
To conclude our derivation, we need to determine a closed form for both $V^\alpha$ and $\cD^{\alpha\beta}$. This is done by solving the cell problem of Eq.~\eqref{eq:cell-problem} in the next section.
\subsection{The cell problem}
\label{subsec:cell-problem}
In order to solve the cell problem, we first isolate $\Phi^\alpha$ by applying the inverse backward generator $\cL_0^{-1}$ to Eq.~\eqref{eq:cell-problem}.
This is useful because, when acting on a centered function as $\cV_0^{0\alpha}$, the inverse generator can be shown to satisfy the following formal identity:
\begin{equation}
    \cL_0^{-1} = -\int_0^\infty\rmd t\,e^{\cL_0t}\,.
    \label{eq:identity-invereseL0}
\end{equation}
A derivation of Eq.~\eqref{eq:identity-invereseL0} can be found in Appendix~\eqref{app:identity-L0}.
This leads to:
\begin{equation}
    \Phi^\alpha(\bfR,\Theta) = \int_0^\infty\rmd t\,e^{\cL_0t}\cV_0^{0\alpha}(\bfR,\Theta)\,.
    \label{eq:Phi-cell_problem}
\end{equation}
We recall here that $\cL_0$ is the generator of the fast dynamics of order $\cO(\varepsilon^{-2})$, which evolves the observables associated with the corresponding fast process. 
In our case, it is useful to consider the following observable $A$ parametrized by the center of mass $R^\alpha$:
\begin{equation}
    A(\Theta,t ;\bfR) = \mathbb{E} [\cV_0^{0\alpha}(\bfR,\Theta(t)) | \Theta(0)=\Theta ] \;,
    \label{eq:O-cell-problem}
\end{equation}
where we define $\mathbb{E} [ \cdot | \Theta(0)=\Theta]$ as the average over the realizations of the $\Theta$-dynamics constrained on the initial value for $\Theta$.
Since the observable $A$ of Eq.~\eqref{eq:O-cell-problem} evolves via the backward Kolmogorov equation as $A(t) = e^{\cL_0 t} A(0)$, the solution Eq.~\eqref{eq:Phi-cell_problem} of the cell problem finally reads:
\begin{align}
    \Phi^\alpha(\bfR,\Theta) =\int_0^\infty\rmd t\, \mathbb{E}[ \cV_0^{0\alpha}(\bfR,\Theta(t)) | \Theta(0)=\Theta ]\,.
    \label{eq:Phi-cell_problem-v2}
\end{align}
\subsection{The large-scale dynamics}
\label{eq:large-scale-dynamics}
Now that we have a formal expression for the vector field $\Phi^\alpha$, we can compute all contributions appearing in the definition of the effective drift $V^\alpha$ and diffusion tensor $\cD^{\alpha \beta}$ in Eqs.~\eqref{eq:drift-Phi} and~\eqref{eq:diffusion-Phi}.
First, we note that $\Phi^\alpha$ does not depend on the higher-order Rouse modes $\chi$ and therefore all drift terms of the type $\langle \cV_j^{1\beta} \partial^\beta_{\bchi_j} \Phi^\alpha \rangle_\Psi$ vanish.
Then we have:
\begin{align}
    \big\langle \cV_0^{0\beta} \partial^\beta  \Phi^\alpha  \big\rangle_\Psi &= \big\langle \cV_0^{0\beta} \partial^\beta  \Phi^\alpha  \big\rangle_\psi \\&=\int_0^\infty\rmd t\big\langle \cV_0^{0\beta} (0)\partial^\beta \cV_0^{0\alpha} (t)\big\rangle\;, \n
\end{align}
where in the first equality we used the fact that neither $\cV_0^{0\alpha}$ nor $\Phi^\alpha$ depend on $\chi$, and in the second one the law of total expectations.
We recall here that $\langle \cdot \rangle$ without subscript denotes the average over realizations of the $\Theta$-dynamics at steady state, as introduced in Sec.~\ref{sec:model}.
Analogously, for the diffusion term we get:
\begin{equation}
    \big\langle \Phi^\alpha \cV_0^{0\beta}    \big\rangle_\Psi = \int_0^\infty\rmd t\big\langle  \cV_0^{0\alpha} (t) \cV_0^{0\beta} (0)\big\rangle\;.
\end{equation}
All in all, we obtain the following expression for the large scale drift and diffusion tensor:
\begin{eqnarray}
    \label{eq:drift_formal}
    V^\alpha &=& \langle \cV_0^{1\alpha} \rangle_{\Psi} + \int_0^\infty \rmd t\langle  \cV_0^{0\beta}(0) \partial^{\beta}  \cV_0^{0\alpha}(t)\rangle \;, \\[0.2cm]
    \cD^{\alpha\beta} &=& \frac{D_{\rm t}}{N} \delta^{\alpha\beta} + \int_0^\infty \rmd t\langle \cV_0^{0\alpha}(t) \cV_0^{0\beta}(0)  \rangle \;.
    \label{eq:diff_formal}
\end{eqnarray}
We observe that the expression for the diffusion tensor $\mathcal{D}^{\alpha \beta}$ takes the form of a generalized Green-Kubo relation \cite{green1954markoff, kubo2012statistical}, relating diffusion to the velocity autocorrelation function.
In the present case, however, only the contribution $\cV_0^{0\alpha}$ of leading order in $\varepsilon$ appears in the formula.
Finally, using the definition of the dimensionless variables and parameters introduced in Sec.~\ref{subsec:adimensional_eom}, we can restore the physical dimensions in our equations. \par
The large-scale dynamics in Eq.~\eqref{eq:forward-FP-Phi} can be specialized to the case in which the active coupling is of the form ${\rm v}_i^\alpha = v(\bfr_i^\alpha)u_i^\alpha$. This yields:
\begin{align}
\label{eq:drift_final}
    V^{\alpha} &= \frac{\partial^\beta v^2(\bfR)}{2N}\sum_{ijk}  \varphi_{ji} \varphi_{jk}  \int_0^\infty \rmd t\,  e^{-\lambda_j t} \mathbb{C}_{ik}^{\alpha \beta}(t)\;, \\
    \cD^{\alpha\beta}  &= \frac{D_{\rm t}}{N} \delta^{\alpha\beta} + \frac{v^2(\bfR)}{N^2}\sum_{ij}   \int_0^\infty \rmd t\,  \mathbb{C}_{ij}^{\alpha \beta}(t)  \;. 
\label{eq:diffusion_final}
\end{align}
These expressions are insightful because they clarify how the fast variables, once goarse-grained, affect the large-scale dynamics of the center of mass $R^\alpha$.
Equations~\eqref{eq:drift_final} and~\eqref{eq:diffusion_final} show that the only relevant quantity arising from the orientational degrees of freedom $\Theta$ is the time integral of the stationary autocorrelation function $\mathbb{C}_{ij}^{\alpha \beta}(t)$.
In contrast, the internal structure of the polymer enters the drift term $V^\alpha$ through the relaxation rates $\{\lambda_i\}$ of the Rouse modes $\chi$ and the eigenvectors $\varphi_{ij}$, thereby encoding how the microscopic configurational dynamics affects the macroscopic behavior.\par
We conclude this section by pointing out a subtle aspect related to the expression of the diffusion tensor $\cD^{\alpha\beta}$ reported in Eq.~\eqref{eq:diff_formal}.
As shown in the above derivation, our multiscale approach leads to the effective evolution of the probability density $p(\bfR,t)$, given by Eq.~\eqref{eq:forward-FP-Phi}.
However, we know that the dynamics in Eq.~\eqref{eq:forward-FP-Phi} is invariant under the gauge transformation $\cD^{\alpha\beta} \to \cD^{\alpha\beta} + \mathcal{A}^{\alpha\beta}$, with $\mathcal{A}^{\alpha\beta}$ being any anti-symmetric tensor.
While the choice of the gauge does not affect the probability density $p(\bfR,t)$ at any time $t$, it does have an impact on particle fluxes.
Nevertheless, it is possible to fix $\cA^{\alpha\beta}=0$ by requiring $\cD^{\alpha\beta}$ to be compatible with the generalized Green-Kubo formula derived by Hargus and co-authors~\cite{hargus2021odd}.
In the following we adopt this gauge choice, and thoroughly test it against particle-based simulations of chiral active particles and polymers.

\newpage 

\section{Generalization of the adiabatic elimination method}
\label{sec:generalization}
The adiabatic elimination technique described in Sec.~\ref{sec:adiabatic_elimination} is a versatile tool that can be readily extended to account for more general cases than the one analyzed up to this point of the discussion.
The aim of this section is to illustrate some of the possible generalizations.
In particular, we focus on the following cases: in Sec.~\ref{subsec:non-markov} we show how the previous derivation can be adapted to systems whose rotational degrees of freedom $\Theta$ follow a dynamics with memory, yet can still be cast in a Markovian form through the introduction of auxiliary degrees of freedom (e.g., via a Markovian embedding).
In Sec.~\ref{subsec:adiabatic_elimination-taxis}, we examine the case of tactic motility regulation, for which the self-propulsion speed of the individual monomers is modulated by the alignment between their orientation and the gradient of an external biasing field.
Finally, in Sec.~\ref{subsec:heteropolymers}, we extend the derivation to heteropolymers, namely systems characterized by heterogeneous mobilities.

\subsection{Non-Markov fast process with Markovian embedding}
\label{subsec:non-markov}
The derivation presented in Sec.~\ref{sec:adiabatic_elimination} was restricted to Markov processes for the orientational degrees of freedom $\Theta$, described by a Markov generator $\cL_\Theta$ for the backward Kolmogorov equation. This assumption excludes physically relevant cases in which the orientational dynamics is effectively non-Markovian, such as systems with rotational inertia~\cite{scholz2018inertial,lowen2020inertial,caprini2022role,lisin2022motion,sprenger2023dynamics} or intermittent activity
\cite{berg2004coli,detcheverry2017generalized,shaebani2022kinematics,datta2024random,caraglio2024learning,santra2024dynamics}.

To overcome this limitation, here we consider a non-Markov dynamics for $\Theta$ which admits a Markovian representation upon introducing a set of auxiliary variables $Z$. As before, we assume that the dynamics of $(\Theta, Z)$ has no dependence on the position variables $\{r_i^\alpha\}$. Specific examples are discussed in Sec.~\ref{subsec:ABPinertia}--~\ref{subsec:multistate}. 
We extend the Markov generator $\cL_{\Theta}$ of Sec.~\ref{sec:adiabatic_elimination} to a generator $\cL_{\Theta,Z}$ that evolves the observables $W(\Theta, Z)$ via the associated backward Kolmogorov equation:
\begin{equation}
    \partial_t W(\Theta,Z) = \cL_{\Theta,Z} W(\Theta,Z) \;.
\end{equation}
Furthermore, we assume that the dynamics of the additional degrees of freedom $Z$ occurs on the same, fast timescale $\tau$ as $\Theta$. Formally, this implies that, within the diffusive scaling of Sec.~\ref{subsec:diffusive-regime}, the generator can be written as $\cL_{\Theta,Z} = \tau \tilde\cL_{\Theta,Z}$, where $\tilde\cL_{\Theta,Z} \sim \cO(1)$. To lighten the notation, in the following we omit the superscript $\sim$ for the non-dimensionalized quantities, but implicitly assume it throughout the rest of the derivation.

The non-dimensionalized backward-Kolmogorov equation for the full process directly extends Eq.~\eqref{eq:backward-kolmogorov} as:
\begin{equation}
\partial_{t} W(\bfR, \chi,\Theta,Z, t) = [\varepsilon^{-2}
    \cL_0 + \varepsilon^{-1} \cL_1+  \cL_2] W\;,
    \label{eq:backward-kolmogorov2}
\end{equation}
where $\cL_1, \>\cL_2$ are still given by Eqs.~\eqref{eq:backward-operatorsL1},~\eqref{eq:backward-operators}, while $\cL_0$ reads:
\begin{align}
    \cL_0 &= \cL_{\Theta,Z} + \sum_{j=1}^{N-1}\cV_j^{0\alpha}\partial^\alpha_{\bchi_j} +\sum_{j=1}^{N-1}D_{\rm t}\partial^\alpha_{\tilde\bchi_j}\partial^\alpha_{\tilde\bchi_j}\;.
\end{align}
Note that, under our assumption for the timescale of evolution of $Z$ and $\Theta$, the auxiliary variables $Z$ enter only at the fastest order $\varepsilon^{-2}$, and hence only affect the operator $\cL_0$. Consequently, all expressions of $\cV_{i}^{0\alpha}$ and $\cV_i^{1\alpha}$ are still given by Eqs.~\eqref{eq:expansion_cV0}--\eqref{eq:expansion_cVi}.

Upon re-definition of $\Psi(\chi, \Theta, Z; \bfR)$ as the invariant measure associated with $\cL^{\dagger}_0$, all the steps of the derivation carried out in Sec.~\ref{sec:adiabatic_elimination} up to Eqs.~\eqref{eq:drift-Phi}--\eqref{eq:diffusion-Phi} thus remain valid. 
The solution $\Phi^{\alpha}$ to the cell problem, Eq.~\eqref{eq:cell-problem}, is now a function of both $(\Theta,Z)$. Hence, the derivative of $\Phi^{\alpha}$ with respect to the Rouse modes $\chi$ is still equal to zero. Finally, using the representation of $\Phi^{\alpha}$ in Eq.~\eqref{eq:Phi-cell_problem-v2}, we obtain:
\begin{eqnarray}
    \label{eq:drift_formal_nonmarkov}
    V^\alpha &=& \langle \cV_0^{1\alpha} \rangle_{\Psi} + \int_0^\infty \rmd t\langle  \cV_0^{0\beta}(0) \partial^\beta \cV_0^{0\alpha}(t) \rangle \;, \\[0.2cm]
    \cD^{\alpha\beta} &=& \frac{D_{\rm t}}{N} \delta^{\alpha\beta} + \int_0^\infty \rmd t \langle  \cV_0^{0\beta}(0) \cV_0^{0\alpha}(t) \rangle \;.
    \label{eq:diff_formal_nonmarkov}
\end{eqnarray}
Note that, in this case, the average $\langle \cdot \rangle$ is to be interpreted under the stationary path probability associated with the fast process for $(\chi, \Theta, Z)$.

\subsection{Tactic interactions}
\label{subsec:adiabatic_elimination-taxis}

The goal of this section is to extend our derivation to tactic motility regulation, whereby active agents adapt their motility to the gradients of some external cue $c(\bfr)$. Taxis is widespread in biological systems, from motile bacteria like \textit{E. Coli}~\cite{budrene1991complex,schnitzer1993theory,berg2004coli} to phototactic algae~\cite{foster1980light} and cells responding to pressure gradients via durotaxis~\cite{ji2023durotaxis}. In the context of synthetic active matter, self-phoretic colloids provide an example of how tactic motility regulation can drive the dynamics of active agents~\cite{howse2007self,soto2014self,saha2014clusters}. 

From the statistical-mechanics perspective, the comparison between space-dependent self-propulsion speed and taxis raises the question on whether different strategies of motility regulation can give rise to equivalent macroscopic descriptions, and under which conditions this occurs~\cite{dinelli2024fluctuating}. Furthermore, previous works in the literature highlighted the existence of a large-scale equivalence between tactic active particles and equilibrium Brownian particles~\cite{o2020lamellar}. While these studies have considered the single-particle case, it remains unclear to which extent such large-scale mappings are valid, in particular for more complex scenarios such as the one of active polymers.

To tackle these questions, we consider an active polymer whose dynamics is described by Eq.~\eqref{eq:SDE-monomers}, where the active terms ${\rm v}_i^\alpha(\bfr_i,\bfu_i)$ are given by~\cite{o2020lamellar}:
\begin{equation}
    {\rm v}_i^\alpha(\bfr_i,\bfu_i) = [v_0 - v_1 u_i^\beta \partial_{\bfr_i}^\beta c(\bfr_i)] u^\alpha_i \;,
    \label{eq:taxis-definition}
\end{equation}
with $c(\bfr)$ being a (non-dimensional) external biasing field. 
According to Eq.~\eqref{eq:taxis-definition}, particles with $v_1>0$ decrease their self-propulsion speed when pointing in the direction of $\nabla_{\bfr_i} c$, and increase their speed otherwise.
For the sake of simplicity, in the following we take $v_0, v_1$ to be constant, but note that extensions of our coarse-graining method to space-dependent parameters is straightforward.

Our aim is now to derive a large-scale Fokker-Planck description for the center of mass $\bfR$ of a tactic polymer. The only difference with respect to the derivtion of Sec.~\ref{sec:adiabatic_elimination} now lies in how $ {\rm v}_i^\alpha(\bfr_i,\bfu_i) $ enters the diffusive scaling. 
More precisely, we assume that $c(\bfr)$ varies over a lengthscale $\delta$ of the order of the system size, $\delta \sim L$, which is assumed to be much larger than typical microscopic lengthscale $\ell$ (see Sec.~\ref{subsec:diffusive-regime}). Our scale separation thus implies that the definition of the small parameter $\varepsilon$ in Eq.~\eqref{eq:scaling} remains valid.
Consequently, the active term now contains both a contribution of order $\cO(v_0)$ and a tactic correction of order $\cO(\varepsilon v_0)$.

To proceed, we thus non-dimensionalize the dynamics by introducing the variables $\tilde{R}^\alpha=R^\alpha/\delta$, $\tilde{\chi}_i^\alpha=\chi_i^\alpha/\ell$ and $\tilde{t} = t/\cT$. In the active term, we introduce $\tilde{v}_0 = (\tau/\ell) v_0$, $\tilde{v}_1 = (\tau/\ell^2) v_1$, and $\tilde c(\bfr_i/\delta) = c(\bfr_i)$. Upon rescaling the active force as $\tilde{{\rm v}}_i^{\alpha}(\bfr_i/\delta, \bfu_i) \equiv (\tau/\ell){\rm v}_i^\alpha(\bfr_i, \bfu_i)$, we eventually obtain:
\begin{align}
    &\tilde{{\rm v}}_i^{\alpha}(\bfr_i/\delta, \bfu_i) = [ \tilde{v}_0 - \varepsilon \> \tilde{v}_1 u_i^\beta \partial^\beta \tilde{c}(\tilde\bfR) ] u^\alpha_i+ \cO(\varepsilon^2)\;,
    \label{eq:expansion-tactic-coupling}
\end{align}
where we used the notation $\tilde{\partial}^\alpha \equiv \partial^\alpha_{\tilde{\bfR}}$.

As we did for the case of space-dependent self-propulsion speed, we multiply Eq.~\eqref{eq:SDE-com} by $\cT/\delta = \varepsilon^{-1} \tau/\ell$ and use Eq.~\eqref{eq:expansion-tactic-coupling} to obtain the dimensionless dynamics of the center of mass:
\begin{equation}
\partial_{\tilde{t}}\tilde{R}^\alpha = \frac{1}{\varepsilon}\mathcal{V}_0^{0\alpha}+ \mathcal{V}_0^{1\alpha} + \sqrt{2\tilde{D}_{\rm t}N^{-1}}\tilde{\eta}^\alpha_0 +\cO(\varepsilon)\;,
\label{eq:dimensionless_SDE_com-taxis}
\end{equation}
where:
\begin{align}
\label{eq:expansion_cV0-taxis}
\mathcal{V}_0^{0\alpha}(\Theta) &= \frac{\tilde{v}_0}{N} \sum_{i=0}^{N-1} u^\alpha_i\;,\\
\mathcal{V}_0^{1\alpha}(\tilde{\bfR}, \Theta) &= - \frac{\tilde{v}_1}{N} \sum_{i=0}^{N-1} u_i^\alpha u_i^\beta \tilde\partial^\beta \tilde{c}\;.
\label{eq:expansion_cV1-taxis}
\end{align}
Here, the dimensionless translational diffusion coefficient is $\tilde{D}_{\rm t}=D_{\rm t}\tau/\ell^2$ and the dimensionless noise $\tilde{\eta}^\alpha_0 = \eta^\alpha_0\cT^{1/2}$. Non-dimensionalization of the equation of motion~\eqref{eq:SDE-chi} for the Rouse modes $\chi$ yields, in the tactic case:
\begin{equation}
    \partial_{\tilde{t}}\tilde{\chi}_i^{\alpha} = \frac{1}{\varepsilon^2}\mathcal{V}_i^{0\alpha} + \frac{1}{\varepsilon}\mathcal{V}_i^{1\alpha} +\mathcal{V}_i^{2\alpha}(\tilde{\bfR},\tilde{\chi}, \Theta)+ \frac{1}{\varepsilon}\sqrt{2\tilde{D}_{\rm t}}\tilde{\eta}^\alpha_i + \cO(\varepsilon)
    \label{eq:dimensionless-SDE-chi-taxis}
\end{equation}
with:
\begin{align}
    \mathcal{V}_i^{0\alpha}(\tilde{\chi}, \Theta) &= -\tilde{\lambda}_i\tilde{\chi}_i^{\alpha} + \sum_{j=0}^{N-1}\varphi_{ij} \tilde{v}_0 u^\alpha_j \;, \\
    \mathcal{V}_i^{1\alpha}(\tilde{\chi}, \Theta) &= -\sum_{j=0}^{N-1} \varphi_{ij} \tilde{v}_1 u_j^\alpha u_j^\beta \tilde\partial^\beta \tilde{c} \;.
    \label{eq:expansion_cVi-taxis}
\end{align}
We note that the explicit expression of $\mathcal{V}_i^{2\alpha}(\tilde{\bfR},\tilde{\chi}, \Theta)$ is not relevant for the rest of the computation, but can be obtained by a higher-order perturbative expansion of the active force. In the following, we omit the $\sim$ everywhere, but implicitly assume non-dimensionalized variables unless stated otherwise. 

The adiabatic elimination of the fast variables $(\Theta, \chi)$ proceeds as in Sec.~\ref{subsec:multiscale-expansion}: we write down the backward Kolmogorov equation up to $\cO(1)$ and solve for each order in $\varepsilon$:
\begin{equation}
    \partial_t W = [\varepsilon^{-2} \cL_0 + \varepsilon^{-1} \cL_1 + \cL_2] W \;,
\end{equation}
where:
\begin{align}
    \label{eq:backward-operatorsL0-taxis}
    \cL_0 &= \cL_{\Theta} + \sum_{j=1}^{N-1}\cV_j^{0\alpha}\partial^\alpha_{\bchi_j} +\sum_{j=1}^{N-1}D_{\rm t}\partial^\alpha_{\tilde\bchi_j}\partial^\alpha_{\tilde\bchi_j}\;,
    \\
    \label{eq:backward-operatorsL1-taxis}
    \cL_1 & = \cV_0^{0\alpha}\partial^\alpha + \sum_{j=1}^{N-1}\cV_j^{1\alpha}\partial^\alpha_{\bchi_j}\;, \\
    \cL_2 & = \cV_0^{1\alpha}\partial^\alpha + \sum_{j=1}^{N-1}\cV_j^{2\alpha}\partial^\alpha_{\bchi_j} + \frac{D_{\rm t}}{N}\partial^\alpha\partial^\alpha\;.
    \label{eq:backward-operators-taxis}
\end{align}
As a consequence, the formal solution obtained in Eqs.~\eqref{eq:drift_formal},~\eqref{eq:diff_formal} remains valid. \if{, yielding:
\begin{eqnarray}
    \label{eq:drift_formal_taxis}
    V^\alpha &=& \langle \cV_0^{1\alpha} \rangle_{\Psi} + \int_0^\infty \rmd t\langle  \cV_0^{0\beta}(0) \partial^{\beta}  \cV_0^{0\alpha}(t)\rangle \;, \\[0.2cm]
    \cD^{\alpha\beta} &=& \frac{D_{\rm t}}{N} \delta^{\alpha\beta} + \int_0^\infty \rmd t\langle  \cV_0^{0\beta}(0) \cV_0^{0\alpha}(t) \rangle \;.
    \label{eq:diffusion_formal_taxis}
\end{eqnarray}
}\fi
By using the explicit expressions~\eqref{eq:expansion_cV0-taxis},~\eqref{eq:expansion_cV1-taxis}, we eventually obtain a closed expression for the macroscopic drift and diffusion tensor of a tactic polymer: 
\begin{eqnarray}
    \label{eq:drift_final_taxis}
    V^\alpha &=& -\frac{v_1}{N} \sum_{i=0} \mathbb{C}^{\alpha\beta}_{ii}(0) \partial^\beta c \;, \\[0.2cm]
    \cD^{\alpha\beta} &=& \frac{D_{\rm t}}{N} \delta^{\alpha\beta} +\frac{v_0^2}{N^2}  \sum_{i,j=0}^{N-1} \int_0^\infty \rmd t \>\mathbb{C}^{\alpha\beta}_{ij}(t) \;.
    \label{eq:diffusion_final_taxis}
\end{eqnarray}
The macroscopic description of the dynamics of the center of mass is thus complete, together with the associated Fokker-Planck equation for the probabilty distribution $p(\bfR)$:
\begin{equation}
    \partial_t p(\bfR,t) = -\partial^\alpha [V^\alpha p - \partial^\beta(\cD^{\alpha\beta} p)]\;.
    \label{eq:FP_macroscopic-taxis}
\end{equation}
By re-instating dimensional units, the set of equations~\eqref{eq:drift_final_taxis}-\eqref{eq:FP_macroscopic-taxis} takes the same form.

Our macroscopic equations reveal a significant difference between space-dependent self-propulsion speed and taxis at the level of the drift, as can be seen by comparing Eq.~\eqref{eq:drift_final} and Eq.~\eqref{eq:drift_final_taxis}. In the former case, the drift is affected by the time integral of the auto-correlation function, and depends on the structure of the polymer via both the cross-correlations for different $i \neq j$ and the eigenvectors $\varphi_{ij}$ of the connectivity matrix $M_{ij}$. In the case of taxis, instead, only the single-monomer orientational covariance enters the drift, which is thus not sensitive to the full time-dependence of $\mathbb{C}^{\alpha\beta}(t)$. While in the simplest examples of single-particle dynamics ($N=1$), the emergent phenomenology is equivalent for the two modes of motility regulation, such an equivalence is violated for active polymers. We explore the consequences of such inequivalence in Sec.~\ref{sec:equilibrium}, where we study emergent equilibrium descriptions for space-dependent self-propulsion speed and taxis, and for explicit coarse-grained descriptions of active Rouse systems in Sec.~\ref{sec:cg_polymer}.

\subsection{Heteropolymers}
\label{subsec:heteropolymers}
Thus far, our coarse-graining framework has been restricted to polymer models in which all monomers experience the same friction coefficient. This simplifying assumption excludes all systems where the microscopic units differ in size, namely heteropolymers. For instance, active filaments transporting attached loads~\cite{wang2012nano,winkler2017active} and microswimmer–cargo assemblies~\cite{jain2022cargo,vuijk2021chemotaxis} naturally combine propulsion with elements of different size and drag. 

As we detail in this section, our coarse-graining method enables us to directly assess the macroscopic drift $V^\alpha$ and diffusion tensor $\cD^{\alpha\beta}$ for this class of systems. Note that the only specificity of this problem enters in the derivation of the diffusive dynamics to the correct orders in $\varepsilon$. Once this is done, the formal results of Sec.~\ref{sec:adiabatic_elimination} can be directly used to derive $V^\alpha$ and $\cD^{\alpha\beta}$, allowing us to study the properties of heteropolymers and the impact of passive loads on their emergent dynamics.

We start by modifying the dynamics Eq.~\eqref{eq:SDE-monomers} to account for different friction coefficients $\{\gamma_i\}$ for each bead $i$:
    \begin{eqnarray}
    \dot r^\alpha_i &=& -\gamma_i^{-1} \partial^\alpha_{\bfr_i}\mathcal{H} +{\rm v}^\alpha_i(\bfr_i,\bfu_i) + \sqrt{\frac{2 k_{\rm B} T}{\gamma_{i}} } \xi^\alpha_i(t) \;,
\label{eq:SDE_multi_friction}
\end{eqnarray}
where the Hamiltonian is given by Eq.~\eqref{eq:Hamiltonian}, with $M_{ij}$ being the symmetric connectivity matrix of the system. 
In the overdamped dynamics Eq.~\eqref{eq:SDE_multi_friction}, the only coordinate which does not undergo the effect of internal forces is the center of friction $\bfR$, defined as:
\begin{equation}
    R^\alpha = \frac{\sum_i \gamma_i r^\alpha_i}{\sum_i \gamma_i} = \frac{1}{\Gamma} \sum_{i=0}^{N-1} \gamma_i r^\alpha_i \;,
\end{equation}
where $\Gamma = \sum_i \gamma_i$ is the total friction coefficient. 
This coordinate is thus the slow mode whose large scale dynamics we want to derive. Note that, when all particles are equal, $\bfR$ coincides with the standard definition of the center of mass. Its dynamics reads:
\begin{equation}
\dot{R}^\alpha = \sum_{i=0}^{N-1} \frac{\gamma_i}{\Gamma} \, {\rm v}_i^\alpha(\bfr_i,\bfu_i) + \sqrt{\frac{2 k_{\rm B} T}{\Gamma}} \eta^\alpha_0(t) \;.
\label{eq:center-of-friction-dyn}
\end{equation}
We now introduce the Rouse modes $\{\chi\}$ for this model, i.e. the fast variables describing the internal structure of the polymer. Since the dynamical matrix $\kappa \gamma_i^{-1} M_{ij}$ is not symmetric, it is convenient to introduce the auxiliary coordinates $w_i^\alpha = \sqrt{ N \gamma_i/\Gamma} \> r_i^\alpha$, for which the dynamics induced by the internal forces reads $\dot w_i^\alpha = - \frac{N\kappa}{\Gamma} \sum_j \sqrt{\frac{\Gamma}{N\gamma_i}} M_{ij} \sqrt{\frac{\Gamma}{N\gamma_j}} w^\alpha_j$. The rescaled matrix:
\begin{equation}
    \tilde{M}_{ij} := \sqrt{\frac{\Gamma}{N \gamma_i}} M_{ij} \sqrt{\frac{\Gamma}{N \gamma_j}}
\end{equation}
is the symmetric generator of the internal dynamics for $w_i$, and can thus be diagonalized by an orthonormal matrix $\varphi_{ij}$:
\begin{equation}
    \sum_{j,k} \varphi_{ij} \tilde{M}_{jk} \varphi_{lk} = \sigma_i \delta_{il} \;, \quad \text{with} \quad \sum_{j=0}^{N-1} \varphi_{ij} \varphi_{kj} = \delta_{ik} \;.
    \label{eq:phi-def-multifriction}
\end{equation}
Note that the eigenvalue $\lambda_0 = 0$ has $\varphi_{0j} = \sqrt{\gamma_j / \Gamma}$. The Rouse modes can then be built from the auxiliary variables following Eq.~\eqref{eq:rouse-modes-def}, and read:
\begin{equation}
    \chi^\alpha_i = \sum_{j=0}^{N-1} \varphi_{ij} w^\alpha_j = \sum_{j=0}^{N-1} \varphi_{ij} \sqrt\frac{N \gamma_j}{\Gamma} r^\alpha_j \;.
\end{equation}  
We note that the center of friction is given by $R^\alpha = \chi^\alpha_0/\sqrt{N}$, as can be verified by direct substitution.  
Using the orthonormality condition in Eq.~\eqref{eq:phi-def-multifriction}, we can invert the relation between $r$ and $\chi$:
\begin{equation}
    r^\alpha_i = \sqrt{\frac{\Gamma}{N \gamma_i}} \sum_{j=0}^{N-1} \varphi_{ji} \chi^\alpha_j = R^\alpha + \sqrt{\frac{\Gamma}{N \gamma_i}} \sum_{j=1}^{N-1} \varphi_{ji} \chi^\alpha_j\;.
\end{equation}
Applying the operator $\varphi_{ij} \sqrt{N \gamma_j/\Gamma}$ to Eq.~\eqref{eq:SDE_multi_friction} finally yields
the dynamics of the Rouse modes:
\begin{equation}
    \dot{\chi^\alpha_i} = - \lambda_i \chi^\alpha_i + \sum_{j=0}^{N-1} \varphi_{ij} \sqrt{\frac{N \gamma_j}{\Gamma}} {\rm v}_j + \sqrt{\frac{2 N k_{\rm B} T}{\Gamma}} \eta^\alpha_i(t) \;,
    \label{eq:chi_dynamics-multi_friction}
\end{equation}
where the relaxation rates are given by $\lambda_i = \frac{N\kappa}{\Gamma} \sigma_i$.

Now that we have defined the dynamics of the center of friction, Eq.~\eqref{eq:center-of-friction-dyn}, and of the Rouse modes, Eq.~\eqref{eq:chi_dynamics-multi_friction}, we can apply the diffusive scaling presented in Sec.~\ref{subsec:diffusive-regime} and obtain the forward Kolmogorov operators to different order in $\varepsilon$. In the following, we focus on the case of space-dependent self-propulsion speed ${\rm v}_i(\bfr_i,\bfu_i) = v(\bfr_i) \bfu_i$ and derive the corresponding macroscopic dynamics. 
The perturbative coarse-graining procedure then strictly follows the one provided in Sec.~\ref{sec:adiabatic_elimination}, yielding the same formal result for the drift velocity and the diffusion tensor:
\begin{eqnarray}
    \label{eq:drift_formal_multi-friction}
    V^\alpha &=& \langle \cV_0^{1\alpha} \rangle_{\Psi} + \int_0^\infty \rmd t\langle  \cV_0^{0\beta}(0) \partial^{\beta}  \cV_0^{0\alpha}(t)\rangle \;, \\[0.2cm]
    \cD^{\alpha\beta} &=& \frac{k_{\rm B} T}{\Gamma} \delta^{\alpha\beta} + \int_0^\infty \rmd t\langle \cV_0^{0\beta}(0)  \cV_0^{0\alpha}(t)\rangle \;,
    \label{eq:diff_formal_multi-friction}
\end{eqnarray}
where:
\begin{eqnarray}
    \label{eq:V1_multiftiction}
    \cV_0^{1\alpha} &=& \sum_{i=0}^{N-1} \sum_{j=1}^{N-1} \sqrt{\frac{\gamma_i}{N \Gamma}} \varphi_{ji} \chi^\beta_j\partial^\beta {\rm v}^\alpha_i(\bfR, \bfu_i)  \;, \\[0.2cm]
    \cV_0^{0\alpha}  &=&  \sum_{i=0}^{N-1} \frac{\gamma_i}{\Gamma} {\rm v}^\alpha_i(\bfR, \bfu_i) \;,
    \label{eq:V0_multiftiction}
\end{eqnarray}
where we recall that $\Psi(\chi,\Theta)$ is the steady-state distribution associated to the fast dynamics generated by $\cL_0$. Under our assumption of space-dependent self-propulsion speed, i.e. ${\rm v}_i(\bfr_i,\bfu_i) = v(\bfr_i) \bfu_i$, the diffusion tensor directly follows from Eq.~\eqref{eq:diff_formal_multi-friction} and reads:
\begin{align}
    \cD^{\alpha\beta}(\bfR) &= \frac{k_{\rm B} T}{\Gamma} \delta^{\alpha\beta} + v^2(\bfR)  \sum_{ij} \frac{\gamma_i \gamma_j}{\Gamma^2} \int_0^{\infty} \rmd t  \> \mathbb{C}^{\alpha\beta}_{ij}(t) \;.
    \label{eq:meso_diffusion_polymer-multi-friction}
\end{align}
To write down the explicit form of the drift $V^\alpha$ we need to evaluate the equal-time covariance $\langle\chi_j^\beta u_i^\alpha \rangle_{\Psi} $ in the steady-state of the fast process. The latter can be computed from the Rouse-mode dynamics of Eq.~\eqref{eq:chi_dynamics-multi_friction} and reads:
\begin{equation}
    \langle \chi_j^\beta u^\alpha_i \rangle_{\Psi}  =v(\bfR) \sum_{k=0}^{N-1} \varphi_{jk} \sqrt{\frac{N \gamma_k}{\Gamma}} \int_0^\infty  e^{- \lambda_j s} \mathbb{C}_{ik}^{\alpha\beta}(s) \rmd s \;.
    \label{eq:ss_covariance-multi-friction}
\end{equation}
Combining Eqs.~\eqref{eq:drift_formal_multi-friction},~\eqref{eq:V1_multiftiction}--\eqref{eq:V0_multiftiction} and Eq.~\eqref{eq:ss_covariance-multi-friction} we obtain:
\begin{eqnarray}
    V^{\alpha} &=& \frac{\partial^\beta v^2}{2} \sum_{i,k=0}^{N-1} \sum_{j=1}^{N-1} \varphi_{ji} \varphi_{jk} \frac{\sqrt{\gamma_i \gamma_k}}{\Gamma}\int_0^\infty  e^{-\lambda_j t} \mathbb{C}_{ik}^{\alpha\beta}(t) \rmd s \notag \\
    &&+ \frac{\partial^\beta v^2}{2} \sum_{i,j =0}^{N-1} \frac{\gamma_i \gamma_j}{\Gamma^2} \int_0^\infty \mathbb{C}^{\alpha\beta}_{ij}(t) \rmd t \;.
    \label{eq:meso_drift_polymer2-mf}
\end{eqnarray}
Finally, using the fact that $\varphi_{0i} \varphi_{0k} = \sqrt{\gamma_i \gamma_k}/\Gamma$, 
we note that the missing contribution from $j=0$ in the first term of $V^\alpha$ coincides with the second term of Eq.~\eqref{eq:meso_drift_polymer2-mf}. All in all, we thus obtain the final expression for the large-scale drift of the center of friction:
\begin{equation}
    V^{\alpha}(\bfR) = \frac{\partial^\beta v^2}{2} \sum_{i, j ,k=0}^{N-1} \varphi_{ji} \varphi_{jk} \frac{\sqrt{\gamma_i \gamma_k}}{\Gamma}\int_0^\infty  e^{- \lambda_j s} \mathbb{C}_{ik}^{\alpha\beta}(t) \rmd t \;.
    \label{eq:meso_drift_polymer-multi-friction}
\end{equation}
Equations~\eqref{eq:meso_diffusion_polymer-multi-friction} and~\eqref{eq:meso_drift_polymer-multi-friction} thus extend the results of Sec.~\ref{sec:adiabatic_elimination} to heteropolymers.

\newpage

\section{Macroscopic effective equilibrium regimes}
\label{sec:equilibrium}
In this section, we study the conditions that guarantee the existence of an effective equilibrium description at the large scale level. To do so, we impose the coarse-grained steady-state current $J^\alpha = V^\alpha p - \partial^\beta (\cD^{\alpha\beta} p)$ to be zero and derive general constraints on the transport coefficients. When such conditions are satisfied, the system exhibits zero entropy production rate at this level of description, and the steady-state distribution $p_s(\bfR)$ of the center-of-mass position is described by a Boltzmann distribution with effective potential $U(\bfR)$. 

Since our coarse-graining method does not rely on any specific model for the orientational dynamics, we derive general conditions on the auto-correlation function $\mathbb{C}^{\alpha\beta}_{ij}$ for such an equilibrium description to emerge. We first analyze the single-particle case ($N=1$) and then extend our discussion to active polymers. In either cases, we distinguish between space-dependent self-propulsion speed and taxis. 

One of the central results of this section is that, if the auto-correlation tensor $\mathbb{C}^{\alpha\beta}_{ij}$ is proportional to the identity $\delta^{\alpha\beta}$, then the system admits an equilibrium macroscopic description, both for space-dependent self-propulsion and for taxis. From a microscopic perspective, this constraint is satisfied by any system in which the orientational dynamics is isotropic and achiral. 

\subsection{Single-particle with space-dependent self-propulsion}
\label{subsec:equil_singlepart_space}
We start by considering the large-scale diffusion tensor and drift of a single particle, corresponding to $N=1$ in Eqs.~\eqref{eq:diffusion_final},~\eqref{eq:drift_final}. They read, respectively:
\begin{align}
\label{eq:singlepart_diff}
    \cD^{\alpha\beta} &= D_{\rm t} \delta^{\alpha\beta} +v^2(\bfR) \int_0^{\infty} \rmd t  \> \mathbb{C}^{\alpha\beta}(t) \;, \\[0.2cm] V^{\alpha} &= \frac{\partial^\beta v^2(\bfR)}{2} \int_0^\infty \rmd t\, \mathbb{C}^{\alpha \beta}(t)  \;.
\label{eq:singlepart_drift}
\end{align}
In what follows, we assume that the diffusion tensor is non-singular, so that the inverse $(\cD^{-1})^{\alpha\beta}$ is well defined.
We first note that $V^\alpha$ and $\cD^{\alpha\beta}$ are not independent, since the former is proportional to the divergence of the latter. It is thus convenient to express this relation as:
\begin{equation}
    V^{\alpha} = \frac{1}{2} \partial^\beta \cD^{\alpha\beta} \;.
    \label{eq:rel_V_D}
\end{equation}
The corresponding particle current then reads:
\begin{equation}
    J^{\alpha} = \frac{1}{2} p \partial^\beta \cD^{\alpha\beta} - \partial^\beta (\cD^{\alpha\beta} p)  \;.
\end{equation}
To determine the equilibrium conditions we impose $J^\alpha = 0$ at steady state, obtaining:
\begin{equation}
    \cD^{\alpha\beta} \partial^\beta p_s = -\frac{p_s}{2} \partial^\beta \cD^{\alpha\beta} \;.
\end{equation}
We now introduce an effective potential $U(\bfR)$ such that $p_s(\bfR) = \exp[-U(\bfR)]$, leading to:
\begin{equation}
    \cD^{\alpha\beta} \partial^\beta U(\bfR) = \frac{1}{2} \partial^\beta \cD^{\alpha\beta} \;.    
     \label{eq:effective_pot_1}
\end{equation}
Under the hypothesis of non-singular $\cD^{\alpha\beta}$, the effective potential is thus directly related to the diffusion tensor, and hence to the auto-correlation of the orientations, via:
\begin{equation}
    \partial^\alpha U(\bfR) = \frac{1}{2} (\cD^{-1})^{\alpha\gamma} \partial^\beta \cD^{\gamma\beta} \;
    \label{eq:gradU_def1}
\end{equation}
for all $\alpha \in \{1, \dots, d\}$. 

To check whether such an effective-equilibrium description is valid, i.e., whether Eq.~\eqref{eq:gradU_def1} is solvable by a scalar field $U(\bfR)$, we rely on the Schwarz theorem to impose the equality of the second cross derivatives of $U(\bfR)$~\cite{o2020lamellar,o2022time,dinelli2023nonreciprocity,o2024geometric,o2025geometric,duan2025phase}:
\begin{equation}
    \partial^\alpha \partial^\beta U = \partial^\beta \partial^\alpha U \;.
\end{equation}
Applying the Schwarz condition to Eq.~\eqref{eq:gradU_def1}, we obtain a necessary and sufficient condition on the diffusion tensor to yield an effective macroscopic equilibrium description:
\begin{equation}
      \forall\> \alpha,\beta : \quad \partial^\alpha \left[ (\mathcal{D}^{-1})^{\beta\gamma} \, \partial^\delta \mathcal{D}^{\gamma\delta} \right]
= \partial^\beta \left[ (\mathcal{D}^{-1})^{\alpha\gamma} \, \partial^\delta \mathcal{D}^{\gamma\delta} \right]\;.
\label{eq:generalzied_schwarz1}
\end{equation}
Alternatively, by defining the tensor:
\begin{equation}
    \mathcal{Q}^{\alpha\beta} := \partial^\alpha \left[ (\mathcal{D}^{-1})^{\beta\gamma} \, \partial^\delta \mathcal{D}^{\gamma\delta} \right]
\end{equation}
the Schwarz condition given by Eq.~\eqref{eq:generalzied_schwarz1} can be formulated as a symmetry condition on $\mathcal{Q}^{\alpha\beta}$: 
\begin{equation}
    J^\alpha = 0 \quad \Leftrightarrow \quad \mathcal{Q}^{\alpha \beta} = \mathcal{Q}^{\beta \alpha} \;.
\label{eq:generalzied_schwarz}
\end{equation}

\subsubsection{Athermal particles.}
For the sake of concreteness, we now consider specific classes of models for which Eq.~\eqref{eq:generalzied_schwarz} is satisfied. We start by considering the athermal case where $k_{\rm B} T = 0$. Denoting by $\tau^{\alpha\beta} = d \int_0^\infty \mathbb{C}^{\alpha\beta} \rmd t$ the microscopic timescales, the diffusion tensor is simply given by $\cD^{\alpha\beta}=v^2(\bfR) \tau^{\alpha\beta}/d$. In this case, $\mathcal{Q}^{\alpha\beta}$ simplifies into:
\begin{equation}
    \mathcal{Q}^{\alpha\beta} = \partial^\alpha \left[ \frac{\delta^{\beta\delta}}{v^2} \partial^\delta v^2 \right] = \partial^\alpha \partial^\beta \log v^2 \;,
\end{equation}
which is always symmetric, hence always satisfying Eq.~\eqref{eq:generalzied_schwarz}. As such, any athermal model of self-propelled particles with space-dependent self-propulsion speed $v(\bfR)$ admits an equilibrium macroscopic description with zero particle current. Using Eq.~\eqref{eq:gradU_def1} we can derive the corresponding effective potential $U(\bfR)$, which reads:
\begin{equation}
    U(\bfR) = \log v(\bfR) \;.
\end{equation}
The resulting steady-state distribution is then given by the well-known expression:
\begin{equation}
    p_s \propto \frac{1}{v(\bfR)} \;,
\end{equation}
which implies accumulation in region of smaller activity~\cite{schnitzer1993theory,tailleur2008statistical,cates2013when,arlt2018painting,frangipane2018dynamic,metzger2024revisiting,metzger2025exceptions}.
On a final note, we observe that the Schwarz condition Eq.~\eqref{eq:generalzied_schwarz} implies zero entropy production to leading order in gradient truncation at the mesoscopic scale. Notably, this feature has recently been shown to hold to all orders in gradient expansions for run-and-tumble (RTPs) and active Brownian particles (ABPs) in an activity landscape~\cite{metzger2024revisiting,metzger2025exceptions}.

\subsubsection{Finite temperature}
In the general case of finite $T>0$, Eq.~\eqref{eq:generalzied_schwarz} is satisfied only by some specific classes of models, which we now present. The first trivial example is the one of flat activity profile $v(\bfR) = v_0$, which always yields a constant steady-state particle distribution $p_s(\bfR)$. More interestingly, Eq.~\eqref{eq:generalzied_schwarz} is generically satisfied in the isotropic achiral case, corresponding to $\mathbb{C}^{\alpha\beta} = C(t) \delta^{\alpha\beta}$. For this class of models, the diffusion tensor resulting from Eq.~\eqref{eq:singlepart_diff} is also isotropic, and reads:
\begin{equation}
    \cD^{\alpha\beta} = \cD_0 \delta^{\alpha\beta} \;, \quad \text{with} \quad \cD_0 = D_{\rm t} + \frac{v^2(\bfR) \tau_{\rm p}}{d}\;.
\end{equation}
The tensor $\mathcal{Q}^{\alpha\beta}$ is thus given by:
\begin{equation}
    \mathcal{Q}^{\alpha\beta} = \partial^\alpha \left[ \frac{\delta^{\beta\gamma}}{\cD_0} \delta^{\gamma\delta} \partial^\delta {\cD_0}\right] = \partial^\alpha \partial^\beta \log \cD_0 \;,
\end{equation}
thus satisfying the Schwarz symmetry condition, Eq.~\eqref{eq:generalzied_schwarz}. We thus conclude that, in the thermal case, any isotropic achiral process for the orientational degrees of freedom gives rise to an effective equilibrium description. The corresponding effective potential $U(\bfR)$ is then obtained by solving Eq.~\eqref{eq:effective_pot_1}:
\begin{equation}
    \partial^\alpha U(\bfR) = \frac{1}{2} \partial^\alpha \log \cD_0 \;,
\end{equation}
leading to:
\begin{equation}
    U(\bfR) = \frac{1}{2} \log\cD_0 \;.
\label{eq:steady_state_singlepart}
\end{equation}
The corresponding steady-state particle distribution eventually reads:
\begin{equation}
    p_s(\bfR) \propto \frac{1}{\cD_0^{1/2}} \propto \left( D_{\rm t} + \frac{v^2(\bfR) \tau_{\rm p}}{d}\right)^{-1/2} \;.
    \label{eq:singlepart_dist}
\end{equation}

\subsection{Single-particle with taxis}
\label{subsec:singlepart_taxis}

Next, we consider the diffusion-drift dynamics of a single particle undergoing taxis. In this case, the self-propulsion speed of each particle is described by $v(\bfr_i) = v_0 - v_1 u^\alpha \partial^\alpha c(\bfr_i)$, where $c$ is the external tactic field. The transport properties obtained from Eqs.~\eqref{eq:drift_final_taxis} and~\eqref{eq:diffusion_final_taxis} are:
\begin{align}
\label{eq:singlepart_taxis-diff}
    V^{\alpha} &= - v_1 \partial^\beta c(\bfR) \, \mathbb{C}^{\alpha\beta}(0) \;, \\[0.2cm]
    \cD^{\alpha\beta} &= D_{\rm t} \delta^{\alpha\beta} + v_0^2 \int_0^{\infty} \rmd t  \> \mathbb{C}^{\alpha\beta}(t)   \;.
\label{eq:singlepart_taxis-drift}
\end{align}
We remark that, in this case, the diffusion tensor is always constant. As before, to derive the equilibrium condition we impose $J^\alpha = 0$ and look for an effective potential $U(\bfR)$ such that $p_s \propto \exp[ - U(\bfR)]$. Straightforward algebra then leads to:
\begin{equation}
    \partial^\alpha U =  (\cD^{-1})^{\alpha\gamma} v_1 \mathbb{C}^{\gamma\delta}(0) \partial^\delta c \;.
    \label{eq:effective_pot_2}
\end{equation}
Imposing the equality of the cross second derivatives of $U$ then leads to a generalized Schwarz condition for taxis:
\begin{equation}
    J^\alpha = 0 \quad \Leftrightarrow \quad \mathcal{Q}^{\alpha\beta}_{\rm taxis} = \mathcal{Q}^{\beta\alpha}_{\rm taxis}
    \label{eq:generalzied_schwarz_taxis}
\end{equation}
where the tensor $ \mathcal{Q}_{\rm taxis}^{\alpha\beta}$ is defined as:
\begin{equation}
      \mathcal{Q}_{\rm taxis}^{\alpha\beta} :=  (\mathcal{D}^{-1})^{\alpha\gamma}  \mathbb{C}^{\gamma\delta}(0) \partial^\beta \partial^\delta c \;.
\end{equation}
While such an equilibrium condition is not generically satisfied, we shall now restrict our analysis to the case of isotropic achiral dynamics for the orientation vector $\bfu$. In this case, the steady-state covariance is simply $\mathbb{C}^{\alpha\beta}(0) = \delta^{\alpha\beta}/d$, while the diffusion tensor is given by:
\begin{equation}
    \cD^{\alpha\beta} = \cD_0 \delta^{\alpha\beta} \;, \quad \text{with} \quad \cD_0 = D_{\rm t} + \frac{v_0^2 \tau_{\rm p}}{d} \;.
\end{equation}
The Schwarz condition Eq.~\eqref{eq:generalzied_schwarz_taxis} is then directly satisfied, and the effective potential resulting from Eq.~\eqref{eq:effective_pot_2} reads:
\begin{equation}
    U(\bfR) = \frac{v_1}{d\cD_0} c(\bfR) \;. 
    \label{eq:effective_pot_taxis}
\end{equation}
The associated steady-state distribution $p_s(\bfR)$ thus corresponds to accumulation in regions of low $c$ for the chemo-repulsive case $v_1 > 0$, and accumulation in regions of high $c$ for chemo-attraction $v_1 < 0$~\cite{o2020lamellar}.

\subsection{Active polymer with space-dependent self-propulsion}
\label{subsec:equil_Rouse_space}
We now study the general case of an active polymer with $N > 1$ monomers subjected to space-dependent active forces, whose large-scale dynamics is captured by the diffusion tensor in Eq.~\eqref{eq:diffusion_final} and the drift in Eq.~\eqref{eq:drift_final}.
In contrast with the single-particle example, we note that no simple Schwarz criterion for equilibrium can be derived in the most general case. 
Nonetheless, if we restrict ourselves to isotropic achiral dynamics for the orientational degrees of freedom, the auto-correlation matrix simplifies as $\mathbb{C}^{\alpha\beta}_{ij}(t) =  C_{ij}(t) \delta^{\alpha\beta}$, and the resulting diffusion tensor is proportional to the identity: $\cD^{\alpha\beta} = \cD_0 \delta^{\alpha\beta}$. The corresponding diffusivity reads:
\begin{equation}
    \cD_0 = \frac{D_{\rm t}}{N} + \frac{v^2(\bfR)}{N^2} \sum_{ij} \int_0^{\infty} C_{ij} (t) \rmd t \;.
\end{equation}
Note that, by virtue of Wiener–Khinchin theorem~\cite{gardiner2004handbook}, $\sum_{ij} \int_0^{\infty} C_{ij} (t) \rmd t \geq 0$, and hence $\cD_0$ is also non-negative. 
It is then possible to express the drift velocity as:
\begin{equation}
    V^{\alpha} = \left( 1 - \frac{\varepsilon}{2} \right)\partial^\alpha \cD_0 \;,
    \label{eq:eps_relation}
\end{equation}
where $\epsilon$ is a coefficient that depends on the specific details of the model. Note that Eq.~\eqref{eq:eps_relation} extends the relation~\eqref{eq:rel_V_D} to $\epsilon \neq 1$.  To compute the explicit value of $\epsilon$, one can compare Eq.~\eqref{eq:eps_relation} with Eqs.~\eqref{eq:drift_final},~\eqref{eq:diffusion_final}, eventually obtaining:
\begin{equation}
    \epsilon = 2 - \dfrac{N \sum_{ijk} \varphi_{ji} \varphi_{jk} \int_0^\infty \rmd t e^{-\lambda_j t} C_{ij}(t)}{\sum_{ij} \int_0^\infty \rmd t \> C_{ij}(t)} \;.
    \label{eq:eps_value}
\end{equation}
The relation given by Eqs.~\eqref{eq:eps_relation}--\eqref{eq:eps_value} allows us to show that the resulting coarse-grained description is always an equilibrium one. To see this, we impose $J^\alpha = 0$ and, using Eq.~\eqref{eq:eps_relation}, we look for a Boltzmann-like steady-state distribution $p_{\rm s} = \exp[-U(\bfR)]$ for the center-of-mass position. We then obtain:
\begin{eqnarray}
    0 &=& p_{\rm s} \left(1-\frac{\epsilon}{2}\right) \partial^\alpha \cD_0 - \partial^\alpha (\cD_0 p_{\rm s}) \;,
\end{eqnarray}
so that:
\begin{eqnarray}
    \cD_0 \partial^\alpha p_{\rm s} &=& \frac{\epsilon}{2} \partial^\alpha \cD_0 \;.
    \label{eq:eps_2_d0}
\end{eqnarray}
Equation~\eqref{eq:eps_2_d0} is finally solved by
\begin{equation}
    p_s(\bfr) \propto e^{-U(\bfR)} \quad \text{with} \quad 
    U(\bfR) = \frac{\epsilon}{2} \log \cD_0 \;.
    \label{eq:effective_pot_selfprop}
\end{equation}
Interestingly, the sign of the prefactor $\epsilon$ dictates whether the polymer accumulates in regions where the activity---captured by $\cD_0$---is small ($\epsilon > 0$), or whether accumulation at high activity is preferred ($\epsilon < 0$). From Eq.~\eqref{eq:eps_relation}, we notice that the prefactor $\epsilon$ explicitly depends on the auto-correlation function $C_{ij}$, the number of monomers $N$ and the structure of the polymer via $\varphi_{ij}$. Consequently, multiple transition pathways are possible from accumulation in high-activity to low-activity regions. For instance, Ref.~\cite{muzzeddu2024migration} explored one transition route involving $N$ and $\varphi_{ij}$, which we recapitulate briefly in Sec.~\ref{subsec:polymer_simple}. Furthermore, in Sec.~\ref{subsec:polymer_correlated} we propose a new transition mechanism that relies on the properties of the auto-correlation function of the monomer orientations $C_{ij}$.

\subsection{Active polymer with taxis}
\label{subsec:Rouse_taxis}
We conclude this section by illustrating the large-scale behavior of tactic polymers, whose diffusion tensor and drift are described by Eqs.~\eqref{eq:drift_final_taxis} and~\eqref{eq:diffusion_final_taxis}, which we report here for clarity:
\begin{align*}
    V^\alpha(\bfR) &= -\frac{v_1}{N} \sum_{i=0}^{N-1} \mathbb{C}^{\alpha\beta}_{ii}(0) \partial^\beta c(\bfR)\;,\\
    \cD^{\alpha\beta} &= \frac{k_{\rm B}T}{\gamma N} \delta^{\alpha\beta} +\frac{v_0^2}{N^2} \sum_{ij}\int_0^{\infty} \rmd t  \> \mathbb{C}^{\alpha\beta}_{ij}(t) \;.
\end{align*}
The conditions for the emergence of an effective equilibrium regime can be determined by following the same strategy as in Sec.~\ref{subsec:Rouse_taxis}.
Imposing $J^\alpha = 0$ and looking for a steady-state distribution $p_{\rm s} = \exp[-U(\bfR)]$ yields the following equation for the effective potential:
\begin{equation}
    \partial^\alpha U = (\cD^{-1})^{\alpha\gamma} \frac{v_1 }{N} \sum_{i=1}^N \mathbb{C}_{ii}^{\gamma\delta}(0)\partial^\delta c(\bfR)
\end{equation}
The Schwarz equality of the cross second derivatives of $U$ finally implies the following condition for large-scale equilibrium:
\begin{equation}
    J^\alpha = 0 \quad \Leftrightarrow \quad \mathcal{Q}^{\alpha\beta}_{N,\,\rm taxis} = \mathcal{Q}^{\beta\alpha}_{N,\,\rm taxis}
    \label{eq:generalzied_schwarz_Ntaxis}
\end{equation}
where the tensor $ \mathcal{Q}_{N,\,\rm taxis}^{\alpha\beta}$ is defined as:
\begin{equation}
      \mathcal{Q}_{N,\,\rm taxis}^{\alpha\beta} :=  (\mathcal{D}^{-1})^{\alpha\gamma}  \sum_{i=1}^N \mathbb{C}_{ii}^{\gamma\delta}(0) \partial^\beta \partial^\delta c \;.
\end{equation}
To provide a simple example where the effective potential can be found, we consider the isotropic achiral case $\mathbb{C}^{\alpha\beta}_{ii}(t) = \delta^{\alpha\beta}/d$ for all monomers $i$. The resulting diffusion tensor then reads:
\begin{equation}
    \cD^{\alpha\beta} = \cD_0 \delta^{\alpha\beta} \;, \quad \text{with} \quad \cD_0 = \frac{D_{\rm t}}{N} + \frac{v_0^2}{d N^2} \sum_{ij}  \tau_{ij} \;,
    \label{eq:D0_Ntaxis}
\end{equation}
where we have assumed that $\mathbb{C}^{\alpha\beta}_{ij} = C_{ij}(t) \delta^{\alpha\beta}$, and hence the correlation time reads $\tau_{ij} = d \int_0^\infty C_{ij} \rmd t$. 
The Schwarz condition Eq.~\eqref{eq:generalzied_schwarz_Ntaxis} is then satisfied, and the effective potential resulting from Eq.~\eqref{eq:effective_pot_Ntaxis} reads:
\begin{equation}
    U(\bfR) = \frac{v_1}{d\cD_0} c(\bfR) \;. 
    \label{eq:effective_pot_Ntaxis}
\end{equation}
Since the Wiener–Khinchin theorem~\cite{gardiner2004handbook} guarantees that $\cD_0$ is always non-negative, the location of the minima of the effective potential are here dictated by the sign of $v_1$ only: chemo-repulsion for $v_1 > 0$, chemo-attraction for $v_1 < 0$.
When compared to the steady-state distribution in Eq.~\eqref{eq:effective_pot_selfprop} for active polymers with space-dependent self-propulsion speed, 
the behavior of tactic polymers revealed by Eq.~\eqref{eq:effective_pot_Ntaxis} is far less rich: indeed, no phase transition from high-activity to low-activity accumulation can be observed in the latter upon tuning the number of monomers $N$, the polymer structure via $\varphi_{ij}$ nor the auto-correlation $C_{ij}$ of the orientations.
Notably, previous studies on coarse-graining of active particles with motility regulation had highlighted a macroscopic equivalence between taxis and space-dependent self-propulsion for single-particle systems~\cite{schnitzer1993theory,o2020lamellar,dinelli2024fluctuating}, see also Eqs.~\eqref{eq:steady_state_singlepart} and~\eqref{eq:effective_pot_taxis}. Our result reveals how such an equivalence is violated in the context of active polymers, where spatial modulation of the self-propulsion speed offers a broader range of transition routes.

\newpage

\section{Orientational auto-correlation across active-particle models}
\label{sec:correlations}
Our coarse-graining method of Sec.~\ref{sec:adiabatic_elimination} shows how the auto-correlation tensor of orientations, $\mathbb{C}^{\alpha\beta}(t) = \langle u^{\alpha}(t) u^{\beta}(0) \rangle$, controls the macroscopic drift and the diffusion tensor for active particles and active polymers undergoing motility regulation.
In particular, the properties of $\mathbb{C}^{\alpha\beta}(t)$ determine whether the system admits an equilibrium description at the macroscopic scale in the presence of motility regulation, as discussed in Sec.~\ref{sec:equilibrium}. 

Here we provide the expression of $\mathbb{C}^{\alpha\beta}$ for a number of stochastic processes of the orientational degrees of freedom $\Theta$, together with the associated persistence times $\tau^{\alpha\beta} = d\int_0^\infty \mathbb{C}^{\alpha\beta} (t) \rmd t$.
Our presentation mostly recapitulates pre-existing results in the literature.
The only exception is detailed in Sec.~\ref{subsec:multistate}, where we study a new example of active dynamics alternating between $M>2$ motility states.
To the best of our knowledge, this case has not been investigated in the literature, and we thus complement our theoretical predictions with numerical measurements of the auto-correlation tensor.

\subsection{Run-and-tumble particles}
\label{subsec:RTPcorr}
In run-and-tumble particles (RTPs), the orientation unit vector $\bfu$ undergoes a Poisson jump process from $\bfu \to \bfu'$ with a tumbling rate $\alpha$. In the simplest setup, $\bfu'$ is drawn uniformly from the unit sphere in $d$ dimensions $\mathbb{S}^{d-1}$~\cite{schnitzer1993theory,berg2004coli,kurzthaler2024characterization}:
\begin{equation}
    \bfu \to \bfu' \in \mathbb{S}^{d-1} \quad \text{with Poisson rate } \alpha .
    \label{eq:RTPdynamics}
\end{equation}
The corresponding auto-correlation function reads~\cite{schnitzer1993theory}: 
\begin{equation}
    \mathbb{C}^{\alpha\beta} = \frac{\delta^{\alpha\beta}}{d} \exp [ -\alpha t]\;,
\end{equation}
and the persistence time is $\tau_{\rm p} = \int_0^\infty \exp [ -\alpha t] = \alpha^{-1}$.

\subsection{Active Brownian particles}
\label{subsec:ABPcorr}
In Active Brownian particles (ABPs), the orientation unit vector $\bfu$ diffuses over the $d$-dimensional unit sphere~\cite{golestanian2007designing,jiang2010active,theurkauff2012dynamic,palacci2013living}. The associated It\^o dynamics in $d$ dimensions reads:
\begin{equation}
    \dot{u}_\alpha = -(d-1) D_r u^\alpha + \sqrt{2 D_r} (\delta^{\alpha\beta} - u^\alpha u^\beta) \eta^\beta(t) \;,
    \label{eq:ABPdynamics}
\end{equation}
where $\eta_\beta(t)$ is a Gaussian white noise with zero mean and correlations $\langle \eta^\alpha(t) \eta^\beta(t') \rangle = \delta^{\alpha\beta} \delta(t-t')$. The dynamics~\eqref{eq:ABPdynamics} ensures that the norm of $u^\alpha u^\alpha = 1$ at all times, as can be shown directly via It\^o calculus. Computing the auto-correlation tensor from Eq.~\eqref{eq:ABPdynamics} yields, as in the RTP case, an exponential decorrelation:
\begin{equation}
    \mathbb{C}^{\alpha\beta} = \frac{\delta^{\alpha\beta}}{d} \exp [ -(d-1)D_r t]
\end{equation}
with persistence time $\tau_{\rm p} = [(d-1) D_r]^{-1}$.

\subsection{Active Ornstein Uhlenbeck particles}
\label{subsec:AOUPcorr}
Active Ornstein Uhlenbeck particles (AOUPs) are endowed with an orientation vector $\bfu$ whose modulus is also allowed to fluctuate~\cite{sepulveda2013collective,szamel2014self,wittmann2017effective,wittmann2017effective2,martin2021statistical}. More precisely, the dynamics of $\bfu$ is described by a $d$-dimensional Ornstein-Uhlenbeck process:
\begin{equation}
    \dot{u}_\alpha = - \frac{u^\alpha}{\tau_0} + \sqrt{\frac{2}{d \tau_0}} \eta^\alpha(t) \;,
    \label{eq:AOUPdynamics}
\end{equation}
where $\eta_\alpha(t)$ is a Gaussian white noise with zero mean and correlations $\langle \eta^\alpha(t) \eta^\beta(t') \rangle = \delta^{\alpha\beta} \delta(t-t')$. Our choice of the noise amplitude ensures that the squared norm of $\bfu$ fluctuates around a mean value of $1$. As in the previous cases, the dynamics described by Eq.~\eqref{eq:AOUPdynamics} leads to an exponential decorrelation of the orientations:
\begin{equation}
    \mathbb{C}^{\alpha\beta} = \frac{\delta^{\alpha\beta}}{d} \exp [ - t/\tau_0]
\end{equation}
with persistence time $\tau_{\rm p}=\tau_0$.

\subsection{Chiral active particles}
\label{subsec:chiralABP}
The minimal models introduced above have been further extended to account for spatial symmetry breaking induced by the intrinsic chirality of the individual active agents, see, e.g.,~\cite{liebchen2022chiral,pattanayak2024impact, kalz2024field}.
Both theoretical and experimental studies on synthetic systems have demonstrated that the interplay between activity and chirality gives rise to a remarkably rich phenomenology, including the formation of finite-size rotating clusters~\cite{massana2021arrested}, the emergence of coherent vortices~\cite{caprini2024self} and enhanced flocking~\cite{liebchen2017collective}, among others.
Chiral effects are also known to play a crucial role in the microscale dynamics of biological systems, such as bacteria~\cite{lauga2006swimming} and sperm cells~\cite{friedrich2007chemotaxis}.
In this section, we compute the temporal autocorrelation function of the rotational degrees of freedom $\Theta$ in the case where the latter are affected by an active torque.
We start by considering the case of ABPs and extend Eq.~\eqref{eq:ABPdynamics} to account for chirality in the orientation dynamics. To do so, we introduce the anti-symmetric infinitesimal generator of rotations $\Omega^{\alpha\beta} = -\Omega^{\beta\alpha}$. The dynamics of the unit orientation vector then reads:
\begin{equation}
    \dot{u}_\alpha = -(d-1) D_r u^\alpha - \Omega^{\alpha\beta} u^\beta+ \sqrt{2 D_r} (\delta^{\alpha\beta} - u^\alpha u^\beta) \eta^\beta(t) \;,
    \label{eq:chiralABPdynamics}
\end{equation}
where $\eta_\beta(t)$ is a Gaussian white noise with zero mean and correlations $\langle \eta^\alpha(t) \eta^\beta(t') \rangle = \delta^{\alpha\beta} \delta(t-t')$. The steady-state auto-correlation tensor $\mathbb{C}^{\alpha\beta}(t)$ can be straightforwardly computed using It\^o's lemma~\cite{pattanayak2024impact}. Its dynamics follows directly from Eq.~\eqref{eq:chiralABPdynamics} and reads:
\begin{equation}
    \partial_t \mathbb{C}^{\alpha\beta} = -(d-1) D_r \mathbb{C}^{\alpha\beta}  - \Omega^{\alpha\gamma} \mathbb{C}^{\gamma\beta} \;,
    \label{eq:Cdyn_chiral}
\end{equation}
with $\mathbb{C}^{\alpha\beta}(0) = \frac{1}{d}\delta^{\alpha\beta}$, corresponding to the fact that, on average, all components of $\bfu$ have squared-modulus $1/d$ at steady state. Solving Eq.~\eqref{eq:Cdyn_chiral} with the given initial condition then yields:
\begin{eqnarray}
    \mathbb{C}^{\alpha\beta} &=& \frac{1}{d} e^{ -(d-1) D_r t} \exp[-\boldsymbol{\Omega}t ]^{\alpha\beta} \;.
    \label{eq:Corr_chiral}
\end{eqnarray}
For most practical purposes, it is useful to consider a special case of Eq.~\eqref{eq:Cdyn_chiral} where a single axis of rotation is defined. In the following, we accordingly choose our coordinate system such that $\Omega^{\alpha\beta}$ is the infinitesimal generator of rotation in the $xy$ plane. Under this hypothesis, the only two non-zero components of $\Omega^{\alpha\beta}$ are: $\Omega^{xy} = -\Omega^{yx} = \omega$. It is then useful to define the projector operator $\Pi^{\alpha\beta}$ over the $xy$ plane as $\Pi^{\alpha\beta} := \delta^{\alpha x} \delta^{\beta x} + \delta^{\alpha y} \delta^{\beta y} $, the orthogonal projector $Q^{\alpha\beta} := \delta^{\alpha\beta} - \Pi^{\alpha\beta}$, and the rescaled matrix $\hat{\Omega}^{\alpha\beta} := \Omega^{\alpha\beta}/\omega$. The matrix $\Omega^{\alpha\beta}$ then satisfies the following properties:
\begin{align}
\label{eq:Omega_prop1}
    \Omega^{\alpha\gamma} \Pi^{\gamma\beta} &= \omega \hat\Omega^{\alpha\beta} \;, \\[0.2cm]
    [\boldsymbol{\Omega}^{2n} ]^{\alpha\beta} &= (-1)^{n} \omega^{2n} \Pi^{\alpha\beta} \;, \\[0.2cm]
     [\boldsymbol{\Omega}^{2n-1} ]^{\alpha\beta} &= (-1)^{n-1} \omega^{2n-1} \hat\Omega^{\alpha\beta}\;.
\label{eq:Omega_prop3}
\end{align}
for integer $n \geq 1$. By Taylor-expanding the exponential $\exp[-\boldsymbol{\Omega}t ]^{\alpha\beta}$ and using the properties~\eqref{eq:Omega_prop1}--\eqref{eq:Omega_prop3} we
obtain:
\begin{align*}
    &\exp[-\Omega t]^{\alpha\beta} = \sum_{n=0}^{\infty} \frac{ t^{2n}}{(2n)!} [\boldsymbol{\Omega}^{2n}]^{\alpha\beta} - \sum_{n=0}^{\infty} \frac{ t^{2n+1}}{(2n+1)!} [\boldsymbol{\Omega}^{2n+1}]^{\alpha\beta} \notag \\[0.1cm]
    &= Q^{\alpha\beta} +\Pi^{\alpha\beta} \sum_{n=0}^{\infty} \frac{ (-1)^n (\omega t)^{2n}}{(2n)!} - \hat{\Omega}^{\alpha\beta}\sum_{n=0}^{\infty} \frac{(-1)^n (\omega t)^{2n+1}}{(2n+1)!}  \notag \\[0.1cm]
    &= Q^{\alpha\beta} + \Pi^{\alpha\beta} \cos(\omega t) - \hat{\Omega}^{\alpha\beta} \sin(\omega t) \;.
\end{align*}
We can thus rewrite Eq.~\eqref{eq:Corr_chiral} for chiral ABPs with a single axis of rotation as:
\begin{multline}
    \mathbb{C}^{\alpha \beta}(t)= \frac{e^{-(d-1) D_r t}}{d} \left[Q^{\alpha\beta} + \Pi^{\alpha\beta} \cos(\omega t) \right. \\
    \left. - \hat{\Omega}^{\alpha\beta} \sin(\omega t) \right]\,.
    \label{eq:chiral_correlations_3}
\end{multline}
Equation~\eqref{eq:chiral_correlations_3} decouples the three distinct contributions to the auto-correlation tensor. The orthogonal projector $Q^{\alpha\beta}$ is the identity in the space orthogonal to $xy$, and zero otherwise; it thus ensures that all terms $\mathbb{C}^{\alpha\beta}$ with either $\alpha$ or $\beta$ different from $x,y$ decorrelate with the usual ABP persistence time, $[(d-1) D_r]^{-1}$. The two remaining terms act exclusively on the $xy$-subspace. On the one hand, the projector term $\Pi^{\alpha\beta}$, which is proportional to the identity in the $xy$ susbpace, generates cosinusoidal oscillations in the diagonal components $\mathbb{C}^{xx}, \> \mathbb{C}^{yy}$ with frequency $\omega$. On the other hand, the purely antisymmetric term $\hat{\Omega}^{\alpha\beta}$ induces anti-correlation between the $x$ and $y$ components, so that $\mathbb{C}^{xy}(t) = - \mathbb{C}^{yx}(t)$ at any time. Explicitly, Eq.~\eqref{eq:chiral_correlations_3} thus reads:
\begin{equation}
\mathbb{C}(t)
=\frac{e^{-(d-1)D_r t}}{d}
\begin{pmatrix}
\cos(\omega t)  & -\sin(\omega t) & 0 & \cdots & 0\\
\sin(\omega t)  & \cos(\omega t)  & 0 & \cdots & 0\\
0 & 0 & 1 &  & 0\\
\vdots & \vdots &  & \ddots & \\
0 & 0 & 0 &  & 1
\end{pmatrix}\;.
\end{equation}

A straightforward generalization of the above results can be obtained for chiral Active Ornstein-Uhlenbeck particles (AOUPs). Their orientation dynamics extends Eq.~\eqref{eq:AOUPdynamics} to include an additional torque:
\begin{equation}
    \dot u^{\alpha} =  - \frac{u^\alpha}{\tau_0} - \Omega^{\alpha\beta} u^\beta + \sqrt{\frac{2}{d \tau_0}} \eta^\alpha(t) \;,
    \label{eq:chiralAOUPdynamics}
\end{equation}
where $\Omega^{\alpha\beta}$ is an anti-symmetric tensor. The auto-correlation tensor can be obtained by applying the same method as for ABPs, yielding:
\begin{eqnarray}
    \mathbb{C}^{\alpha\beta}(t) &=& \frac{1}{d} e^{ -t/\tau_0} \exp[-\boldsymbol{\Omega}t ]^{\alpha\beta} \;.
    \label{eq:Corr_chiral_AOUP}
\end{eqnarray}
In the simpler case where a single axis of rotation is present---chosen to be orthogonal to the $xy$ plane---Eq.~\eqref{eq:Corr_chiral_AOUP} directly simplifies into Eq.~\eqref{eq:chiral_correlations_3}, upon replacing $(d-1) D_r$ with $\tau_0^{-1}$.

\subsection{Active particles with rotational inertia}
\label{subsec:ABPinertia}
While the overdamped limit is typically appropriate for biological active matter at low Reynolds number~\cite{purcell2014life}, macroscopic synthetic self-propelled particles, such as vibrobots, can exhibit significant inertial effects due to their finite mass and angular momentum~\cite{scholz2018inertial,lowen2020inertial,caprini2022role,lisin2022motion,sprenger2023dynamics}. To broaden our discussion and study the effects of inertia on the angular dynamics, we first consider a $2d$-ABP model where the angular dynamics is subjected to an underdamped evolution via:
\begin{eqnarray}
    \label{eq:thetadot}
    \dot\theta &=& \omega(t) \\[0.1cm]
    I\dot\omega &=& -\omega + \sqrt{2 D_r} \xi_\omega(t)\;.
    \label{eq:Iomega}
\end{eqnarray}
In the dynamics~\eqref{eq:Iomega}, the term $\xi_\omega(t)$ represents a Gaussian white noise term with zero mean and unit variance $\langle \xi_\omega(t) \xi_\omega(s)\rangle = \delta(t-s)$. Note that we have set the time units so that the moment of inertia $I$ has the dimensions of time: $[I]=[D_r^{-1}]=[t]$. The expression of the auto-correlation tensor $\mathbb{C}^{\alpha\beta}$ for the orientations was found by Ref.~\cite{sprenger2023dynamics} and reads:
\begin{eqnarray}
\mathbb{C}^{\alpha\beta} &=& \frac{\delta^{\alpha\beta}}{2} \exp\left[ -D_rt - ID_r (1-e^{-t/I})\right]\;. 
\end{eqnarray}
Introducing $z=I D_r$ and the incomplete Gamma function $\Gamma(x,a,b) = \int_a^b t^{x-1} e^{-t} \rmd t$, the resulting persistence time then reads:
\begin{equation}
    \tau_{\rm p} = 2 \int_0^\infty \mathbb{C}^{xx}(t) \rmd t =  I e^{z} z^{-z} \Gamma(z,0,z) \;.
\end{equation}

A second example of active dynamics with rotational inertia is the one of inertial AOUPs, whose detailed study can be found in~\cite{sprenger2023dynamics}. The corresponding dynamics for the orientation vector reads:
\begin{align}
    \dot u^\alpha &= - \tau_0^{-1} u^\alpha + \sqrt{\tau_0^{-1}} w^\alpha(t) \;, \\
    I \dot w^\alpha &= -  w^\alpha + \sqrt{2 D_{w}} \>\xi_w^\alpha(t) \;,
\end{align}
where $\xi_w^\alpha(t)$ is a zero-mean Gaussian white noise term with unit variance.
The auto-correlation tensor associated to this process reads~\cite{sprenger2023dynamics}:
\begin{eqnarray}
\mathbb{C}^{\alpha\beta}(t) &=& \frac{\delta^{\alpha\beta}}{d} \frac{2 D_w \tau_0}{\tau_0^2 - I^{2}} \left[ \tau_0 e^{-t/\tau_0} - I e^{-t/I}\right]\;.
\end{eqnarray}
Note that the requirement of $\langle |\bfu|^2 \rangle = 1$ can be enforced by assuming that $D_w = (\tau_0 + I)/(2\tau_0)$~\cite{sprenger2023dynamics}. Under this hypothesis,
the resulting persistence time is then given by:
\begin{equation}
    \tau_{\rm p} = \frac{1}{\tau_0 + I} \;.
\end{equation}

\subsection{Active particles switching between active and arrested states}
\label{subsec:2state}
In the remaining sections we determine the autocorrelation function and persistence time for active particles (RTPs, ABPs, AOUPs) endowed with multiple internal motility states~\cite{detcheverry2017generalized,shaebani2022kinematics,datta2024random,caraglio2024learning,santra2024dynamics}. This is motivated by agents with intermittent dynamics, alternating between a passive and an active state~\cite{datta2024random,caraglio2024learning,santra2024dynamics,lefranc2025synthetic}, but it also allows us to model the dynamics of \textit{E. Coli} bacteria, accounting for a finite residence time in the arrested state~\cite{berg2004coli,tailleur2008statistical, curatolo2020cooperative, santra2024dynamics}. 

It is instructive to start by considering only $2$ states of motion: an active one with finite persistence time and a passive one with $\bfu=0$.
Generalization to an arbitrary number of states $K$ with distinct persistence times is discussed in Sec.~\ref{subsec:multistate}.
We assume that switching events from the active to the passive state occur with a Poisson arrest rate $\alpha$, while passive particles resume their motion at a resuming rate $\beta$. When particles resume their motion, their new orientation is sampled from the steady-state distribution $\psi(\Theta)$. This implies that, for any pair of spatial components $(\gamma,\delta)$, we have $\langle u^\gamma(t) u^\delta(0)\rangle = 0$ if the particle underwent at least one state switch in the time window $[0,t)$~\footnote{We here exclude all processes in which the orientations are drawn from a discrete set (including the case $d=1$), where there is a finite probability that the newly sampled orientation is equal to the one before the transition.}. 

In the autocorrelation function $\mathbb{C}^{\gamma\delta}(t)$, particles in the arrested state provide zero contribution, since the corresponding value of the orientation $\bfu$ is zero at all times. Consequently, $\mathbb{C}^{\gamma \delta}(t)$ takes finite contributions only from particles that are in the run phase at time $0$, and that undergo no arrest events in $[0,t]$:
\begin{align}
    \mathbb{C}^{\gamma \delta}(t) =& \cP\{\text{active state at $t=0$}\} \times \n\\
    &\langle u^\gamma(t) u^\delta(0) |{\text{no arrest in $[0,t)$}}\rangle\> \times \notag\\
    & \cP\{\text{no arrest in $[0,t)$}\} \;,
\end{align}
where $\cP\{A\}$ denotes the probability of the event $A$ while $\langle \cdot |A\rangle$ is the conditional average constrained on the realization of the event $A$. Since $\mathbb{C}^{\gamma \delta}(t)$ is computed at steady state, $\cP\{\text{active state at $t=0$}\} = \beta/(\alpha + \beta)$. The exponential distribution of the arrest times also gives: $\cP\{\text{no arrest in $[0,t)$}\} = e^{-\alpha t}$. Finally, the conditional auto-correlation function $\langle u^\gamma(t) u^\delta(0) |{\text{no arrest in $[0,t)$}}\rangle$ reads, respectively for RTPs, ABPs, AOUPs:
\begin{eqnarray}
    \frac{\delta^{\gamma \delta}}{d}\begin{cases}
        & 1 \hspace{2.95cm} \text{(RTP)} \\
        & \exp[-(d-1) D_r t] \hspace{0.5cm} \text{(ABP)} \\
        & \exp[-t/\tau_0] \hspace{1.55cm} \text{(AOUP)}\\
    \end{cases}
\end{eqnarray}
so that, finally, the auto-correlation function reads:
\begin{eqnarray}
    \mathbb{C}^{\gamma \delta}(t) = \frac{\beta}{\alpha + \beta}e^{-\alpha t} \frac{\delta^{\gamma \delta}}{d}\begin{cases}
        & 1 \hspace{1.7cm} \text{(RTP)} \\
        & e^{-(d-1) D_r t} \hspace{0.3cm} \text{(ABP)} \\
        & e^{-t/\tau_0} \hspace{1.0cm} \text{(AOUP)}\\
    \end{cases}
    \label{eq:autocorr_multistate}
\end{eqnarray}
Eq.~\eqref{eq:autocorr_multistate} finally allows us to compute the persistence times for these processes:
\begin{eqnarray}
    \tau_{\rm p} = \frac{1}{1 + \alpha/\beta} \begin{cases}
        & \alpha^{-1} \hspace{2.45cm} \text{(RTP)} \\
        & [(d-1) D_r + \alpha]^{-1} \hspace{0.3cm} \text{(ABP)} \\
        & [\tau_0^{-1} + \alpha]^{-1} \hspace{1.2cm} \text{(AOUP)}\\
    \end{cases} \;.
    \label{eq:autocorr_2state}
\end{eqnarray}

\subsection{Active particles switching between $M$ motility states}
\label{subsec:multistate}

We conclude this section by studying the general case in which an active particle can switch between $M$ distinct motility states.
Here, we derive the expression of the auto-correlation tensor $\mathbb{C}^{\alpha\beta}(t)$ and the persistence time $\tau_{\rm p}$ for multi-state AOUPs.
As we show in the end, our computation can be straightforwardly extended to other types of motion such as ABPs.

We consider an AOUP dynamics endowed with $M$ distinct motility states $\{\mu_i\}_{i=1}^M$, each of which is characterized by an internal persistence time for the active motion $\tau_{\mu}>0$. In a given state $\mu$, the time evolution of the orientation vector $\bfu$ thus reads:
\begin{equation}
    \tau_\mu \dot u^\alpha(t) = - u^\alpha(t) + \sqrt{\frac{2 \tau_\mu}{d}} \eta^\alpha(t) \;,
    \label{eq:multistate_udyn}
\end{equation}
where $\eta^\alpha(t)$ is a Gaussian white noise with unit variance. The Poisson switching rates from state $\mu$ to $\nu$ are denoted by $k_{\mu\to \nu}$, while the total escape rate from state $\mu$ is simply indicated as $k_\mu \equiv \sum_\nu k_{\mu \to \nu}$.
During a switching event occurring at time $t$, $\tau_\mu \to \tau_\nu$ while the orientation vector $u^{\alpha}$ is unaltered.
At steady state, the probability for the particle to be in state $\mu$ is denoted by $p_\mu$.
Finally, the present discussion can be directly extended to include passive states where $\tau_\mu=0, u^\alpha=0$. In this case, whenever the particle switches from the arrested to the motile state, its orientation is drawn from the steady state distribution associated to the new active state.

We aim at expressing the steady-state autocorrelation function $\mathbb{C}^{\alpha\beta}(t)$ as:
\begin{equation}
    \mathbb{C}^{\alpha\beta}(t) = \sum_{n=0}^{\infty} c_n^{\alpha\beta}(t)
\end{equation}
where $c_n^{\alpha\beta}(t)$ contains the contribution to $\mathbb{C}^{\alpha\beta}$ corresponding to trajectories where exactly $n$ switching events occur in the time interval $[0,t)$.
For example, for $n=0$ we have:
\begin{equation}
    c_0^{\alpha\beta}(t) = 
    \frac{\delta^{\alpha\beta}}{d} \sum_{\mu=1}^M p_\mu e^{- \zeta_\mu t} \;,
\end{equation}
where we introduced the effective decay rates $\zeta_\mu \equiv k_\mu + \frac{1}{\tau_\mu}$. The next contribution, $c^{\alpha\beta}_1(t)$, can be found as:
\begin{eqnarray}
    c^{\alpha\beta}_1(t) &= & \sum_{\mu=1}^M p_\mu \int_0^t \rmd t_1 \> \sum_{\nu=1}^M \cP\{\text{$\mu \to \nu$ at  $t=t_1$}\} \notag \\[0.2cm]
    && \times \langle u^\alpha(t) u^\beta(0) | \text{ $\mu \to \nu$ at $t=t_1$}\rangle \notag \\[0.2cm]
    && \times \cP\{\text{no transition in $[t_1,t]$}\} \;.
    \label{eq:n_1_trans_step1}
\end{eqnarray}
To compute Eq.~\eqref{eq:n_1_trans_step1} we first remark that the probability density to jump from state $\mu$ to $\nu$ at time $t=t_1$ is given by the escape probability density times the probability to transition into state $\nu$, i.e.:
\begin{equation}
    \cP\{\text{$\mu \to \nu$ at  $t=t_1$}\} = \frac{k_{\mu\to\nu}}{k_\mu} k_\mu e^{-k_\mu t_1} = k_{\mu\to\nu} e^{-k_\mu t_1} \;.
\end{equation}
We then need to compute the autocorrelation function conditional on the occurrence of a single transition from $\mu \to \nu$ at time $t=t_1$. To do so, we use the propagator $G(t,0)$ of the dynamics~\eqref{eq:multistate_udyn} to express $\bfu(t)$ at time $t$:
\begin{align}
u^\alpha(t) = G(t,0) u^\alpha(0) + \int_0^t G(t,s) \sqrt\frac{2 d}{\tau_\mu(s)} \eta^\alpha(s) \notag \\
    \text{where} \quad G(t,s) = \exp\left[-\int_s^t \frac{1}{\tau_\mu(z)} \rmd z \right] \;.
\end{align}
For the purpose of computing the conditional correlations in Eq.~\eqref{eq:n_1_trans_step1}, the trajectory $\tau_\mu(t)$ is fixed and the propagator is deterministic~\footnote{Note that, in the limit where $\tau_\mu=0$, the process becomes passive: the propagator is equal to zero and, consequently, $\bfu(t)=0$ for any $t>0$.}. We can thus write the conditional auto-correlation function as:
\begin{multline}
    \langle u^\alpha(t) u^\beta(0) | \text{$\mu \to \nu$ at $t=t_1$}\rangle =  \frac{\delta^{\alpha\beta}}{d} G(t,0) \\
    = \frac{\delta^{\alpha\beta}}{d} \exp\left[ -\frac{t_1}{\tau_\mu}-\frac{t-t_1}{\tau_\nu}\right] \;.
\end{multline}
All in all, Eq.~\eqref{eq:n_1_trans_step1} explicitly reads:
\begin{eqnarray}
    c^{\alpha\beta}_1(t)  
    &=& \frac{\delta^{\alpha\beta}}{d} \sum_{\mu,\nu=1}^M   p_\mu k_{\mu \to \nu} \int_0^t \rmd t_1 \> e^{-\zeta_\mu t_1}  e^{-\zeta_\nu (t-t_1)} \;.\notag
\end{eqnarray}
Similarly, for $n=2$ transitions occurring in $[0,t]$:
\begin{eqnarray}
    c^{\alpha\beta}_2(t) &=&  \frac{\delta^{\alpha\beta}}{d} \sum_{\mu=1}^M p_\mu \int_0^t \rmd t_1 \int_{t_1}^t \rmd t_2 \> e^{-t_1/\tau_\mu} \notag \\
    &&\times \sum_{\nu=1}^M k_{\mu \to \nu} e^{-k_\mu t_1} e^{-(t_2-t_1)/\tau_\nu} e^{-k_\nu (t_2-t_1)}  \notag \\[0.2cm]
    && \times \sum_{\omega=1}^M k_{\nu \to \omega} e^{-(t-t_2)/\tau_\omega} e^{-k_\omega(t-t_2)} \notag \\[0.2cm]
    &=& \frac{\delta^{\alpha\beta}}{d} \sum_{\mu,\nu,\omega=1}^M  p_\mu k_{\mu \to \nu}k_{\nu \to \omega} \int_0^t \rmd t_1 \> e^{-\zeta_\mu t_1} \notag \\
    &&\times \int_{t_1}^t \rmd t_2 \>  e^{-\zeta_\nu (t_2-t_1)} \> e^{-\zeta_\omega(t-t_2)}\;.
\end{eqnarray}

This line of reasoning can be extended to compute any term $c^{\alpha\beta}_n$. All in all, we thus express the autocorrelation function as a series in $n$ via:
\begin{widetext}
\begin{eqnarray}
    \mathbb{C}^{\alpha\beta}(t) 
    &=& \frac{\delta^{\alpha\beta}}{d} \sum_{n=0}^\infty \sum_{\mu_1=1}^M \cdots \sum_{\mu_{n+1}=1}^M p_{\mu_1} k_{\mu_1 \to \mu_2} \cdots  k_{\mu_{n} \to \mu_{n+1}} \int_{t_0}^t \rmd t_1 \cdots \int_{t_{n-1}}^t \rmd t_n \prod_{k=1}^{n} \> e^{-\zeta_k(t_k-t_{k-1})} e^{-\zeta_{n+1}(t-t_{n})} \;, 
    \label{eq:Dyson}
\end{eqnarray}
\end{widetext}
where we defined $t_0 =0$.
To make analytical progress, we now show that Eq.~\eqref{eq:Dyson} corresponds to a Dyson series~\cite{vankampen1992stochastic}.
We first define the off-diagonal operator $\hat K_{\mu\nu} = k_{\mu \to \nu}$ for $\mu \neq \nu$, the diagonal operator $\hat Z_{\mu\nu} = \zeta_\mu \delta_{\mu\nu}$, and the exponential operator $\hat E(t) = e^{-Zt}$.
For a given term $n$ of the series, the integrand results from successive applications of the operators $\hat E$ and $\hat K$ to the row-vector $\mathbf{p}=(p_1,\dots,p_M)$.
This operation yields a vector $\boldsymbol{\ell}_n$ that reads:
\begin{multline}
    \boldsymbol{\ell}_n = \sum_{\mu_1=1}^M \cdots \sum_{\mu_n=1}^M p_{\mu_1} k_{\mu_1 \to \mu_2} \cdots  k_{\mu_{n} \to \mu_{n+1}} \times \\ \prod_{k=1}^{n} \> e^{-\zeta_k(t_k-t_{k-1})} e^{-\zeta_{n+1}(t-t_{n})} \\
    = \mathbf{p} \hat E(t_1) \hat K \hat E(t_2-t_1) \hat K \cdots \hat K \hat{E}(t-t_n) \;.
\end{multline}
The sum over the final state $\mu_{n+1}$ is finally obtained by projecting $ \boldsymbol{\ell}_n$ onto the column vector $\mathbf{1} = (1,\dots,1)^{\rm t}$. Our Dyson series can thus be re-written as:
\begin{multline}
    \mathbb{C}^{\alpha\beta} = \frac{\delta^{\alpha\beta}}{d} \sum_{n=0}^\infty a_n  \;, \\ \text{where} \quad a_n =\mathbf{p} \left(\int_{0 \leq t_1 \leq \dots \leq t_n}^t\prod_{i=1}^n \hat E(t_i-t_{i-1}) \hat{K} \rmd t_k\right) \mathbf{1} \;.
    \label{eq:DysonCorr}
\end{multline}
Next, we rely on the Duhamel expansion to write:
\begin{multline}
    e^{(-\hat Z+ \hat K)t} = e^{-\hat Zt}  + \int_0^t e^{-\hat Z s} \hat K e^{(-\hat Z + \hat K)(t-s)} \rmd s \\
    = \hat{E}(t) + \int_0^t \hat{E}(s) \hat K e^{(-\hat Z + \hat K)(t-s)} \rmd s \;.
    \label{eq:Duhamel}
\end{multline}
Recursive applications of Eq.~\eqref{eq:Duhamel} exactly yields the Dyson series:
\begin{multline}
    \bfp e^{(-\hat Z+ \hat K)t} = \bfp \hat{E}(t) + \int_0^t \rmd t_1 \bfp \hat{E}(t_1) \hat{K} \hat{E}(t-t_1) \\
    + \int_0^t \rmd t_1 \int_{t_1}^t \rmd t_2 \> \bfp \hat{E}(t_1) \hat{K} \hat{E}(t_2-t_1) \hat{K} \hat{E}(t-t_2) + \dots \;.
    \label{eq:DysonDuhamel}
\end{multline}
All in all, we can use the expansion Eq.~\eqref{eq:DysonDuhamel} to express Eq.~\eqref{eq:DysonCorr} in a compact form.
Upon defining the effective propagator: 
\begin{equation}
    \hat{\mathcal{L}} := \hat{Z} - \hat{K}
\end{equation}
we can finally write the autocorrelation function as:
\begin{eqnarray}
    \mathbb{C}^{\alpha\beta}(t) &=& \frac{\delta^{\alpha\beta}}{d} \mathbf{p} e^{- \hat{\mathcal{L}} t} \mathbf{1} \;, \\ \text{with} \quad
    \hat{\mathcal{L}}_{\mu\nu} &=& \begin{cases}
        -k_{\mu \to \nu} \hspace{0.8cm} \text{for } \mu \neq \nu \\
        k_\mu + \frac{1}{\tau_\mu} \qquad \text{for } \mu=\nu
    \end{cases} \quad \;.
    \label{eq:correlation_multistate}
\end{eqnarray}
The persistence time is then obtained integrating Eq.~\eqref{eq:correlation_multistate} over time, and yields:
\begin{equation}
    \tau_{\rm p} = \mathbf{p} \hat{\mathcal{L}}^{-1} \mathbf{1} = \sum_{\mu\nu}^M p_\mu (\hat{\mathcal{L}}^{-1})_{\mu\nu} \;.
    \label{eq:persistencetime_multistate}
\end{equation}

To test the validity of the above expression, one can show that Eq.~\eqref{eq:persistencetime_multistate} allows to recover the examples of $2$-state switching dynamics discussed in Sec.~\ref{subsec:2state}.
Furthermore, when $\tau_\mu \equiv \tau_0$ for all states, i.e. when all states are equivalent, Eq.~\eqref{eq:persistencetime_multistate} simply yields $\tau_{\rm p} = \tau_0$ as one would expect. 

We note that Eq.~\eqref{eq:persistencetime_multistate} is valid also for ABPs switching between $M$ states of motion with internal rotational diffusivities $D_{r,\mu}$, upon replacing $\tau_\mu \to (d-1) D_{r,\mu}$.
Indeed, the only difference with respect to the AOUP-computation lies in the autocorrelation function conditional on the occurrence of one transition $\mu \to \nu$ at time $t=t_1$, i.e. $\langle u^\alpha(t) u^\beta(0) | \text{ $\mu \to \nu$ at $t=t_1$}\rangle$.
To compute this object, we rely on the formal integration of the ABP dynamics Eq.~\eqref{eq:ABPdynamics} to write:
\begin{multline}
    u^\alpha(t) = G(t) u^\alpha(0)  \\ + \int_0^t \rmd s G(t-s) \sqrt{2 D_{r,\mu}(s)} [\delta^{\alpha\beta}-u^\alpha(s) u^\beta(s)] {\eta}^\beta(s)\;,
\end{multline}
where the propagator reads:
\begin{equation}
    G(t) = \exp\left[ -\int_0^t \frac{\rmd s}{(d-1) D_{r,\mu}(s)} \right].
\end{equation}
When computing the autocorrelation function constrained on a state trajectory, $D_{r,\mu}(s)$ is a deterministic function, so that we eventually obtain:
\begin{multline}
    \langle u^\alpha(t) u^\beta(0) | \mu \to \nu \text{ at }t=t_1\rangle = G(t) \frac{\delta_{\alpha\beta}}{d} \\
    = \exp \left[ -\frac{t_1}{\tau_{\mu}} -\frac{t-t_1}{\tau_{\nu}} \right]\;,
\end{multline}
where $\tau_\mu \equiv [(d-1) D_{r,\mu}]^{-1}$. The rest of the computation thus follows the same passages as for AOUPs. 

\begin{figure}
    \centering
    \includegraphics[width=1.0\columnwidth]{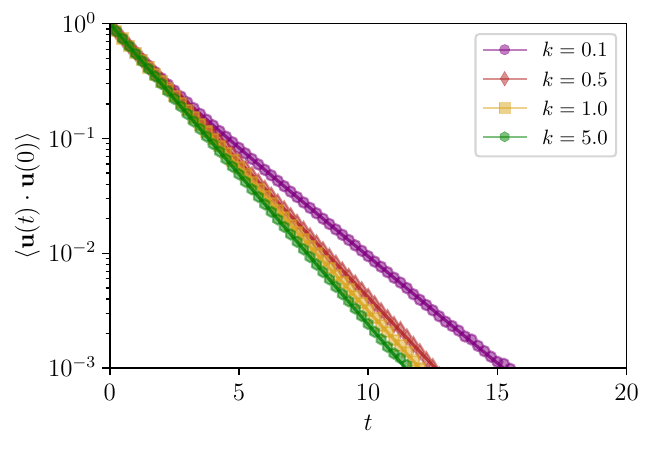}
    \caption{Steady-state auto-correlation function of the orientations $\langle \bfu(t) \cdot \bfu(0) \rangle$ for a $3$-state ABP process: comparison between theory (solid line) and numerical simulations in $d=2$ (symbols). In our simulations, an ABP undergoes a cycle between $3$ motility states $0 \to 1 \to 2 \to 0$, all with a Poisson rate $k$. The rotational diffusivity in each state is $D_{r,0}=1$, $D_{r,1}=1/2$, $D_{r,2}=1/3$, respectively. Parameters: $L_x=50$, $L_y=10$, $v_0=1$.}
    \label{fig:multistate_correlations}
\end{figure}

To test our result of Eq.~\eqref{eq:correlation_multistate} we performed particle-based simulation of an ABP that can cycle between $M=3$ different activity states, characterized by internal persistence times $\tau_0 = 1/[(d-1) D_{r,0}],\; \tau_1= 1/[(d-1) D_{r,1}]$ and $\tau_2= 1/[(d-1) D_{r,2}]$, respectively. In our cycle, we take the only non-zero switching rates to be $k_{0\to 1},\;k_{1\to 2},\;k_{2 \to 0}$ , and assume that they are all equal to a constant $k$. Equation~\eqref{eq:persistencetime_multistate} then yields the following result:
\begin{equation}
    \tau_{\rm p} = \frac{(\tau_0+\tau_1 + \tau_2) + 3k(\tau_0\tau_1 + \tau_1\tau_2 + \tau_2\tau_0) + 9 k^2 \tau_0 \tau_1 \tau_2}{3[1+k (\tau_0 +\tau_1 + \tau_2) + k^2 (\tau_0 \tau_1 + \tau_1 \tau_2 + \tau_2 \tau_0)]} \;.
    \label{eq:tau_3state_cycle}
\end{equation}
In Fig.~\eqref{fig:multistate_correlations} we show the comparison between the auto-correlation function as predicted by Eq.~\eqref{eq:correlation_multistate} and the one measured from numerical data, showing an excellent match between theory and simulations.

\subsection{Active particles with fractional Brownian noise}
\label{subsec:ABPfrac}
As a final example, we present a non-Markovian orientation process which does not admit a Markovian embedding via the addition of auxiliary variables.
While our coarse-grained theory is, in principle, not directly applicable to this case, we show in our companion paper~\cite{dinelli2026PRL} that 
the integrability of the auto-correlation tensor still allows to obtain excellent predictions for the large-scale drift and diffusion.

More specifically, we consider a $2d$ active Brownian particle whose angular dynamics undergoes a fractional Brownian motion~\cite{mandelbrot1968fractional,gomez2020active}:
\begin{equation}
    \dot\theta = \sqrt{2D_H} \xi_\theta(t) \;, \quad \theta\in [0, 2 \pi) \;,
\end{equation}
where $\xi_\theta$ has zero mean and covariance:
\begin{equation}
    \langle \xi_\theta(t) \xi_\theta(s)\rangle = H |t-s|^{2H-1} \left[\frac{2H-1}{|t-s|}+2\delta(t-s) \right] \;.
    \label{eq:autocorr_angle_frac}
\end{equation}
The parameter $H$ is called the Hurst parameter, and can vary between 0 and 1. 
For $H<1/2$, the angular dynamics features long-range negative autocorrelations, and the angle $\theta(t)$ is characterized by a subdiffusive behavior.
On the contrary, when $H>1/2$, Eq.~\eqref{eq:autocorr_angle_frac} dictates long-range positive autocorrelations, giving rise to angular superdiffusion.
The limit case of $H=1/2$ corresponds to standard Brownian motion.

The corresponding auto-correlation function for the orientation of the ABP, $\bfu = (\cos\theta, \sin\theta)$, was computed in~\cite{gomez2020active}. It corresponds to a stretched exponential and reads:
\begin{equation}
    \mathbb{C}^{\alpha\beta}(t) = \frac{\delta^{\alpha\beta}}{2} C(t) \;, \quad C(t) = \frac{1}{2}\delta^{\alpha\beta} \exp[-D_H t^{2H}] \;.
    \label{eq:correlator_fABP}
\end{equation}
From Eq.~\eqref{eq:correlator_fABP} we can compute the effective persistence time. It reads:
\begin{equation}
    \tau_{\rm p} = 2 \int_0^\infty C(t) \rmd t = \frac{1}{2H} \Gamma\left( \frac{1}{2H}\right) D_H^{-\frac{1}{2H}} \;.
\end{equation}

\newpage

\if{
\begin{table*}[ht]
\centering
\renewcommand{\arraystretch}{4}
\begin{tabular}{|c|c|c|}
\hline
\textbf{Orientation dynamics} & \textbf{Correlation function} $\mathbb{C}^{\alpha\beta}(t)$ & \makecell{\textbf{Persistence times} \\ $\tau^{\alpha\beta} = d \int_0^\infty \mathrm{d} t\, \mathbb{C}^{\alpha\beta}(t)$} \\
\hline
\hline
\makecell[c]{$d$-dimensional RTP \\ $\bfu \overset{\alpha}{\longrightarrow} \bfu' \in \mathbb{S}^{d-1}$} 
& \makecell[c]{$\dfrac{\delta^{\alpha\beta}}{d} \exp[-\alpha t]$} 
& \makecell[c]{$\alpha^{-1} \delta^{\alpha\beta} $} \\
\hline
\makecell{$d$-dimensional ABP \\[0.1cm] 
$\dot\bfu = - (d-1) D_r \bfu + \sqrt{2 D_r} (\bfI-\bfu^{\otimes 2}) \cdot \boldsymbol{\eta}(t)$} & $\dfrac{\delta^{\alpha\beta}}{d} \exp[-(d-1)D_r t]
$ 
& $[(d-1) D_r]^{-1} \delta^{\alpha\beta} $
\\
\hline
\makecell{$d$-dimensional AOUP \\[0.1cm] 
$\tau \dot\bfu = -\bfu + \sqrt{2\tau/d} \>\boldsymbol{\eta}(t)$} & $\dfrac{\delta^{\alpha\beta}}{d} \exp[-t/\tau]
$
& $\tau \delta^{\alpha\beta} $ \\
\hline
\makecell{$2d$-ABP with rotational inertia \\[0.1cm] 
$\dot\theta = \omega(t)$, 
$\quad I \dot\omega = -\omega + \sqrt{2 D_r} \xi_\omega(t)$} & $\dfrac{\delta^{\alpha\beta}}{2} \exp\left[ -D_rt - ID_r (1-e^{-t/I})\right]$
& $I e^{z} z^{-z} \Gamma(z,0,z) $, $\quad z = I D_r$ \\
\hline
\makecell{$2d$-ABP with chirality \\[0.1cm] 
$\dot\theta = \omega + \sqrt{2 D_r}\eta(t)$} & \makecell{$\dfrac{1}{2} e^{-D_r t} \begin{pmatrix}
        \cos(\omega t) & -\sin(\omega t) \\
        \sin(\omega t) & \cos(\omega t)  
\end{pmatrix}_{\alpha\beta}$}
& \makecell{$\dfrac{1}{D_r^2 + \omega^2} \begin{pmatrix}
        D_r & -\omega \\
        \omega & D_r 
    \end{pmatrix}_{\alpha\beta}$} \\
\hline
\makecell{$2d$-ABP with fractional Brownian noise \\[0.1cm] 
$\dot\theta = \sqrt{2 D_H}\eta(t)$, $\quad$ with\\[0.1cm] 
$\langle \eta(t) \eta(s)\rangle = H |t-s|^{2H-1} \left[\frac{2H-1}{|t-s|}+2\delta(t-s) \right]$}
& $\dfrac{\delta^{\alpha\beta}}{2} \exp[-D_H t^{2H}]$ & $\dfrac{1}{2H} \Gamma\left( \dfrac{1}{2H}\right) D_H^{-1/2H} \delta^{\alpha\beta}$ \\
 \hline
\makecell{$d$-dimensional RTP with\\ running and stopped state \\[0.1cm] 
$\bfu \overset{\alpha}{\longrightarrow} \boldsymbol{0} \overset{\beta}{\longrightarrow}  \bfu'  \in \mathbb{S}^{d-1}$ }
& $\dfrac{\delta^{\alpha\beta}}{d} \> \dfrac{\beta}{\alpha + \beta} \> \exp[{-\alpha t}]$ & $\alpha^{-1}\dfrac{\delta^{\alpha\beta}}{1+\alpha/\beta} $ \\
 \hline
 
\makecell{$d$-dimensional ABP with\\ running and stopped state \\[0.1cm] 
$\bfu \overset{\alpha}{\longrightarrow} \boldsymbol{0} \overset{\beta}{\longrightarrow}  \bfu'  \in \mathbb{S}^{d-1}\;,$\\
$\dot\bfu = - (d-1) D_r \bfu + (\bfI-\bfu^{\otimes 2}) \cdot \boldsymbol{\eta}(t)$}
& $\dfrac{\delta^{\alpha\beta}}{d} \> \dfrac{\beta}{\alpha + \beta} \> \exp\{{-[\alpha+D_r(d-1)] t}\}$ & $[\alpha+D_r(d-1)]^{-1}\dfrac{\delta^{\alpha\beta}}{1+\alpha/\beta} $ \\
 \hline
 
\makecell{$d$-dimensional AOUP with\\ running and stopped state \\[0.1cm] 
$\bfu \overset{\alpha}{\longrightarrow} \boldsymbol{0} \overset{\beta}{\longrightarrow}  \bfu'  \in \mathbb{S}^{d-1}\;,$\\
$\tau \dot\bfu = -\bfu + \sqrt{2\tau/d} \>\boldsymbol{\eta}(t)$}
& $\dfrac{\delta^{\alpha\beta}}{d} \> \dfrac{\beta}{\alpha + \beta} \> \exp\{{-[\alpha+\tau^{-1}] t}\}$ & $[\alpha+\tau^{-1}]^{-1}\dfrac{\delta^{\alpha\beta}}{1+\alpha/\beta} $ \\
 \hline
 
\makecell{$d$-dimensional AOUP with\\ $M$ motility states \\[0.1cm] 
State dynamics: $\mu \overset{k_{\mu \to \nu}}{\longrightarrow} \nu $\\
$\tau_\mu \dot\bfu = -\bfu + \sqrt{2\tau_\mu/d} \>\boldsymbol{\eta}(t)$}
& \makecell{$\dfrac{\delta^{\alpha\beta}}{d} \mathbf{p} e^{- \hat{\mathcal{L}} t} \mathbf{1} \;,$ \\[0.2cm]
$ \text{with} \>\>
\hat{\mathcal{L}}_{\mu\nu} = \begin{cases}
        -k_{\mu \to \nu} \hspace{1.55cm} \text{for } \mu \neq \nu \\
        \sum_\nu k_{\mu\to\nu} + \tau_\mu^{-1
        } \quad \text{for } \mu=\nu
    \end{cases},$\\[0.3cm]
$\bfp: \text{stationary distribution of states}$,
\\[0.2cm] $\boldsymbol{1}=(1,\dots,1)^{\rm t}$} 
& $\delta^{\alpha\beta} \sum_{\mu,\nu} p_\mu (\hat{\cL})^{-1}_{\mu\nu}$ \\
\hline
\end{tabular}
\caption{Correlation functions of different processes.}
\label{tab:correlations}
\end{table*}
}\fi

\newpage 

\section{Coarse-grained theory of single particles}
\label{sec:cg_particle}
In this section we rely on the expressions of the auto-correlation functions derived in Sec.~\ref{sec:correlations} to write down explicit coarse-grained dynamics for several models of single active particles subjected to motility regulation. For each of these cases, we then compute the associated steady-state particle distribution $p_{\rm s}(\bfr)$ and, when macroscopically out-of-equilibrium, the particle current $J^{\alpha}(\bfr)$.

\subsection{Passive limit}
We start by considering the case of passive particles, which can be obtained as a limit for $\tau_{\rm p} \to 0$ of the dynamics of active Ornstein-Uhlenbeck particles~\cite{martin2021statistical}. In this limit, the auto-correlation tensor $\mathbb{C}^{\alpha\beta} = 0$ identically, so that we obtain a purely diffusive theory with:
\begin{equation}
    \mathcal{D}^{\alpha\beta} = D_{\rm t} \delta^{\alpha\beta}
\end{equation}
and zero drift, as expected.

\subsection{Active particles with isotropic achiral orientational dynamics}
We consider a class of systems where the auto-correlation function of the orientations is proportional to the identity tensor, namely:
\begin{equation}
    \mathbb{C}^{\alpha\beta}(t) = C(t) \delta^{\alpha\beta} \;,
\end{equation}
where the associated persistence time is given by $\tau_{\rm p} = d \int_0^\infty C(t) \rmd t > 0$. 
This condition is generically satisfied by orientational processes that are isotropic and achiral. It thus includes the standard RTP, ABP and AOUP models studied in Sec.~\ref{subsec:RTPcorr}---\ref{subsec:AOUPcorr}, but also the other classes of systems discussed in Sec.~\ref{subsec:ABPinertia}---\ref{subsec:ABPfrac}. As we now show, our result for the large-scale drift and diffusion tensor thus extend previously known results for RTPs, ABPs and AOUPs~\cite{cates2013when,solon2015active,martin2021statistical,dinelli2024fluctuating} to a much broader class of systems.

\subsubsection{Space-dependent self-propulsion}
In the presence of space-dependent self-propulsion $v(\bfr)$, our coarse-grained equations~\eqref{eq:drift_final_taxis},~\eqref{eq:diffusion_final_taxis} yield the following drift and diffusion terms:
\begin{eqnarray}
    V^{\alpha} &=& \frac{\partial^\alpha v^2(\bfr)}{2 d} \tau_{\rm p} \;, \\[0.2cm] \cD^{\alpha\beta} &=& \left[ D_{\rm t} + \frac{v^2(\bfr) \tau_{\rm p}}{d}  \right] \delta^{\alpha\beta}\;.
\end{eqnarray}
The corresponding coarse-grained, Boltzmann-like steady-state distribution $p_{\rm s} (x)$ is then given by Eq.~\eqref{eq:singlepart_dist}, see also Ref.~\cite{cates2013when,solon2015active,martin2021statistical,dinelli2024fluctuating}.

\subsubsection{Taxis}
In the presence of tactic modulation of the self-propulsion speed, $v = v_0 - v_1 \bfu \cdot \nabla c$, our coarse-grained equations~\eqref{eq:drift_final_taxis} and~\eqref{eq:diffusion_final_taxis} give the following drift and diffusion terms:
\begin{eqnarray}
    V^{\alpha} &=& -\frac{v_1 \partial^\alpha c(\bfr)}{d} \;, \\[0.2cm] \cD^{\alpha\beta} &=& \left[ D_{\rm t} + \frac{v_0^2 \tau_{\rm p}}{d}  \right] \delta^{\alpha\beta}\;.
\end{eqnarray}
At the large-scale level, the coarse-grained particle distribution is an equilibrium distribution $p_{\rm s} (x) = \exp[-U(\bfr)]$ where the effective potential $U$ is given by Eq.~\eqref{eq:effective_pot_taxis}, see also Ref.~\cite{dinelli2024fluctuating}.

\subsection{Chiral active particles}
We consider chiral ABPs in $d$ dimensions subjected to motility regulation. The corresponding orientational dynamics is given by Eq.~\eqref{eq:chiralABPdynamics} and yields the auto-correlation matrix $\mathbb{C}^{\alpha\beta}$ in Eq.~\eqref{eq:chiral_correlations_3}, under the hypothesis that the anti-symmetric generator of rotations $\Omega^{\alpha\beta}$ is zero everywhere except for $\Omega^{xy} = -\Omega^{yx} = \omega$. This special case corresponds to a $d$-dimensional system in which a single axis of rotation, orthogonal to the $xy$ plane, is defined. Note that the following discussion can be extended directly to chiral AOUPs upon indentification of $(d-1) D_r \to \tau_0$, see Sec.~\ref{subsec:chiralABP}.

Before proceeding, we find it convenient to define the following effective persistence timescales:
\begin{eqnarray}
    \tau^{||} &:=& d \int_0^{\infty} \mathbb{C}^{xx} \rmd t = \frac{(d-1) D_r}{(d-1)^2 D_r^2 + \omega^2} \;,\\[0.2cm]
    \tau^{\rm{a}} &:=&  d \int_0^{\infty} \mathbb{C}^{yx} \rmd t = \frac{\omega}{(d-1)^2 D_r^2 + \omega^2} \;, \\
    \tau^{\perp} &:=&  d \int_0^{\infty} \mathbb{C}^{zz} \rmd t = \frac{1}{(d-1) D_r} \;.
\end{eqnarray}

\subsubsection{Space-dependent self-propulsion}
\label{subsec:single_particles_chiral_CG}
We first study the coarse-grained theory for chiral ABPs in the presence of space-dependent self-propulsion $v(\bfr)$, as described by Eqs.~\eqref{eq:diffusion_final} and~\eqref{eq:drift_final}.
The diffusion tensor resulting from Eqs.~\eqref{eq:diffusion_final} and~\eqref{eq:chiral_correlations_3} eventually reads:
\begin{equation}
\mathcal{D}(\bfr)
= D_{\rm t}\boldsymbol{1} + \frac{v^2(\bfr)}{d}
\begin{pmatrix}
\tau^{||}  & -\tau^{\rm a}  & 0 & \cdots & 0\\
\tau^{\rm a}  & \tau^{||} & 0 & \cdots & 0\\
0 & 0 & \tau^{\perp} &  & 0\\
\vdots & \vdots &  & \ddots & \\
0 & 0 & 0 &  & \tau^{\perp}
\end{pmatrix}\;.
\end{equation}
Similarly, the drift term reads:
\begin{equation}
    V^{\alpha} = \frac{\partial^\beta v^2 (\bfr)}{2} \begin{pmatrix}
\tau^{||}  & -\tau^{\rm a}  & 0 & \cdots & 0\\
\tau^{\rm a}  & \tau^{||} & 0 & \cdots & 0\\
0 & 0 & \tau^{\perp} &  & 0\\
\vdots & \vdots &  & \ddots & \\
0 & 0 & 0 &  & \tau^{\perp}
\end{pmatrix}^{\alpha\beta} = \frac{1}{2}\partial^{\beta} \cD^{\alpha\beta}\;.
\end{equation}
For simplicity, we consider here the case of an activity landscape $v(\bfr)= v(x)$ that only varies along the spatial direction $x$.
In this case, the steady-state solution $p_{\rm s}(\bfr) = p_{\rm s}(x)$ is translationally invariant along all the directions orthogonal to $x$.
As we show below, the corresponding steady-state current $J^\alpha$ then vanishes for all $\alpha \neq y$, while the system sustains a finite particle current $J^y$ along $y$.

\begin{figure}
    \centering
    \includegraphics[width=1.0\columnwidth]{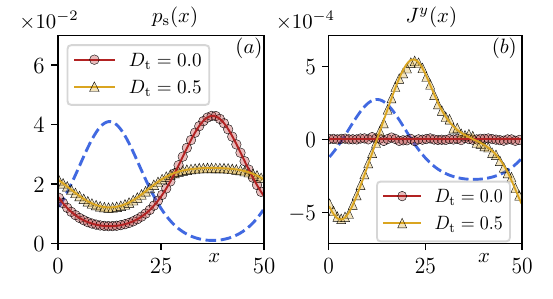}
    \caption{Steady-state distribution $p_{\rm s}(x)$ (a) and particle current $J^y(x)$ (b) of chiral ABPs subjected to space-dependent self-propulsion speed $v(x)$: comparison between theory (solid line, Eqs.~\eqref{eq:ps_chiralABP_vr},~\eqref{eq:Jy_chiralABP_vr}) and numerical simulations in $2d$ (symbols). The space-dependent self-propulsion speed $v(x) = v_0 \exp[A \sin( \frac{2\pi}{L_x}x)]$ is represented by a dashed blue line and, for the purpose of visualization, it has been rescaled and shifted. Parameters: $L_x=50$, $L_y=10$, particle density $\rho_0 = 10$, $A=1$, $v_0=1$, $D_r=1$, $\omega=1$.}
    \label{fig:chiral_ABP_vx}
\end{figure}

We start by solving the condition of zero flux along $x$, $J^x = 0$, which provides us with the steady-state particle distribution:
\begin{eqnarray}
    0 &=& \frac{1}{2} \partial^x \cD^{xx} p_{\rm s}(x) - p_{\rm s}(x) \partial_x \cD^{xx} - \cD^{xx} \partial_x p_{\rm s}(x) \notag
\end{eqnarray}
resulting in:
\begin{equation}
    p_{\rm s}(x) = \frac{1}{Z} \frac{1}{\sqrt{\cD^{xx}(x)}} =\frac{1}{Z}  \left( D_{\rm t} + \frac{v^2(x)}{d} \tau^{||} \right)^{-1/2} \;,
    \label{eq:ps_chiralABP_vr}
\end{equation}
where $Z$ is a normalization factor.
The current along any direction $\alpha \neq x$ then reads:
\begin{equation}
    J^{\alpha}(x) = -\frac{1}{2} (\partial^x \cD^{\alpha x} ) p_{\rm s} - (\partial^x p_{\rm s}) \cD^{\alpha x} \;.
    \label{eq:Jalpha_chiralABP}
\end{equation}
For $\alpha \neq x, y$, Eq.~\eqref{eq:Jalpha_chiralABP} is directly zero, since the corresponding value of $\cD^{\alpha x} = 0$. The only non-zero current thus emerges along the $y$ direction: 
\begin{eqnarray}
    J^y &=& -\frac{1}{2} (\partial^x \cD^{y x} ) p_{\rm s} - (\partial^x p_{\rm s}) \cD^{y x}\;,
\end{eqnarray}
which finally is given by:
\begin{equation}
\label{eq:Jy_chiralABP_vr}
    J^y(x) = -\frac{\tau^{\rm a}}{2 d} p_{\rm s}(x) \partial_x v^2 \left[ 1 - \frac{\tau^{||} v^2(x)}{d \; D_{\rm t} + \tau^{||} v^2(x)}\right]\;.
\end{equation}
As expected, we observe that, when $\tau^{\rm a} = 0$, no current is observed in the system.
Furthermore, in the limit of athermal particles $D_{\rm t}=0$, we retrieve the equilibrium solution with $J^\alpha = 0$.
In Fig.~\ref{fig:chiral_ABP_vx}, we show a very good agreement between our theoretical predictions given by Eqs.~\eqref{eq:ps_chiralABP_vr},~\eqref{eq:Jy_chiralABP_vr} and numerical simulations of a $2$-dimensional chiral ABP.

\subsubsection{Taxis}
We conclude our discussion by considering tactic modulation of the self-propulsion speed $v = v_0 - v_1 \bfu \cdot \nabla c(\bfr)$, under the presence of an external field $c(\bfr)$. The diffusion matrix resulting from Eq.~\eqref{eq:diffusion_final_taxis} is constant and reads:
\begin{eqnarray}
\mathcal{D}
= D_{\rm t}\boldsymbol{1} + \frac{v^2}{d}
\begin{pmatrix}
\tau^{||}  & -\tau^{\rm a}  & 0 & \cdots & 0\\
\tau^{\rm a}  & \tau^{||} & 0 & \cdots & 0\\
0 & 0 & \tau^{\perp} &  & 0\\
\vdots & \vdots &  & \ddots & \\
0 & 0 & 0 &  & \tau^{\perp}
\end{pmatrix}\;,
\end{eqnarray}
while the space-dependent drift obtained from Eq.~\eqref{eq:drift_final_taxis} reads:
\begin{equation}
    V^\alpha(\bfr) =  -\frac{v_1 \partial^\alpha c(\bfr)}{d} \;.
\end{equation}
As is the case for chiral particles with space-dependent self-propulsion speed, finding a general solution for the steady-state particle current $J^\alpha$ and distribution $p_{\rm s}(\bfr)$ is not trivial. Here, we restrict ourselves to the case of $c(\bfr) = c(x)$, so that the drift term is only present along $x$.
In this case, it is possible to find a solution for $p_{\rm s}$ that depends only on $x$, resulting in a current $J^x = 0$:
\begin{equation}
    0 = \frac{-v_1 \partial^{x} c}{d} p_{\rm s}(x) - \cD^{xx} \partial^x p_{\rm s}\;,
\end{equation}
which gives:
\begin{equation}
    p_{\rm s}(x) = \frac{1}{Z} \exp \left[ - \frac{v_1 c}{d \cD^{xx}}\right] \;,
    \label{eq:chiral_taxis}
\end{equation}
where the prefactor $Z$ ensures the normalization of $p_{\rm s}$ over the periodic domain $x \in [0, L)$.
The solution Eq.~\eqref{eq:chiral_taxis} can be used to determine the current orthogonal to $x$, i.e. $J^\alpha$ for $\alpha \neq x$. For $\alpha \neq x,y$, the current $J^{\alpha}$ can be directly shown to vanish. On the contrary, along the $y$ direction we obtain a finite current:
\begin{equation}
    J^y =  - \cD^{yx} \partial^x p_{\rm s}(x) \;.
    \label{eq:taxis_current}
\end{equation}
Differently from the case of active particles with spatially modulated self-propulsion, discussed in the previous section, the athermal limit $D_{\rm t} = 0$ still yields a finite particle current.

\begin{figure}
    \centering
    \includegraphics[width=1.0\columnwidth]{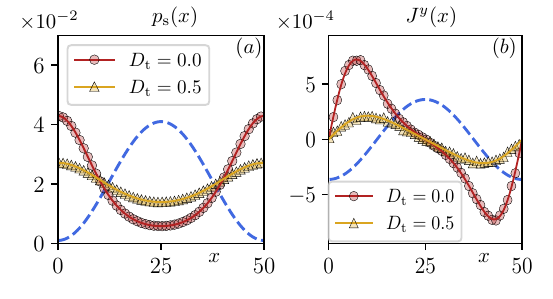}
    \caption{Steady-state distribution $p_{\rm s}(x)$ (a) and particle current $J^y(x)$ (b) of chiral ABPs subjected to tactic regulation of the self-propulsion speed: comparison between theory (solid lines, Eqs.~\eqref{eq:chiral_taxis},~\eqref{eq:taxis_current}) and numerical simulations in $2d$ (symbols). The external tactic field $c(x) = -A \cos( \frac{2\pi}{L_x}x)$ is represented by a dashed blue line and, for the purpose of visualization, it has been rescaled and shifted. Parameters: $L_x=50$, $L_y=10$, particle density $\rho_0 = 10$, $A=1$, $v_0=1$, $v_1 = 0.5$, $D_r=1$, $\omega=1$.}
    \label{fig:chiral_ABP_taxis}
\end{figure}

To test our coarse-grained theory, we measure the chiral current $J^y(x)$ and the steady-state particle distribution $p_{\rm s}(x)$ in particle-based simulations of tactic chiral ABPs.
In our simulations we modulate the self-propulsion speed according to $v=v_0 - v_1 \partial^\alpha c(x) u^\alpha$, where is $c(x) = -A \cos( \frac{2\pi}{L_x}x)$, $L_x$ being the system size along the $x$ direction.
In Fig.~\ref{fig:chiral_ABP_taxis} we report an excellent agreement between our measurements and theoretical predictions of Eqs.~\eqref{eq:chiral_taxis}---\eqref{eq:taxis_current}, both in the athermal and finite-temperature case.

\newpage

\section{Coarse-grained theory of active polymers}
\label{sec:cg_polymer}
In this section, we directly apply our results for the macroscopic drift and diffusion tensor to obtain 
coarse-grained theories of active polymers. We first show that our general expressions for drift and diffusion succesfully retrieve previous results present in the literature across different systems~\cite{vuijk2021chemotaxis,muzzeddu2022active,muzzeddu2023taxis,muzzeddu2024migration}. Furthermore, we propose and study new models for the dynamics of active polymers, ranging from tactic (Sec.~\ref{subsec:polymer_simple}) to chiral (Sec.~\ref{subsec:polymer-chiral}) polymers, and also studying the effect of orientational synchronization between monomers (Sec.~\ref{subsec:polymer_correlated}). All predictions for these new cases are complemented with corresponding numerical measurements of the steady-state center-of-mass distribution and, when non-vanishing, of the steady-state current.

Most of the following discussion will focus on the behavior of active polymers with space-dependent self-propulsion speed, due to the interesting transition from high-activity to low-activity accumulation.
Our study highlights several possible transition pathways, which may rely on the number of monomers, the topology of the polymer, the presence of chiral units or synchronizaton between the orientation vectors. 

\subsection{Active polymer with uncoupled monomer orientations}
\label{subsec:polymer_simple}
As a first example, we consider the case of an active polymer where the orientations $\Theta$ of the monomers are independent of one another, and each of them decorrelate exponentially with the same characteristic time $\tau_{\rm p}$, i.e., 
\begin{align}
    \mathbb{C}_{ij}^{\alpha \beta}(t)=\frac{\delta^{\alpha\beta} \delta_{ij}}{d}\exp(-t/\tau_{\rm p})\,.
    \label{eq:Cijab_exp}
\end{align}

\subsubsection{Space-dependent self-propulsion}
The physics of this system has recently been studied by Refs.~\cite{muzzeddu2024migration,ravichandir2025transport} for AOUP- and ABP-monomers, respectively.
Under the hypothesis Eq.~\eqref{eq:Cijab_exp}, the drift term reads:
\begin{align}
    V^{\alpha}(\bfR) &= \frac{\partial^\beta v^2(\bfR)}{2N}\sum_{ijk}  \varphi_{ji} \varphi_{jk}  \int_0^\infty \rmd t\,  e^{-\lambda_j t} \mathbb{C}_{ik}^{\alpha \beta}(t) \notag \\
    &= \delta^{\alpha\beta} \frac{\partial^\beta v^2(\bfR)}{2dN}\sum_{ijk}  \delta_{ik} \varphi_{ji} \varphi_{jk}  \int_0^\infty \rmd t\,  e^{-(\lambda_j+1/\tau_{\rm p}) t} \notag \\[0.2cm]
    &= \delta^{\alpha\beta} \frac{\partial^\beta v^2(\bfR)}{2dN}\sum_{ij} \varphi_{ji} \varphi_{ji} \frac{1}{\lambda_j + 1/\tau_{\rm p}} \;,
\end{align}
where we remind that $\lambda_i = \tau_{\rm r}^{-1} \sigma_i$, $\{\sigma_i\}$ are the eigenvalues of the connectivity matrix and $\tau_{\rm r} = \gamma/\kappa$ is the typical relaxation time-scale for the elastic interactions.
Using the fact that $\varphi_{ji}$ is orthogonal, i.e., $\sum_{k}\varphi_{jk} \varphi_{ik} = \delta_{ij} $, we have $\sum_{j} \varphi_{ji} \varphi_{ji} = 1$ and:
\begin{align}
    V^{\alpha}(\bfR) &=  \frac{\tau_{\rm p} \partial^\alpha v^2(\bfR)}{2 d N} \sum_{i=0}^{N-1} \frac{1}{1 + \tau_{\rm p} \lambda_i}\;.
    \label{eq:drift_polymer}
\end{align}
Next, we look at the diffusion matrix $\cD^{\alpha\beta}$, defined via Eq.~\eqref{eq:diffusion_final}. Since the process is isotropic and achiral, i.e. $\mathbb{C}_{ij}^{\alpha \beta}(t) \propto \delta^{\alpha\beta}$, the diffusion matrix reads $\cD^{\alpha\beta} = \cD_0 \delta^{\alpha\beta}$, where:
\begin{equation}
    \cD_0(\bfR) = \frac{D_{\rm t}}{N}  +\frac{\tau_{\rm p} v^2(\bfR)}{dN}  \;.
    \label{eq:diff_active_polymer}
\end{equation}
We know from Sec.~\ref{subsec:equil_Rouse_space} that such a macroscopic description corresponds to an effective equilibrium regime, characterized by the Boltzmann steady-state particle distribution $p_{\rm s} = \exp[- U(\bfR)]$ given by Eq.~\eqref{eq:effective_pot_selfprop}. For this class of models, the effective potential $U(\bfR)$
 reads:
 \begin{equation}
     U(\bfR) = \frac{\epsilon}{2} \log \cD_0\;,
 \end{equation}
where $\epsilon$ is given by Eq.~\eqref{eq:eps_value}:
\begin{equation}
    \epsilon = 2 -\sum_{i=0}^{N-1} \frac{1}{1 + \tau_{\rm p} \lambda_i} =  1-\sum_{i=1}^{N-1} \frac{1}{1 + \tau_{\rm p} \lambda_i}\;.
    \label{eq:eps_active_polymers}
\end{equation}
Notably, when $\epsilon < 0$ the minima of the effective potential $U$ correspond to the maxima of the diffusivity, $\cD_0$, so that the center of mass of the polymer accumulates in regions of high-activity. 
Equation~\eqref{eq:eps_active_polymers} thus recovers the results of Refs.~\cite{muzzeddu2024migration,ravichandir2025transport}, revealing the transition from low-activity to high-activity accumulation upon changing the polymer topology, and hence the eigenvalues $\{\sigma_i\}$, or the number of monomers $N$.

\subsubsection{Taxis}
Next, we consider an active polymer whose monomers undergo a tactic active process. More specifically, the self-propulsion speed of monomer $i$ adapts to local chemical gradients as $v(\bfr_i) = v_0 - v_1 u^\alpha_i \partial^\alpha c(\bfr_i)$. To the best of our knowledge, the phenomenology emerging from this model has not been reported in the literature. 

\begin{figure}
    \centering
    \includegraphics[width=\columnwidth]{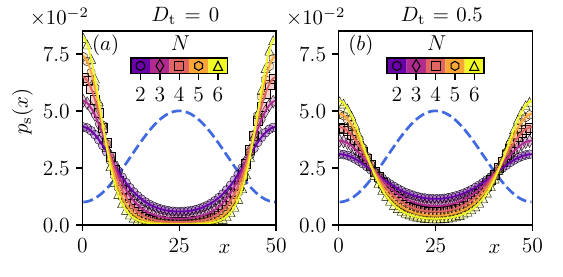}
    \caption{Steady-state distribution $p_{\rm s}(x)$ of a tactic ABP-chain: comparison between theory (solid line, Eq.~\eqref{eq:polymertaxis-steadystate}) and numerical simulations in $2d$ (symbols). ({\bf a}) Athermal particles, $D_{\rm t} = 0$. ({\bf b}) Finite temperature, $D_{\rm t}>0$. The external tactic field $c(x) = -A \cos( \frac{2\pi}{L_x}x)$ is represented by a dashed blue line and, for the purpose of visualization, it has been rescaled and shifted. Different colors and symbols indicate simulations performed for different chain lengths $N$. Parameters: $L_x=50$, $L_y=10$, monomer density $\rho_0 = 10 N$, $A=1$, $v_0=1$, $v_1=0.5$, $D_r=1$.}
    \label{fig:polymertaxis}
\end{figure}

We assume that the auto-correlation function of the orientations is described by Eq.~\eqref{eq:Cijab_exp}. Then, using the results obtained in Sec.~\ref{subsec:Rouse_taxis}, we can directly determine the steady-state distribution $p_s(\bfR)$ of the center-of-mass via Eqs.~\eqref{eq:D0_Ntaxis}, \eqref{eq:effective_pot_Ntaxis}, namely:
\begin{eqnarray}
    p_{\rm s}(\bfR) &=& \exp[ - U(\bfR)]\;, \quad \text{with} \notag \\ U(\bfR)  &=& - \frac{v_1}{d \cD_0} c(\bfR), \qquad \cD_0 = \frac{D_{\rm t}}{N} + \frac{\tau v_0^2}{dN} \;.
    \label{eq:polymertaxis-steadystate}
\end{eqnarray}
In Fig.~\ref{fig:polymertaxis} we verify the theoretical predictions obtained via our coarse-graining method by comparing them with numerical simulations of the microscopic dynamics of an active chain composed of ABP-monomers in $2d$.
We work in the chemo-repelling regime $v_1>0$, favoring accumulation in regions where $c(\bfr)$ is low, and consider an external field $c$ that depends only on the $x$ coordinate.
We observe an excellent agreement between theoretical predictions for $p_{\rm s}(x)$ and numerical sampling of the steady-state probability distribution, both for zero and finite thermal diffusivity $D_{\rm t}$.
Our simulations and theory confirm that, upon increasing the chain size, the effective temperature of the system is lowered and accumulation is enhanced. As expected, no transition from accumulation in high-$c$ to low-$c$ regions is observed upon varying the chain size $N$.

\subsection{Chiral active polymers}
\label{subsec:polymer-chiral}
We now consider the role of chirality in active polymers, which has recently drawn increasing interest for the emergent odd mechanics displayed by these systems ~\cite{epstein2020time,hargus2020time,hargus2021odd}.
In the context of motility-regulated active systems, we show how chirality can both induce steady-state out-of-equilibrium currents and trigger the transition from high-activity to low-activity accumulation~\cite{muzzeddu2022active}. 
In the following discussion, we restrict ourselves to the case of space-dependent self-propulsion speed,  which yields the richest phenomenology.

More precisely, we study a polymer consisting of $N$ active beads, 
each of which undergoes the chiral-ABP (or AOUP) dynamics given by Eqs.~\eqref{eq:chiralABPdynamics},~\eqref{eq:chiralAOUPdynamics}. 
As in previous cases, the following results are equivalent upon mapping $(d-1) D_r$ (ABP) $\to$ $\tau_0^{-1}$ (AOUP).
For each monomer $i$, we denote the associated antisymmetric generator of rotations by $\boldsymbol{\Omega}_i$.
Furthermore, we assume that the only non-zero components of $\boldsymbol{\Omega}_i$ are $\Omega_i^{xy} = -\Omega_i^{yx} = -\omega_i$, 
so that chirality is present only in the $xy$ plane for all monomers.
We can thus use the expression in Eq.~\eqref{eq:chiral_correlations_3} for $\mathbb{C}^{\alpha\beta}_{ij}$ to compute 
the large-scale diffusion tensor:
\begin{equation}
    \cD^{\alpha\beta}(\bfR) = \frac{D_{\rm t}}{N} \delta^{\alpha\beta} + \frac{v^2(\bfR)}{dN^2} 
    \sum_{i=0}^{N-1} [(d-1) D_r \boldsymbol{1} + {\boldsymbol{\Omega}}_i]^{-1}_{\alpha\beta}\;.
    \label{eq:meso_diffusion_final_chiral}
\end{equation}
We note that the matrix $(d-1) D_r \boldsymbol{1}+{\boldsymbol{\Omega}}_i$ has only two non-zero blocks, one corresponding
to the $xy$-submatrix and the second one corresponding to the identity in the orthogonal subspace.
Matrix inversion
can be directly performed yielding:
\begin{equation}
\mathcal{D}(\bfR)
= D_{\rm t}\boldsymbol{1} + \frac{v^2(\bfR)}{d N^2} \sum_{i=0}^{N-1}
\begin{pmatrix}
\tau_i^{||}  & -\tau_i^{\rm a}  & 0 & \cdots & 0\\
\tau_i^{\rm a}  & \tau_i^{||} & 0 & \cdots & 0\\
0 & 0 & \tau^{\perp} &  & 0\\
\vdots & \vdots &  & \ddots & \\
0 & 0 & 0 &  & \tau^{\perp}
\end{pmatrix}\;.
\label{eq:chiral_polymer_diff}
\end{equation}
where:
\begin{eqnarray}
    \tau^{||}_i &:=& \frac{(d-1) D_r}{(d-1)^2 D_r^2 + \omega_i^2} \;,\\[0.2cm]
    \tau^{\rm{a}}_i &:=& \frac{\omega_i}{(d-1)^2 D_r^2 + \omega_i^2} \;, \\
    \tau^{\perp} &:=& \frac{1}{(d-1)D_r} \;.
\end{eqnarray}
Similarly, the drift term reads:
\begin{equation}
    V^{\alpha} = \frac{\partial^\beta v^2 (\bfR)}{2 d N} \sum_{ij} \varphi_{ji}^2 \{ [(d-1) D_r +  \lambda_j] \boldsymbol{1} + {\boldsymbol{\Omega}}_i\}^{-1}_{\alpha\beta} \;.
\end{equation}
As an example, we focus on the simpler case of a dimer ($N=2$). 
We denote by $\omega_0,\omega_1$ the chiralities of each monomer. Furthermore, we note that, in this case,
$\varphi_{ji}^2 = 1/2$ for all $i,j$ and the Rouse eigenvalues $\lambda_j$ are $\tau_{\rm r}^{-1} \cdot \{0,2\}$. Therefore, for an active
dimer the drift term becomes:

\begin{align}
V^\alpha = \frac{\partial^\beta v^2 (\bfR)}{8 d} \sum_{i=0,1}
\begin{pmatrix}
\tilde\tau_i^{||}  & -\tilde\tau_i^{\rm a}  & 0 & \cdots & 0\\
\tilde\tau_i^{\rm a}  & \tilde\tau_i^{||} & 0 & \cdots & 0\\
0 & 0 & \tilde{\tau}^{\perp} &  & 0\\
\vdots & \vdots &  & \ddots & \\
0 & 0 & 0 &  & \tilde{\tau}^{\perp}
\end{pmatrix}^{\alpha\beta}\;.
\label{eq:dimer_chiral_drift}
\end{align}
with:
\begin{eqnarray*}
    \tilde\tau^{||}_i &:=& \frac{(d-1) D_r}{(d-1)^2 D_r^2 + \omega_i^2} + \frac{(d-1) D_r + 2 \tau_{\rm r}^{-1}}{[(d-1) D_r + 2 \tau_{\rm r}^{-1}]^2 + \omega_i^2} \;,\\[0.2cm]
    \tilde\tau^{\rm{a}}_i &:=& \frac{\omega_i}{(d-1)^2 D_r^2 + \omega_i^2} + \frac{\omega_i}{[(d-1) D_r + 2 \tau_{\rm r}^{-1}]^2 + \omega_i^2}  \;, \\
    \tilde{\tau}^{\perp} &:=& \frac{1}{(d-1)D_r} + \frac{1}{(d-1)D_{r} + 2\tau_{\rm r}^{-1}} \;.
\end{eqnarray*}
Interestingly, Eqs.~\eqref{eq:chiral_polymer_diff},~\eqref{eq:dimer_chiral_drift} reveal that, when the two monomers are counter-rotating, i.e. $\omega_0 = -\omega_1$, the resulting drift and diffusion lose their anti-symmetric components: $\sum_i \tilde\tau_i^{\rm a} = \sum_i\tau_i^{\rm a} = 0$. In this limit case, one recovers a macroscopic equilibrium theory, with a transition from high-activity to low-activity accumulation~\cite{muzzeddu2022active}.

\begin{figure}
    \centering
    \includegraphics[width=1.0\columnwidth]{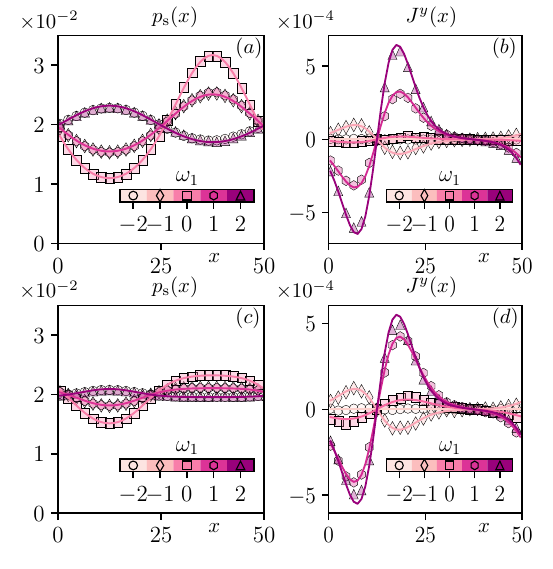}
    \caption{Steady-state distribution $p_{\rm s}(x)$ (a, c) and particle current $J^y(x)$ (b, d) of a dimer ($N=2$) of chiral ABPs subjected to space-dependent self-propulsion speed $v(x)$: comparison between theory (solid line, Eqs.~\eqref{eq:ps_chiralABP-dimer_vr} and~\eqref{eq:Jy_chiralABP-dimer_vr}) and numerical simulations in $2d$ (symbols). ({\bf a, b}) Athermal particles, $D_{\rm t}=0$. ({\bf c, d}) Finite temperature, $D_{\rm t}>0$. In all panels, the space-dependent self-propulsion speed is $v(x) = v_0 \exp[A \sin( \frac{2\pi}{L_x}x)]$. The chirality of the first monomer is fixed to $\omega_0=2$, while the second one, $\omega_1$, is varied, corresponding to different colors and symbols. Parameters: $L_x=50$, $L_y=10$, monomer density $\rho_0 = 20$, $A=1$, $v_0=1$, $D_r=1$, $\omega_0=2$.
    }
    \label{fig:chiral_ABP_dimers_vx}
\end{figure}

To broaden the scope of the discussion beyond the equilibrium limit, we consider an activity $v(\bfR) = v(x)$ and solve for the steady-state particle distribution $p_{\rm s}(x)$ and steady-state current $J^\alpha$. In this case, we note that one can write:
\begin{eqnarray}
    V^x &=& \left( 1 - \frac{\epsilon^{||}}{2}\right) \partial^x \cD^{xx}\;, \\
    V^y &=& \left( 1 - \frac{\epsilon^{a}}{2}\right) \partial^x \cD^{yx} 
\end{eqnarray}
with:
\begin{equation}
    \epsilon^{||} = 2 - \dfrac{\sum_{i=0,1} \tilde\tau_i^{||}}{\sum_{i=0,1} \tau_i^{||}} \;, \quad \epsilon^{a} = 2 - \dfrac{\sum_{i=0,1} \tilde\tau_i^{a}}{\sum_{i=0,1} \tau_i^{a}}\;.
\end{equation}
A steady-state particle distribution $p_{\rm s}(x)$ with zero-flux along $x$, $J^x=0$, can then be found:
\begin{eqnarray}
    0 &=& (1-\epsilon^{||}/2) \partial^x \cD^{xx} p_{\rm s}(x) - p_{\rm s}(x) \partial_x \cD^{xx} - \cD^{xx} \partial_x p_{\rm s}(x) \notag\;, 
\end{eqnarray}
which gives:
\begin{equation}
    p_{\rm s}(x) = \frac{1}{Z} \cD^{xx}(x)^{-\epsilon^{||}/2} =\frac{1}{Z}  \left( \frac{D_{\rm t}}{2} + \frac{v^2(x)}{4 d} \sum_{i=1}^2 \tau_i^{||} \right)^{-\epsilon^{||}/2} \;,
    \label{eq:ps_chiralABP-dimer_vr}
\end{equation}
where $Z$ is a normalization factor. As for single particles, see Sec.~\ref{subsec:single_particles_chiral_CG}, 
the current along any direction $\alpha \neq x, y$ can be directly shown to be $0$. On the contrary, along $y$: 
\begin{eqnarray}
    J^y &=& V^y(x) p_{\rm s}(x) - \partial^x [\cD^{yx}(x) p_{\rm s}(x)] \notag \\
    &=& -\frac{\epsilon^a}{2} (\partial^x \cD^{y x} ) p_{\rm s} - (\partial^x p_{\rm s}) \cD^{y x}\;.
\end{eqnarray}
The resulting expression for the current then reads:
\begin{equation}
    J^y(x) = -\frac{\sum_i \tau_i^{a}}{8 d} p_{\rm s}(x) \partial_x v^2 \left[ \epsilon^{a} - \frac{ \epsilon^{||}\sum_i \tau_i^{||} v^2(x)}{2 d \; D_{\rm t} + \sum_i \tau_i^{||} v^2(x)}\right]\;.
    \label{eq:Jy_chiralABP-dimer_vr}
\end{equation}
As expected from Ref.~\cite{muzzeddu2022active}, we observe that $\sum_i \tau_i^a = 0$, corresponding to counter-rotating monomers, yields zero particle current. Moreover, differently from the single-particle case, a finite current can still be observed for athermal dimers ($D_{\rm t} = 0$), Fig.~\ref{fig:chiral_ABP_dimers_vx}b. Our results are then tested in $2d$ simulations of ABP monomers in an activity landscape $v(x)$. As reported in Fig.~\ref{fig:chiral_ABP_dimers_vx}, our predictions of the particle distribution and current are correctly captured by Eqs.~\eqref{eq:ps_chiralABP-dimer_vr},~\eqref{eq:Jy_chiralABP-dimer_vr}.

\subsection{Active polymers with orientational couplings}
\label{subsec:polymer_correlated}
In the standard active Rouse description, monomers are driven by independent active forces, so activity acts as a spatially uncorrelated noise injected into an otherwise elastic chain. 
Here we extend the model by introducing cross-correlations between the orientations 
of different monomers, allowing for synchronization or 
anti-synchronization along the chain.
We then study the resulting coarse-grained dynamics and
show the impact of synchronization and anti-synchronization on the collective behavior of the polymer. 
In particular, we reveal that, while diffusion is left unchanged by the orientational couplings,
drift can be either enhanced or inhibited by it.
In the rest of this section, we will consider exclusively the case of space-dependent self-propulsion speed $v(\bfr)$
and leave aside the case of tactic interactions.

As a microscopic reference dynamics, we consider a polymer in $d$ dimensions consisting of $N$ AOUPs
whose orientation dynamics is described by:
\begin{equation}
    \dot u^\alpha_i = - \tau_0^{-1} u^\alpha_i - \sum_{j=0}^{N-1} h_{ij} (u^\alpha_i - u^\beta_j) + \sqrt{\frac{2}{d\tau_0}} {\eta_i^\alpha}(t) \;.
\label{eq:correlatedAOUPdynamics}
\end{equation}
Here, the coupling $h_{ij}$ enforces alignment or anti-alignment between monomer $i$ and $j$.
For the sake of simplicity, we focus here on the case $h_{ij} = h A_{ij}$, where $A_{ij}$ is the adjacency matrix of the polymer and all the pairs of interacting orientations are characterized by the same coupling $h$.
In this case, orientational correlations reflect the internal structure of the polymer.
Our goal is now to compute the correlation tensor $\mathbb{C}^{\alpha\beta}_{ij}(t)$ entering the macroscopic drift and diffusion terms. 

We start by rewriting Eq.~\eqref{eq:correlatedAOUPdynamics} as:
\begin{equation}
    \dot u^\alpha_i  = - \tau_0^{-1} u^\alpha_i - h\sum_{j=0}^{N-1} M_{ij} u^\alpha_j + \sqrt{\frac{2}{d\tau}} {\eta}_i^\alpha(t) \;,
    \label{eq:stoch_dyn_correlated_orientations}
\end{equation}
where $M_{ij}$ is the connectivity matrix of the polymer, $M_{ij} = \deg[i] \delta_{ij} - A_{ij}$.  In this case, one can use the same orthogonal transformation $\varphi_{ij}$ defined in~\eqref{eq:rouse-modes-def} to diagonalize Eq.~\eqref{eq:stoch_dyn_correlated_orientations} and obtain the normal modes $w^\alpha_i=\sum_{j}\varphi_{ij} u^\alpha_j$. Their stochastic dynamics will be given by $N$ independent Ornstein-Uhlenbeck processes reading
\begin{equation}
    \dot w^\alpha_i = - (\tau_0^{-1} + h \sigma_i ) w^\alpha_i +\sqrt{\frac{2}{d\tau}} \eta_i^\alpha(t) \;,
    \label{eq:w_dynamics}
\end{equation}
where $\sigma_i$ are the eigenvalues of the connectivity matrix $M_{ij}$. Due to the orthonormality of the rows of ${\varphi_{ij}}$, the amplitude of the noise in Eq.~\eqref{eq:w_dynamics} is the same as in Eq.~\eqref{eq:stoch_dyn_correlated_orientations}. Being the normal modes uncorrelated to each other, it is straightforward to compute their stationary correlations $\mathcal{W}^{\alpha \beta}_{ij}(t-s) := \llangle w_i^\alpha(t) w_j^\beta(s) \rrangle$:
\begin{multline}
     \mathcal{W}^{\alpha \beta}_{ij}(t-s) = \frac{\delta^{\alpha\beta}\delta_{ij}}{d } \frac{\exp \left[-(\tau_0^{-1} + h \sigma_i)(t-s)\right]}{1 + \tau_0  h \sigma_i} \;,
\end{multline}
with $t>s$.
The above expression can be used to compute the stationary correlation $\mathbb{C}_{ij}^{\alpha \beta}(t)$. In particular, using the definition of the normal modes we get:
\begin{align}
    \mathbb{C}_{ij}^{\alpha \beta}(t) &= \sum_{n=0}^{N-1}\sum_{m=0}^{N-1}\varphi^{-1}_{in}\varphi^{-1}_{jm}\mathcal{W}^{\alpha \beta}_{nm}(t) \n \\
    &= \frac{\delta^{\alpha\beta}}{d}\sum_{n=0}^{N-1}\varphi^{-1}_{in}\varphi_{nj}\frac{\exp \left[-(\tau_0 ^{-1} + h \sigma_n)t\right]}{1 + \tau_0 h \sigma_n }\,, \label{eq:C_correlated_u}
\end{align}
where we used the orthogonality of $\varphi_{ij}$. Integrating the correlation in Eq.~\eqref{eq:C_correlated_u} we get the following expression for the effective diffusion:
\begin{align}
    \cD^{\alpha\beta}(\bfx) &= \frac{D_{\rm t}}{N} \delta^{\alpha\beta} + \frac{\delta^{\alpha\beta}}{d}\frac{\tau_0 v^2(\bfx)}{N^2} \sum_n  \frac{\sum_{ij}\varphi_{ni}\varphi_{nj}}{(1 + \tau_0 h \sigma_n)^2}\,.
    \label{eq:diff_correlated_u_1}
\end{align}
The latter expression can be heavily simplified by taking into account further properties of the orthonormal transformation $\varphi_{ij}$. Indeed, since it diagonalizes a connectivity matrix, the elements of its first row are all equal to each other, i.e., $\varphi_{0i}=\varphi^{-1}_{i0}=1/\sqrt{N}$ for all $i$. Thus, the following identity holds:
\begin{equation}
    \sum_{i}\varphi_{ni}=\sqrt{N}\sum_{i}\varphi_{ni}\varphi^{-1}_{i0}=\sqrt{N}\delta_{n0}\,.
    \label{eq:identity_varphi}
\end{equation}
Applying the relation Eq.~\eqref{eq:identity_varphi} to the effective diffusion in Eq.~\eqref{eq:diff_correlated_u_1} finally yields
\begin{equation}
    \cD^{\alpha\beta}(\bfR)=\cD_0(\bfR)\delta^{\alpha\beta}=\left[\frac{D_t}{N} + \frac{\tau_0 v^2(\bfR)}{dN}\right]\delta^{\alpha\beta}
    \label{eq:diff_correlated_u_2}
\end{equation}
where we used $\sigma_0=0$.
The result in Eq.~\eqref{eq:diff_correlated_u_2} shows that adding correlations between orientation vectors of connected monomers does not affect the effective diffusion of the polymer center of mass at the large scale.
Indeed, Eq.~\eqref{eq:diff_correlated_u_2} does not depend on the orientation coupling $h$ and it is formally identical to Eq.~\eqref{eq:diff_active_polymer}. 

Analogously, we can compute the effective drift as:
\begin{align}
    V^{\alpha}(\bfR) 
    &= \frac{\tau_0 \partial^\alpha v^2(\bfR)}{2dN}\sum_{jn}    \frac{\sum_{ik}\varphi_{ji} \varphi_{jk} \varphi^{-1}_{in}\varphi^{-1}_{kn}}{[1 + \tau_0 h \sigma_n][1+\tau_0 (h+\lambda)\sigma_n ]} \n\\
    &= \frac{\tau_0 \partial^\alpha v^2(\bfR)}{2dN}\sum_{j=0}^{N-1} \frac{1}{[1 + \tau_0 h \sigma_j][1+\tau_0 (h+\lambda)\sigma_j]}\,.
\end{align}
As expected from the discussion of Sec.~\ref{subsec:equil_Rouse_space}, also in this case the drift is related to the diffusivity gradient via $V^{\alpha}(\bfR)=(1-\epsilon/2)\partial^\alpha \cD_0(\bfR)$. As a consequence, the large-scale description is an equilibrium one, with the steady-state distribution of the center of mass $p_{\rm s}(\bfR)$ described by Eq.~\eqref{eq:effective_pot_selfprop}. Specifically, our computation shows that the parameter
$\epsilon$ reads:
\begin{equation}
\epsilon = 1-\sum_{j=1}^{N-1}\frac{1}{[1 + \tau_0 h \sigma_j][1+ \tau_0 (h+\lambda)\sigma_j ]}\;.
    \label{eq:eps_correlated_u}
\end{equation}
According to Eq.~\eqref{eq:effective_pot_selfprop}, a positive sign of $\epsilon$ determines accumulation in regions of
low activity, whereas negative $\epsilon$ implies accumulation of the polymer in high-activity regions.
Moreover, in the range of values for $h$ where the dynamical matrix is stable, $\epsilon '(h) > 0$.
As a consequence,
synchronization between monomers $h>0$ tends to drive the system towards regions of low activity, whereas the converse
occurs for the anti-synchronization case ($h<0$). 

\begin{figure}
    \centering
    \includegraphics[width=\columnwidth]{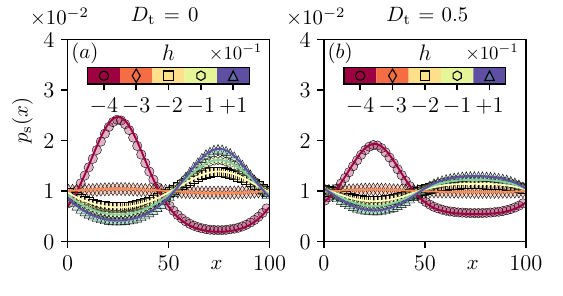}
    \caption{Steady-state distribution $p_{\rm s}(x)$ of active AOUP dimers ($N=2$) with coupled orientations: comparison between theory (solid lines, Eqs.~\eqref{eq:effective_pot_selfprop},~\eqref{eq:eps_correlated_u}) and $2d$ numerical simulations (symbols). ({\bf a}) Athermal particles, $D_{\rm t} = 0$. ({\bf b}) Finite temperature, $D_{\rm t} > 0$. In all panels, the self-propulsion speed $v(\bfr)$ of each monomer is space-dependent and reads $v(x) = v_0 \exp[ A\sin( \frac{2\pi}{L_x}x)]$. Different colors and symbols indicate different values of the synchronization parameter $h$. Parameters: $L_x=100$, $L_y=10$, monomer density $\rho_0 = 10 N$, $A=1$, $v_0=1$, $v_1=0.5$, $D_r=1$.}
    \label{fig:correlatedpolymer}
\end{figure}

Our theoretical predictions are tested against particle-based simulations of active AOUP dimers in $2d$, whose orientations evolve according to Eq.~\eqref{eq:correlatedAOUPdynamics}. We consider a space-dependent self-propulsion speed via $v(x) = v_0 \exp[A \sin(\frac{2\pi}{L_x} x)]$. The results of this comparison are reported in Fig.~\ref{fig:correlatedpolymer}, showing an excellent agreement.

\subsection{Active particle with passive cargo}
\label{subsec:polymer_heteropolymer}
As a final example, we show how the coarse-graining method described in Sec.~\ref{subsec:heteropolymers} for heteropolymers can be applied to the study of a microswimmer with a passive cargo, in the same spirit as Refs.~\cite{vuijk2021chemotaxis,muzzeddu2023taxis}.
Our goal is to derive the large-scale description of the center of friction of the system. Importantly, the relative size of the active carrier and the passive cargo will have an impact on the macroscopic properties of the system, as we now show (See also Refs.~\cite{vuijk2021chemotaxis,muzzeddu2023taxis}).

Specifically, we consider a dimer consisting of a passive monomer ($0$) and an active carrier ($1$), whose friction coefficient are respectively given by $\gamma_0 = q \gamma$ and $\gamma_1= \gamma$. We denote by $\Gamma = \gamma (1+q)$ the total friction coefficient. The orientational dynamics of the active swimmer is characterized by an exponential auto-correlation function $\mathbb{C}^{\alpha\beta}_{11} = \delta^{\alpha\beta} \frac{1}{d} \exp(-t/\tau_{\rm p})$. All elements of the correlation tensor $\mathbb{C}^{\alpha\beta}_{i \neq j}=0$, as well as the auto-correlation term for the passive bead, $\mathbb{C}^{\alpha\beta}_{00}=0$.

In the case of a heterodimer, i.e. when $q \neq 0$, the diffusive slow mode is the center of friction $R^\alpha = \sum_i (\gamma_i/\Gamma) r^\alpha_i$. The large-scale diffusivity $\cD^{\alpha\beta} = \cD_0 \delta^{\alpha\beta}$ is readily obtained from Eq.~\eqref{eq:meso_diffusion_polymer-multi-friction}, and reads:
\begin{equation}
    \cD_0 = \frac{k_{\rm B} T}{\gamma(1+q)} + \frac{v^2(\bfR) \tau}{d (1+q)^2} \;.
\end{equation}
To derive the expression for the large-scale drift of the center of friction we use Eq.~\eqref{eq:meso_drift_polymer-multi-friction}. First, we need to compute the relaxation rates of the Rouse modes, $\lambda_i$, and the associated eigenvectors $\varphi_{ij}$. Given the connectivity matrix of the dimer:
\begin{equation}
    M = \begin{pmatrix}
        1 & -1 \\
        -1 & 1
    \end{pmatrix} \;,
\end{equation}
we define the symmetrized dynamical matrix $\tilde{M}_{ij}$ as in Sec.~\ref{subsec:heteropolymers}:
\begin{equation}
    \tilde{M}_{ij} \equiv \sqrt{\frac{\Gamma}{ N \gamma_i}} M_{ij} \sqrt{\frac{\Gamma}{N \gamma_j}} \quad \Rightarrow \quad \tilde{M} = \frac{1+q}{2} \begin{pmatrix}
        \frac{1}{q}& -\frac{1}{\sqrt{q}} \\
        -\frac{1}{\sqrt{q}} & 1
    \end{pmatrix}\;.
\end{equation}
Denoting by $\sigma_i$ the eigenvalues of $\tilde M$, the relaxation rates of the Rouse modes are given by $\lambda_i = N \kappa/\Gamma \sigma_i$, namely:
\begin{equation}
    \lambda_0 = 0\;, \quad \lambda_1 = \frac{2 \kappa}{\Gamma} \frac{(1 + q)^2}{2 q} =  \frac{\kappa}{\gamma} \frac{1 + q}{q}\;,
\end{equation}
while the matrix $\varphi$ containing the orthonormal eigenvectors reads:
\begin{equation}
\varphi = \begin{pmatrix}
        \sqrt{\frac{q}{1+q}} & \sqrt{\frac{1}{1+q}} \\
        -\sqrt{\frac{1}{1+q}} & \sqrt{\frac{q}{1+q}}
    \end{pmatrix}\;.
\end{equation}
We can now use Eq.~\eqref{eq:meso_drift_polymer-multi-friction} to compute the large-scale drift of the center of friction:
\begin{eqnarray}
    V^{\alpha} &=& \frac{\partial^\beta v^2}{2} \sum_{i, j ,k=0}^{1} \varphi_{ji} \varphi_{jk} \frac{\sqrt{\gamma_i \gamma_k}}{\Gamma}\int_0^\infty  e^{- \lambda_j s} \mathbb{C}_{ik}^{\alpha\beta}(t) \rmd t \notag \\
    &=& \frac{\partial^\beta v^2}{2} \sum_{j=0,1} (\varphi_{j1})^2 \frac{\gamma_1}{\Gamma}\int_0^\infty  e^{- \lambda_j s} \mathbb{C}_{11}^{\alpha\beta}(t) \rmd t \notag \\
    &=& \frac{\partial^\alpha v^2 }{2 d (1+q)} \sum_{j=0, 1} \varphi_{j1}^2 \frac{\tau_{\rm p}}{1 + \lambda_j \tau_{\rm p}} \notag  \\
    &=& \frac{\partial^\alpha v^2 \tau_{\rm p}}{2 d (1+q)} \left[ \frac{1}{1+q} + \frac{q}{1+q} \frac{1}{1 + \frac{1+q}{q} \frac{ \kappa \tau_{\rm p}}{\gamma}} \right] \notag  \\
    &=& \frac{\partial^\alpha v^2 \tau_{\rm p}}{2 d (1+q)^2} \left[ 1 + \frac{q}{1 + \frac{1+q}{q} \frac{ \kappa \tau_{\rm p}}{\gamma}} \right] \;.
\end{eqnarray}
Finally, we remark that:
\begin{equation}
    V^\alpha = \partial^\alpha \cD_0 \left[ 1 -\frac{\epsilon}{2} \right],\;
    \label{eq:rel_VD}
\end{equation}
where
\begin{equation}
    \epsilon = 1- \frac{q}{1 + \frac{1+q}{q} \frac{ \kappa \tau_{\rm p}}{\gamma}} \;.
\end{equation}
This relation, which was first derived by Ref.~\cite{vuijk2021chemotaxis,muzzeddu2023taxis}, finally allows us to determine the steady-state properties of the system. In particular, Eq.~\eqref{eq:rel_VD} ensures the existence of a large scale equilibrium regime where no macroscopic currents are present, according to the discussion of Sec.~\ref{subsec:equil_Rouse_space}. The resulting steady-state Boltzmann distribution for the center of friction, $p_{\rm s}(\bfR)$, thus reads:
\begin{equation}
    p_{\rm s}(\bfR) \propto \cD_0^{-\epsilon/2} \;.
\end{equation}
Note that, once again, the sign of $\epsilon$ determines whether the dimer has a tendency to accumulate in high-activity or low-activity regions.

\newpage

\section{Hydrodynamics of quorum-sensing polymers}
\label{sec:interacting_case}
Thus far, we have restricted our study to non-interacting polymers, revealing the emergence of a unifying large-scale description under general hypotheses for the underlying orientational dynamics. However, many of the most interesting phenomena in scalar active matter arise from interactions between the microscopic constituents~\cite{tailleur2008statistical,fily2012athermal,cates2015motility,soto2014self,zhao2023chemotactic,pisegna2024emergent,caprini2025bubble}. 
In this section, we show that our theory can be directly applied to systems interacting via the mediation of a conserved density field. 
As such, it can be used for studying biological systems where the agents themselves secrete chemical signals that are responsible for motility regulation, as is the case for chemotaxis and quorum sensing in bacteria~\cite{budrene1991complex,liu2011sequential,pohl2014dynamic,curatolo2020cooperative,zhao2023chemotactic}. 
Furthermore, it can be applied to the design self-organizing active materials with density-based interactions, as was recently achieved in colloidal systems with optical feedback-loops~\cite{bauerle2018self}, or for quorum-sensing Quincke rollers~\cite{lefranc2025synthetic}. 

We consider a system of $M$ active polymers, each of which consists of $N$ monomers described by a position vector $r^\alpha_{i, n}$ and an orientation vector $u^\alpha_{i, n}$. Here, $i \in \{0, \dots, N-1\}$ denotes the index of a monomer within a given polymer, while $n \in \{0, \dots, M-1 \}$ specifies which polymer it belongs to. 
In the presence of quorum-sensing (QS) interactions, each monomer modulates its self-propulsion speed $v$ based on the local density of other monomers, i.e., $\rho_{\rm mon}(\bfr):= \sum_{i, n} \delta(\bfr - \bfr_{i, n})$.  
Due to the presence of QS interactions, the overdamped Langevin dynamics of the monomers reads:
\begin{equation}
    \dot r^\alpha_{i,n} = -\gamma^{-1} \partial^\alpha_{\bfr_{i,n}}\mathcal{H} + v(\tilde{\rho}_{\rm mon}(\bfr_{i,n})) u_{i,n}^\alpha + \sqrt{2 D_{\rm t}} \xi^\alpha_{i,n} \;, 
    \label{eq:SDE-QS-monomers}    
\end{equation}
where $\tilde{\rho}_{\rm mon} = K \ast \rho_{\rm mon}$ denotes the convolution product of the density field $\rho_{\rm mon}$ with an isotropic bell-shaped kernel $K$. The latter is characterized by a finite interaction range $\ell_{\rm int}$, accounting for a non-local sampling of the density~\cite{sire2004postcollapse,soto2014self,solon2018generalized,o2020lamellar,dinelli2024fluctuating}.
We will refer to \emph{motility inhibition} whenever the self-propulsion speed is a decreasing function of the density, i.e., $v'(\rho) < 0$, while the opposite case will be referred to as \emph{motility activation}.
In the following, we focus on achiral orientational processes, so that the auto-correlation tensor of the orientations is symmetric at all times $t$ and for all monomer pairs $(i,j)$ belonging to the same polymer, namely $\mathbb{C}_{ij}^{\alpha\beta}(t) = \mathbb{C}_{ij}^{\beta\alpha}(t)$. We assume no cross correlations between the orientations of monomers belonging to different polymers. 

With the help of the coarse-graining method detailed in Sec.~\ref{sec:adiabatic_elimination}, we aim to derive an interacting hydrodynamics for the fluctuating polymer density $\rho(\bfr)$, defined as:
\begin{equation}
    \rho(\bfr) := \sum_{n=0}^{M-1} \delta(\bfr-\bfR_n)
    \label{eq:def_rho}
\end{equation}
where $\bfR_n$ denotes the center of mass of polymer $n$, and the sum runs over all polymers. 
To proceed, we rely on a well-established approach in the context of QS active particles, namely the frozen-field approximation~\cite{tailleur2008statistical,cates2015motility,solon2015active,solon2018generalized,dinelli2023nonreciprocity,obyrne2023introduction,dinelli2024fluctuating}.
Being the density field $\rho$ a conserved quantity, it represents a slow mode of our system. 
Hence, its evolution occurs over spatio-temporal diffusive scales diverging with the linear system size $L$. This implies that the macroscopic diffusive timescale is expected to grow as $L^2$. 
Consequently, over the shorter timescales characterizing the microscopic dynamics of the individual active monomers, the 
density-dependent self-propulsion $v(\tilde{\rho}_{\rm mon})$ can be regarded as a space-dependent activity $v(\bfr_i)$ in a frozen density landscape $\tilde{\rho}_{\rm mon}(\bfr)$. 
Under these conditions, different polymers are effectively independent of each other. 
We thus conclude that, at a mesoscopic timescale $\Delta t$ such that $L^2 \gg \Delta t \sim \cT \gg \tau$, the dynamics of their centers of mass is captured by the Fokker-Planck dynamics Eq.~\eqref{eq:forward-FP-Phi} obtained via the adiabatic elimination.

Over the macroscopic timescales of evolution of the density field, one can thus associate to Eq.~\eqref{eq:forward-FP-Phi} an interacting It\^o-Langevin equation that describes the large-scale stochastic dynamics of the centers of mass $\{R_n^\alpha\}$ through:
\begin{equation}
    \dot{R}_n^\alpha = V^\alpha(\bfR_n, [\rho_{\rm mon}]) + B^{\alpha\beta}(\bfR_n, [ {\rho}_{\rm mon}]) \;\eta_n^\beta(t) \;,
    \label{eq:mesoscopic_CM_dyn}
\end{equation}
where $B^{\alpha\gamma} B^{\beta\gamma} = 2 \cD^{\alpha\beta}$.\footnote{Note that this writing is possible thanks to the absence of chirality, which would have rendered $\cD^{\alpha\beta}$ non symmetric.} Here, the expression of $V^\alpha$ and $\cD^{\alpha\beta}$ are respectively given by Eqs.~\eqref{eq:drift_final},~\eqref{eq:diffusion_final}, and the space dependence has been replaced by the density-dependence of $v$ on $\tilde{\rho}_{\rm mon}(\bfR)$. 
Note that Eq.~\eqref{eq:mesoscopic_CM_dyn} is not closed at this stage, as it requires the knowledge of the monomer density field.
Nonetheless, relying on the scale separation between the typical size of each polymer (microscopic gyration radius $R_g \sim \ell$) and the macroscopic scale of the dynamics~\eqref{eq:mesoscopic_CM_dyn}, we take ${\rho}_{\rm mon} \simeq N \rho$. 
This approximation corresponds to collapsing the whole structure of the polymer onto the centre-of-mass mode.
Therefore, Eq.~\eqref{eq:mesoscopic_CM_dyn} becomes:
\begin{equation}
    \dot{R}_n^\alpha = V^\alpha(\bfR_n, [N \rho]) + B^{\alpha\beta}(\bfR_n, [N {\rho}]) \eta_i^\beta(t) \;,
    \label{eq:mesoscopic_CM_dyn_closed}
\end{equation}
which now is fully closed, as it only depends on the coordinates $\{R_n^\alpha\}$. 

We can now derive the fluctuating hydrodynamics of the polymer density field $\rho$ by adopting Dean's approach~\cite{dean1996langevin,solon2015active,obyrne2023introduction,dinelli2024fluctuating,kuroda2023microscopic,illien2025dean}. In practice, we apply It\^o stochastic calculus to compute the time derivative of $\rho(\bfr,t)$ from Eq.~\eqref{eq:def_rho} and Eq.~\eqref{eq:mesoscopic_CM_dyn_closed}. The computation follows exactly the steps reported in Ref.~\cite{dinelli2024fluctuating}. This finally yields:
\begin{equation}
    \partial_t \rho(\bfr,t) = - \partial^\alpha [ V^\alpha \rho - \partial^\beta (\cD^{\alpha\beta} \rho) + \sqrt{\rho}  B^{\alpha\beta} \Lambda^\beta(\bfr,t)] \;,
    \label{eq:hydroQS_poly}
\end{equation}
where $\Lambda^\alpha(\bfr,t)$ is a Gaussian white noise field with zero mean and $\langle \Lambda^\alpha(\bfr,t)\Lambda^\beta(\bfr',t') \rangle=\delta(\bfr-\bfr') \delta(t-t') \delta^{\alpha\beta}$. Equation~\eqref{eq:hydroQS_poly} describes the general diffusion-drift stochastic hydrodynamics of an active polymer with quorum-sensing interactions, in the absence of chirality. 
It can be used to predict structure factor and intermediate scattering function for stable homogeneous profiles~\cite{dinelli2024fluctuating}, but also the emergence of collective behaviors such as motility-induced phase separation (MIPS)~\cite{tailleur2008statistical,cates2015motility,solon2018generalized,duan2023dynamical}. 
In the following, we detail the conditions for MIPS to arise, and shed light onto a novel type of phase separation that is driven by motility activation.

\subsection{Emergence of anti-MIPS}
\label{subsec:antiMIPS}
To study the emergence of phase separation from Eq.~\eqref{eq:hydroQS_poly}, we consider the specific case of isotropic auto-correlation tensor $\mathbb{C}_{ij}^{\alpha\beta} = C_{ij}(t) \delta^{\alpha\beta}$, which implies that $\cD^{\alpha\beta} = \cD_0 \delta^{\alpha\beta}$. In this case, we know from Sec.~\ref{subsec:equil_Rouse_space} that the drift and the diffusion tensor are related by $V^\alpha = (1 - \epsilon/2) \partial^\alpha \cD_0$, where $\epsilon$ is given by Eq.~\eqref{eq:eps_value}. This allows us to write the hydrodynamics as: 
\begin{equation}
    \partial_t \rho(\bfr,t) = \partial^\alpha [ \cD_0 \rho \partial^\alpha \mu(\bfr,[\rho]) + \sqrt{2 \cD_0 \rho}  \Lambda^\alpha(\bfr,t)] \;,
    \label{eq:hydroQS_poly_effectiveeq}
\end{equation}
where we introduced the effective chemical potential:
\begin{equation}
    \mu(\bfr,[\rho]) = \epsilon \log v( N\tilde{\rho}(\bfr)) + \log \rho(\bfr) \;.
\end{equation}
We note that Eq.~\eqref{eq:hydroQS_poly_effectiveeq} satisfies the Stokes-Einstein relation between the mobility $\cD_0 \rho$ and the noise variance. Consequently, if $\mu$ can be derived from an effective free energy functional $\cF$ via $\mu(\bfr) = \delta \cF / \delta\rho(\bfr)$, then the hydrodynamics~\eqref{eq:hydroQS_poly_effectiveeq} corresponds to a bona-fide equilibrium regime with zero entropy production~\cite{fodor2016far,o2022time,obyrne2023introduction,o2025geometric}. We refer to this condition as integrability of the effective chemical potential.

To make analytical progress, we consider the case of local interactions, by which $v(\tilde\rho(\bfr)) \simeq v(\rho(\bfr))$, corresponding to the limit where the coarsening kernel $K(\bfr) \to \delta(\bfr)$. This approximation, although discarding higher-order gradient contributions to the hydrodynamics~\cite{solon2018generalized,duan2023dynamical,burekovic2026active}, allows us to study bulk phase separation via an equilibrium mapping. Indeed, the local approximation directly guarantees the integrability of $\mu$, as it allows to introduce an effective free energy:
\begin{equation}
    \cF := \int \rmd^d\bfr f(\rho(\bfr)) \;, \>\> f(\rho) := \epsilon \int^\rho \log v(N s) \rmd s + \rho \log \rho 
    \label{eq:free_energy}
\end{equation}
such that $\mu = \delta\cF/\delta\rho = f'(\rho)$. Under the local approximation, the functional $\cF$ plays the role of a Lyapunov functional, which is minimized by the dynamics~\eqref{eq:hydroQS_poly_effectiveeq} in mean field:
\begin{equation}
    \partial_t \langle \cF \rangle \simeq \int \rmd^d \bfr \; \bar\mu \;\partial^\alpha [\bar\cD_0 \bar\rho \partial^\alpha \bar\mu] = - \int \rmd^d \bfr \;\bar\cD_0 \bar\rho (\partial^\alpha \bar\mu)^2 < 0 \;,
    \label{eq:Lyapunov}
\end{equation}
where $\bar g(\rho) = g(\langle \rho \rangle)$ denotes the mean field approximation applied to any function $g$ of the stochastic process $\rho$. Here, the average $\langle \cdot \rangle$ is to be interpreted over all realizations of the noise trajectories $\Lambda^\alpha(\bfr,t)$. Equation~\eqref{eq:Lyapunov} thus implies that the steady-state configurations of the density field minimize the free energy, as in equilibrium. As such, Eq.~\eqref{eq:free_energy} can be used to predict the binodals of motility-induced phase separation via an equilibrium common-tangent construction~\cite{chaikin1995principles,obyrne2023introduction,duan2025phase}, and hence to build the phase diagram in~\cite{dinelli2026PRL}.

Crucially, the free energy density $f(\rho)$ in Eq.~\eqref{eq:free_energy} generalizes the standard expression for single-particle QS to $\epsilon \neq 1$~\cite{cates2015motility,solon2018generalized}. The important role of this prefactor $\epsilon$ can be directly assessed via a linear stability analysis of a homogeneous state at density $\rho_0$. Denoting by $\delta\rho(\bfr) = \rho - \rho_0$, we linearize Eq.~\eqref{eq:hydroQS_poly_effectiveeq} in the mean-field local approximation. This yields an effective diffusion equation:
\begin{equation}
    \partial_t \delta\rho = \cD_{\rm eff} \partial^\alpha \partial^\alpha \delta\rho + \cO(\delta\rho^2)
\end{equation}
where:
\begin{equation}
    \cD_{\rm eff} = \cD_0 \rho_0 \left( N\epsilon \frac{v'(N \rho_0)}{v(N \rho_0)} + \frac{1}{\rho_0}\right) \;.
\end{equation}
The homogeneous profile is thus unstable when the effective diffusivity is negative, i.e. for:
\begin{equation}
    \epsilon \left.\frac{v'(x)}{v(x)}\right|_{x=N \rho_0} < - \frac{1}{N \rho_0} \;.
\end{equation}
For conventional MIPS, $\epsilon=N=1$, so that a linear instability can only occur for motility inhibition, i.e. $v'(\rho_0) < 0$. However, in the polymer case we have shown that $\epsilon$ can also take negative values.
Consequently, it is now possible to induce a linear instability through a sufficiently strong motility \emph{enhancement}, whenever $\epsilon < 0$.
This new type of phase transition, which we call anti-MIPS, drives the formation of dense highly-motile phases that coexist with a dilute gas of polymers at low activity.
As shown in Sec.~\ref{sec:cg_polymer}, there exist a number of distinct transition pathways leading to a negative $\epsilon$.
These might be driven, e.g., by a change in the polymer topology, in the polymer length, or in the synchronization of the monomers' orientation: therefore, we expect anti-MIPS to emerge generically from all these distinct pathways.

In our companion paper~\cite{dinelli2026PRL} we confirm numerically the existence of such phase transition, and predict the phase diagram based on the effective free energy given by Eq.~\eqref{eq:free_energy}. 

\newpage

\section{Discussion and perspectives}

In this work, we explicitly derived a large-scale hydrodynamic description encompassing a broad range of 
scalar active systems with motility regulation, including active polymers. The advantage of this technique is that it does not rely on any specific form of the microscopic orientational dynamics. Our computation thus sheds light on the direct relationship between the auto-correlation tensor of the orientations and the large-scale drift and diffusion tensor. 

In the final part of our work, we built on our results to derive a unifying diffusion-drift fluctuating hydrodynamics of scalar active systems with mediated interactions. 
Our results can be straightforwardly generalized to active mixtures~\cite{curatolo2020cooperative,dinelli2023nonreciprocity,dinelli2025random,duan2023dynamical,brauns2024nonreciprocal,
pisegna2024emergent,greve2025coexistence,pisegna2025spinning}, and are thus expected to provide a comprehensive platform to study collective behaviors in scalar active systems with motility regulation. 
Furthermore, we expect the fluctuating hydrodynamics to capture both static and dynamical properties of homogeneous polymeric systems with motility regulation~\cite{dinelli2024fluctuating,illien2025dean}. 
Non-homogeneous steady-states can also be studied, revealing the existence of a new anti-MIPS phase~\cite{dinelli2026PRL}. 
However, in order to achieve better quantitative descriptions, more refined analysis that capture higher-order gradient terms in the hydrodynamics are required~\cite{solon2018generalized,duan2023dynamical,burekovic2026active}. 
One question is then whether the multi-scale perturbative approach presented here is suitable to derive higher-order gradient corrections to the fluctuating hydrodynamics. 

In this work, we have focused on a dilute regime for the interacting hydrodynamics, neglecting the role of steric repulsion between polymers.
This simplification represents a clear limitation and raises the question of whether the present approach can be extended to derive the corresponding coarse-grained theories. 
Addressing this issue offers an interesting direction for future research, which could benefit from recent advances in the field~\cite{solon2018generalized,omar2023mechanical,nguyen2025contact}.

Finally, in this work we exclusively focused on scalar active matter, thus neglecting all systems exhibiting orientational order at the large scale. 
This is clearly a limitation, especially when describing active polymeric systems for which orientational order can have a major impact~\cite{abaurrea2018collective}. 
In the case of polar and nematic active systems, a large corpus of work already described coarse-graining methods for specific classes of models, see for instance~\cite{bertin2006boltzmann,saintillan2008instabilities,subramanian2009critical,ihle2011kinetic,solon2013revisiting,peshkov2014boltzmann,weady2022thermodynamically}. 
At least in the context of dry active systems, we expect our multi-scale perturbative approach to provide a versatile tool to deriving coarse-grained large-scale descriptions also beyond scalar active systems, upon proper separation between fast and slow hydrodynamic variables. 

\emph{Acknowledgments.---}The authors thank Karsten Kruse for insightful discussions and support. This work was partially funded by Swiss National Science Foundation through grant number $200020\mathrm{E}\_219164$.

\newpage

\appendix

\section{Numerical details}
Particle-based simulations are performed by integrating Eq.~\eqref{eq:SDE-monomers} in $2d$ using an Euler scheme, in a box of size $L_x \times L_y$ with periodic boundary conditions. We initialize our system with $(\rho_0/N) \times L_x \times L_y$ polymer chains of length $N$ uniformly distributed over the domain, and integrate the dynamics up to time $T=10^6$ with time step $dt=0.005$ and $\gamma=1$. The dynamics of the orientation vector $\bfu$ depends on the specific process that is studied, and is detailed in the main text. Specific parameters for each simulations are also reported in the figure captions.

\section{Fredholm alternative}
\label{app:fredholm}
In this Appendix, we discuss how the Fredholm alternative can be applied to our coarse-graining problem. In Sec.~\ref{sec:adiabatic_elimination}, we often need to solve equations of the form:
\begin{equation}
    \cL_0 g = -f\,.
    \label{eq:fredholm1}
\end{equation}
where $\cL_0$ is a differential operator that satisfies ergodicity. Equivalently, we assume that the null space of the adjoint operator $\cL_0^\dagger$ is one-dimensional and is spanned by the unique steady-state distribution $\Psi$: $\cL_0^\dagger \Psi = 0$.
The solvability condition for Eq.~\ref{eq:fredholm1} is obtained by projecting the equation onto $\Psi$. Denoting by $(a,b) := \int_{\mathbb{R}} a(x) b(x) \rmd x$ the scalar product in the space of $\mathbb{L}^1$ functions from $\mathbb{R} \to \mathbb{R}$, we have:
\begin{equation}
    ( \cL_0 g, \Psi) = (g, \cL_0^\dagger \Psi) = 0
\end{equation}
for the steady-state condition on $\Psi$. Projection $(\cdot, \Psi)$ on the right-hand side of Eq.~\eqref{eq:fredholm1} finally gives:
\begin{equation}
    (f, \Psi) = \langle f \rangle_\Psi = 0 \;.
    \label{eq:fredholm2}
\end{equation}
Equation~\eqref{eq:fredholm2} is thus the solvability condition for Eq.~\eqref{eq:fredholm1}.
\section{Identity inverse generator $\cL_0^{-1}$}
\label{app:identity-L0}
In this Appendix we are concerned with proving the formal identity given by Eq.~\eqref{eq:identity-invereseL0}, namely:
\begin{equation}
        \cL_0^{-1} = -\int_0^\infty\rmd t\,e^{\cL_0t}\,.
\end{equation}
To this aim, we consider the fast dynamics of order $\cO(\varepsilon^{-2})$ generated by the backward operator $\cL_0$.
We assume that the associated dynamic variables $\chi(t)$ and $\Theta(t)$ are initialized as $\chi(0)=\chi$ and $\Theta(0)=\Theta$, and we define the observable $A(\chi,\Theta,t)$ as:
\begin{equation}
    A(\chi,\Theta,t) = \mathbb{E}[h(\chi(t), \Theta(t)) | \chi(0)=\chi,\Theta(0)=\Theta]
    \label{eq:observable-A-app}
\end{equation}
which is a generalization of Eq.~\eqref{eq:O-cell-problem} to the case in which the observable also depends on the higher-order Rouse modes $\chi$. 
In Eq.~\eqref{eq:observable-A-app}, $h(\chi,\Theta)$ is a generic function of the  fast variables.
By definition of the generator $\cL_0$, we know that the time evolution of any observable of the type given by Eq.~\eqref{eq:observable-A-app} is governed by $\partial_tA=\cL_0A$.
For simplicity, we assume here that the orientational degrees of freedom $\Theta$ evolve according to a continuous stochastic process, but the derivation can be extended to include the case of jump processes.
Let us consider the function $\Phi^\alpha(\bfR,\chi(t),\Theta(t))$, where the center of mass $R^\alpha$ appears as a parameter whereas the fast variables are fluctuating quantities.
By applying It\^o's lemma and averaging over the path realizations of both $\chi(t)$ and $\Theta(t)$ generated by $\cL_0$, we get:
\begin{equation}
    \mathbb{E}[\dot{\Phi}^\alpha| \chi,\Theta] = \mathbb{E}[ \cL_0\Phi^\alpha| \chi,\Theta]\;.
\end{equation}
Integrating over time from $0$ to $t$ and using the definition of $\Phi^\alpha$ given by the cell problem in Eq.~\eqref{eq:cell-problem} we obtain:
\begin{align}
&\mathbb{E}[ \Phi^\alpha(\bfR,\chi(t),\Theta(t)) ] - \Phi^\alpha(\bfR,\chi,\Theta) \label{eq:identity-Phi-app} \\&= -\int_0^t\rmd s\, \mathbb{E}[ \cV_0^{0\alpha}(\bfR,\chi(s),\Theta(s)) ] \n \\&= -\int_0^t\rmd s\,  e^{\cL_0s}\cV_0^{0\alpha}(\bfR,\chi,\Theta)\;,\n  
\end{align}
where, for the sake of clarity, we kept explicit the argument $\chi$ even if $\cV_0^{0\alpha}$ is independent of the higher-order Rouse modes.
Finally, taking the limit $t \to \infty$, and using the ergodicity property of the fast process generated by $\cL_0$, we have that
\begin{equation}
\lim_{t\to \infty}\mathbb{E}[ \Phi^\alpha(\bfR,\chi(t),\Theta(t)) ] = \big\langle \Phi^\alpha \big\rangle_\Psi=0\;,
\end{equation}
where the average over the steady-state distribution $\Psi$ vanishes due to the centering condition in Eq.~\eqref{eq:cell-problem}.
As a result, we obtain Eq.~\eqref{eq:identity-invereseL0} of the main text.
\section{Explicit drift-diffusion for space-dependent self-propulsion speed}
\label{app:explicit-space-dependent-self-propulsion-speed}
In this Appendix we specialize the expressions for the large-scale drift $V^\alpha$ and diffusion tensor $\cD^{\alpha \beta}$ reported in Eqs.~\eqref{eq:drift_formal} and~\eqref{eq:diff_formal} to the case where the active coupling is of the form ${\rm v_i^\alpha}(\bfR,\bfu_i)=v(\bfR)u_i^\alpha$.
First, by using the definition of $\cV_0^{0\alpha}$ given in Eq.~\eqref{eq:expansion_cV00}, we get:
\begin{equation}
    \big\langle \cV_0^{0\alpha}(t) \cV_0^{0\beta}(0) \big\rangle = \frac{v^2(\bfR)}{N^2}\sum_{ij}\mathbb{C}_{ij}^{\alpha \beta}(t)\;,
\end{equation}
which directly leads to Eqs.~\eqref{eq:diffusion_final} of the main text.
Analogously, from the second drift term in Eq.~\eqref{eq:drift_final} we compute:
\begin{equation}
    \big\langle \cV_0^{0\beta}(0) \partial^\beta\cV_0^{0\alpha}(t) \big\rangle = \frac{\partial^\beta v^2(\bfR)}{2N}\sum_{ik}\varphi_{0i}\varphi_{0k}\mathbb{C}_{ik}^{\alpha \beta}(t)\;,
    \label{eq:cv-dercV-app}
\end{equation}
where we used the property $\varphi_{0i}=1/\sqrt{N}$ for all $i$.
The first term of Eq.~\eqref{eq:drift_final}, given by
\begin{equation}
    \big\langle \mathcal{V}_0^{1\alpha} \big\rangle_\Psi =  \frac{\partial^\beta {\rm v}_i^{\alpha}(\bfR)}{N} \sum_{i=0}^{N-1} \sum_{j=1}^{N-1} \varphi_{ji} \big\langle\chi^\beta_j u_i^\alpha\big\rangle_\Psi\;,
    \label{eq:cV1-app}
\end{equation}
requires to compute the steady-state cross correlations between the orientational degrees of freedom $\Theta$ and the higher order Rouse modes $\chi$.
To this purpose, we first recall that $\Psi$ is the stationary density associated to the fast dynamics of order $\cO(\varepsilon^{-2})$, generated by the backward operator $\cL_0$.
At this scale, the evolution of the Rouse modes $\chi$ is governed by the following Langevin dynamics:
\begin{equation}
    \dot{\chi}_i^\beta = \cV_i^{0\beta}(\bfR,\chi,\Theta) + \sqrt{2D_{\rm t}} \eta_i^{\beta}\;.
    \label{eq:fast-dynamics-chi}
\end{equation}
Due to the linearity in $\chi$, the above equation can be exactly solved leading to:
\begin{align}
    &\chi_j^\beta(t) = e^{-\lambda_j t}\chi_j^\beta(0) \\&+ \int_0^{t} \rmd s\,e^{-\lambda_j(t-s)}\left[ \sum_{k=0}^{N-1}\varphi_{jk}v(\bfR)u_k^\beta(s) + \sqrt{2D_{\rm t}}\eta_j^\beta(s) \right]\n\;.
\end{align}
Multiplying by $u^\alpha_i(t)$, taking the long-time limit $t \to \infty$ and averaging over the stochastic trajectories of the orientational degrees of freedom $\Theta$ yields:
\begin{equation}
    \big\langle\chi^\beta_j u_i^\alpha\big\rangle_\Psi =  \sum_{k=0}^{N-1}\varphi_{jk}v(\bfR)\int_0^{\infty} \rmd s\,e^{-\lambda_j s} \mathbb{C}_{ik}^{\alpha \beta}(s)\;.
    \label{eq:cross-corr-uchi}
\end{equation}
Combining Eqs.~\eqref{eq:cross-corr-uchi},~\eqref{eq:cV1-app} and~\eqref{eq:cv-dercV-app} with the drift definition of Eq.~\eqref{eq:drift_formal}, we finally get Eq.~\eqref{eq:drift_final} of the main text.

\if{
\section{Notation}
\begin{table*}[ht!]
    \centering
    \begin{tabular}{|c|c|}
    \hline
       Monomer positions  & $\bfr_i = \sum_j \varphi_{ji} \bchi_j$ \\
       Orientations & $\bfu_i$ \\
       Rouse modes $(i\geq 1)$ & $\bchi_i = \sum_j \varphi_{ij} \bfr_j$ \\
       Frictions & $\gamma$ \\
       Center of mass / friction & $\bfR = \frac{\bchi_0}{\sqrt{N}}$ \\
       Active/external velocity & $\mathbf{v}_i(\bfr_i,\bfu_i)$ \\
       Dynamics of positions & $\dot{\bfr}_i = - \gamma^{-1} \nabla_i \cH + \mathbf{{v}}_i(\bfr_i,\bfu_i) + \sqrt{2 D_{\rm t}} \boldsymbol{\eta}_i$ \\
       Dynamics of orientations & Only in operator form \\
       Hamiltonian & $\cH = \frac{\kappa}{2} \sum_{ij} M_{ij} \bfr_i \cdot \bfr_j$ \\
       Internal forces & $-\kappa M_{ij} \bfr_j$ \\
       Espansioni che prima si chiamavano $\boldsymbol{f}^{(n)}$ & $\boldsymbol{\mathcal{V}}^{(n)}$ \\
       Persistence times  & $\tau_{ij}^{\alpha\beta} = \int_0^\infty \mathbb{C}_{ij}^{\alpha\beta} \rmd t$ \;, $\quad$ $\tau_{\rm p} = \max |\tau_{ij}^{\alpha\beta}|$ \\
       Relaxation timescale Rouse modes & $\tau_{\rm r} = \gamma/\kappa$\\
       Rouse rates & $\lambda_i=\tau^{-1}_{\rm r} \sigma_i$ \\
        Eigenvalues connectivity & $ \sigma_i$ \\
       Shorthand Rouse modes & $\chi = \{\chi_i^\alpha\}$ \\
       Shorthand orientations & $\Theta = \{u_i^\alpha\}$ \\
       Shorthand fast variables & $Y=(\Theta, \chi)$\\
       Scale of variation of the external control & $\delta$ \\
       Scale of the activity & $\bfv_i \sim v_0$ \\
       Persistence length & $\ell_{\rm p} = v_0\tau_{\rm p}$ \\
       Epsilon & $\varepsilon = \ell/\delta$, $\quad$ $\varepsilon^2 = \tau/\cT$\\
       Microscopic lengthscale & $\ell \sim \ell_{\rm p} \sim R_{\rm g}$\\
       Mesoscopic length-timescale & $\delta, \mathcal{T}$ \\
       Macroscopic scale (system size) & $L \gg \delta \gg \ell$ \\
    \hline
    \end{tabular}
    \caption{Notations that we use. Shift everything to component notation and avoid bold if possible. In general, spatial components are up, monomer indices down.}
    \label{tab:placeholder}
\end{table*}
}\fi

\bibliography{biblio}

\begin{thebibliography}{123}%
\makeatletter
\providecommand \@ifxundefined [1]{%
 \@ifx{#1\undefined}
}%
\providecommand \@ifnum [1]{%
 \ifnum #1\expandafter \@firstoftwo
 \else \expandafter \@secondoftwo
 \fi
}%
\providecommand \@ifx [1]{%
 \ifx #1\expandafter \@firstoftwo
 \else \expandafter \@secondoftwo
 \fi
}%
\providecommand \natexlab [1]{#1}%
\providecommand \enquote  [1]{``#1''}%
\providecommand \bibnamefont  [1]{#1}%
\providecommand \bibfnamefont [1]{#1}%
\providecommand \citenamefont [1]{#1}%
\providecommand \href@noop [0]{\@secondoftwo}%
\providecommand \href [0]{\begingroup \@sanitize@url \@href}%
\providecommand \@href[1]{\@@startlink{#1}\@@href}%
\providecommand \@@href[1]{\endgroup#1\@@endlink}%
\providecommand \@sanitize@url [0]{\catcode `\\12\catcode `\$12\catcode
  `\&12\catcode `\#12\catcode `\^12\catcode `\_12\catcode `\%12\relax}%
\providecommand \@@startlink[1]{}%
\providecommand \@@endlink[0]{}%
\providecommand \url  [0]{\begingroup\@sanitize@url \@url }%
\providecommand \@url [1]{\endgroup\@href {#1}{\urlprefix }}%
\providecommand \urlprefix  [0]{URL }%
\providecommand \Eprint [0]{\href }%
\providecommand \doibase [0]{https://doi.org/}%
\providecommand \selectlanguage [0]{\@gobble}%
\providecommand \bibinfo  [0]{\@secondoftwo}%
\providecommand \bibfield  [0]{\@secondoftwo}%
\providecommand \translation [1]{[#1]}%
\providecommand \BibitemOpen [0]{}%
\providecommand \bibitemStop [0]{}%
\providecommand \bibitemNoStop [0]{.\EOS\space}%
\providecommand \EOS [0]{\spacefactor3000\relax}%
\providecommand \BibitemShut  [1]{\csname bibitem#1\endcsname}%
\let\auto@bib@innerbib\@empty
\bibitem [{\citenamefont {Budrene}\ and\ \citenamefont
  {Berg}(1991)}]{budrene1991complex}%
  \BibitemOpen
  \bibfield  {author} {\bibinfo {author} {\bibfnamefont {E.~O.}\ \bibnamefont
  {Budrene}}\ and\ \bibinfo {author} {\bibfnamefont {H.~C.}\ \bibnamefont
  {Berg}},\ }\bibfield  {title} {\bibinfo {title} {Complex patterns formed by
  motile cells of escherichia coli},\ }\href
  {https://doi.org/https://doi.org/10.1038/349630a0} {\bibfield  {journal}
  {\bibinfo  {journal} {Nature}\ }\textbf {\bibinfo {volume} {349}},\ \bibinfo
  {pages} {630} (\bibinfo {year} {1991})}\BibitemShut {NoStop}%
\bibitem [{\citenamefont {Polin}\ \emph {et~al.}(2009)\citenamefont {Polin},
  \citenamefont {Tuval}, \citenamefont {Drescher}, \citenamefont {Gollub},\
  and\ \citenamefont {Goldstein}}]{polin2009chlamydomonas}%
  \BibitemOpen
  \bibfield  {author} {\bibinfo {author} {\bibfnamefont {M.}~\bibnamefont
  {Polin}}, \bibinfo {author} {\bibfnamefont {I.}~\bibnamefont {Tuval}},
  \bibinfo {author} {\bibfnamefont {K.}~\bibnamefont {Drescher}}, \bibinfo
  {author} {\bibfnamefont {J.~P.}\ \bibnamefont {Gollub}},\ and\ \bibinfo
  {author} {\bibfnamefont {R.~E.}\ \bibnamefont {Goldstein}},\ }\bibfield
  {title} {\bibinfo {title} {Chlamydomonas swims with two “gears” in a
  eukaryotic version of run-and-tumble locomotion},\ }\href
  {https://doi.org/https://doi.org/10.1126/science.1172667} {\bibfield
  {journal} {\bibinfo  {journal} {Science}\ }\textbf {\bibinfo {volume}
  {325}},\ \bibinfo {pages} {487} (\bibinfo {year} {2009})}\BibitemShut
  {NoStop}%
\bibitem [{\citenamefont {Jiang}\ \emph {et~al.}(2010)\citenamefont {Jiang},
  \citenamefont {Yoshinaga},\ and\ \citenamefont {Sano}}]{jiang2010active}%
  \BibitemOpen
  \bibfield  {author} {\bibinfo {author} {\bibfnamefont {H.-R.}\ \bibnamefont
  {Jiang}}, \bibinfo {author} {\bibfnamefont {N.}~\bibnamefont {Yoshinaga}},\
  and\ \bibinfo {author} {\bibfnamefont {M.}~\bibnamefont {Sano}},\ }\bibfield
  {title} {\bibinfo {title} {Active motion of a janus particle by
  self-thermophoresis in a defocused laser beam},\ }\href
  {https://doi.org/https://doi.org/10.1103/PhysRevLett.105.268302} {\bibfield
  {journal} {\bibinfo  {journal} {Phys. Rev. Lett.}\ }\textbf {\bibinfo
  {volume} {105}},\ \bibinfo {pages} {268302} (\bibinfo {year}
  {2010})}\BibitemShut {NoStop}%
\bibitem [{\citenamefont {Palacci}\ \emph {et~al.}(2013)\citenamefont
  {Palacci}, \citenamefont {Sacanna}, \citenamefont {Steinberg}, \citenamefont
  {Pine},\ and\ \citenamefont {Chaikin}}]{palacci2013living}%
  \BibitemOpen
  \bibfield  {author} {\bibinfo {author} {\bibfnamefont {J.}~\bibnamefont
  {Palacci}}, \bibinfo {author} {\bibfnamefont {S.}~\bibnamefont {Sacanna}},
  \bibinfo {author} {\bibfnamefont {A.~P.}\ \bibnamefont {Steinberg}}, \bibinfo
  {author} {\bibfnamefont {D.~J.}\ \bibnamefont {Pine}},\ and\ \bibinfo
  {author} {\bibfnamefont {P.~M.}\ \bibnamefont {Chaikin}},\ }\bibfield
  {title} {\bibinfo {title} {Living crystals of light-activated colloidal
  surfers},\ }\href {https://doi.org/10.1126/science.1230020} {\bibfield
  {journal} {\bibinfo  {journal} {Science}\ }\textbf {\bibinfo {volume}
  {339}},\ \bibinfo {pages} {936} (\bibinfo {year} {2013})}\BibitemShut
  {NoStop}%
\bibitem [{\citenamefont {B{\"a}uerle}\ \emph {et~al.}(2018)\citenamefont
  {B{\"a}uerle}, \citenamefont {Fischer}, \citenamefont {Speck},\ and\
  \citenamefont {Bechinger}}]{bauerle2018self}%
  \BibitemOpen
  \bibfield  {author} {\bibinfo {author} {\bibfnamefont {T.}~\bibnamefont
  {B{\"a}uerle}}, \bibinfo {author} {\bibfnamefont {A.}~\bibnamefont
  {Fischer}}, \bibinfo {author} {\bibfnamefont {T.}~\bibnamefont {Speck}},\
  and\ \bibinfo {author} {\bibfnamefont {C.}~\bibnamefont {Bechinger}},\
  }\bibfield  {title} {\bibinfo {title} {Self-organization of active particles
  by quorum sensing rules},\ }\href
  {https://doi.org/https://doi.org/10.1038/s41467-018-05675-7} {\bibfield
  {journal} {\bibinfo  {journal} {Nat. Commun.}\ }\textbf {\bibinfo {volume}
  {9}},\ \bibinfo {pages} {3232} (\bibinfo {year} {2018})}\BibitemShut
  {NoStop}%
\bibitem [{\citenamefont {Lavergne}\ \emph {et~al.}(2019)\citenamefont
  {Lavergne}, \citenamefont {Wendehenne}, \citenamefont {B{\"a}uerle},\ and\
  \citenamefont {Bechinger}}]{lavergne2019group}%
  \BibitemOpen
  \bibfield  {author} {\bibinfo {author} {\bibfnamefont {F.~A.}\ \bibnamefont
  {Lavergne}}, \bibinfo {author} {\bibfnamefont {H.}~\bibnamefont
  {Wendehenne}}, \bibinfo {author} {\bibfnamefont {T.}~\bibnamefont
  {B{\"a}uerle}},\ and\ \bibinfo {author} {\bibfnamefont {C.}~\bibnamefont
  {Bechinger}},\ }\bibfield  {title} {\bibinfo {title} {Group formation and
  cohesion of active particles with visual perception--dependent motility},\
  }\href {https://doi.org/10.1126/science.aau534} {\bibfield  {journal}
  {\bibinfo  {journal} {Science}\ }\textbf {\bibinfo {volume} {364}},\ \bibinfo
  {pages} {70} (\bibinfo {year} {2019})}\BibitemShut {NoStop}%
\bibitem [{\citenamefont {Miller}\ and\ \citenamefont
  {Bassler}(2001)}]{miller2001quorum}%
  \BibitemOpen
  \bibfield  {author} {\bibinfo {author} {\bibfnamefont {M.~B.}\ \bibnamefont
  {Miller}}\ and\ \bibinfo {author} {\bibfnamefont {B.~L.}\ \bibnamefont
  {Bassler}},\ }\bibfield  {title} {\bibinfo {title} {Quorum sensing in
  bacteria},\ }\href
  {https://doi.org/https://doi.org/10.1146/annurev.micro.55.1.165} {\bibfield
  {journal} {\bibinfo  {journal} {Annu. Rev. Microbiol.}\ }\textbf {\bibinfo
  {volume} {55}},\ \bibinfo {pages} {165} (\bibinfo {year} {2001})}\BibitemShut
  {NoStop}%
\bibitem [{\citenamefont {Hammer}\ and\ \citenamefont
  {Bassler}(2003)}]{hammer2003quorum}%
  \BibitemOpen
  \bibfield  {author} {\bibinfo {author} {\bibfnamefont {B.~K.}\ \bibnamefont
  {Hammer}}\ and\ \bibinfo {author} {\bibfnamefont {B.~L.}\ \bibnamefont
  {Bassler}},\ }\bibfield  {title} {\bibinfo {title} {Quorum sensing controls
  biofilm formation in vibrio cholerae},\ }\href
  {https://doi.org/https://doi.org/10.1046/j.1365-2958.2003.03688} {\bibfield
  {journal} {\bibinfo  {journal} {Mol. Microbiol.}\ }\textbf {\bibinfo {volume}
  {50}},\ \bibinfo {pages} {101} (\bibinfo {year} {2003})}\BibitemShut
  {NoStop}%
\bibitem [{\citenamefont {Daniels}\ \emph {et~al.}(2004)\citenamefont
  {Daniels}, \citenamefont {Vanderleyden},\ and\ \citenamefont
  {Michiels}}]{daniels2004quorum}%
  \BibitemOpen
  \bibfield  {author} {\bibinfo {author} {\bibfnamefont {R.}~\bibnamefont
  {Daniels}}, \bibinfo {author} {\bibfnamefont {J.}~\bibnamefont
  {Vanderleyden}},\ and\ \bibinfo {author} {\bibfnamefont {J.}~\bibnamefont
  {Michiels}},\ }\bibfield  {title} {\bibinfo {title} {Quorum sensing and
  swarming migration in bacteria},\ }\href
  {https://doi.org/https://doi.org/10.1016/j.femsre.2003.09.004} {\bibfield
  {journal} {\bibinfo  {journal} {FEMS Microbiol. Rev}\ }\textbf {\bibinfo
  {volume} {28}},\ \bibinfo {pages} {261} (\bibinfo {year} {2004})}\BibitemShut
  {NoStop}%
\bibitem [{\citenamefont {Tailleur}\ and\ \citenamefont
  {Cates}(2008)}]{tailleur2008statistical}%
  \BibitemOpen
  \bibfield  {author} {\bibinfo {author} {\bibfnamefont {J.}~\bibnamefont
  {Tailleur}}\ and\ \bibinfo {author} {\bibfnamefont {M.~E.}\ \bibnamefont
  {Cates}},\ }\bibfield  {title} {\bibinfo {title} {Statistical mechanics of
  interacting run-and-tumble bacteria},\ }\href
  {https://doi.org/https://doi.org/10.1103/PhysRevLett.100.218103} {\bibfield
  {journal} {\bibinfo  {journal} {Phys. Rev. Lett.}\ }\textbf {\bibinfo
  {volume} {100}},\ \bibinfo {pages} {218103} (\bibinfo {year}
  {2008})}\BibitemShut {NoStop}%
\bibitem [{\citenamefont {Liu}\ \emph {et~al.}(2011)\citenamefont {Liu},
  \citenamefont {Fu}, \citenamefont {Liu}, \citenamefont {Ren}, \citenamefont
  {Chau}, \citenamefont {Li}, \citenamefont {Xiang}, \citenamefont {Zeng},
  \citenamefont {Chen}, \citenamefont {Tang} \emph
  {et~al.}}]{liu2011sequential}%
  \BibitemOpen
  \bibfield  {author} {\bibinfo {author} {\bibfnamefont {C.}~\bibnamefont
  {Liu}}, \bibinfo {author} {\bibfnamefont {X.}~\bibnamefont {Fu}}, \bibinfo
  {author} {\bibfnamefont {L.}~\bibnamefont {Liu}}, \bibinfo {author}
  {\bibfnamefont {X.}~\bibnamefont {Ren}}, \bibinfo {author} {\bibfnamefont
  {C.~K.}\ \bibnamefont {Chau}}, \bibinfo {author} {\bibfnamefont
  {S.}~\bibnamefont {Li}}, \bibinfo {author} {\bibfnamefont {L.}~\bibnamefont
  {Xiang}}, \bibinfo {author} {\bibfnamefont {H.}~\bibnamefont {Zeng}},
  \bibinfo {author} {\bibfnamefont {G.}~\bibnamefont {Chen}}, \bibinfo {author}
  {\bibfnamefont {L.-H.}\ \bibnamefont {Tang}}, \emph {et~al.},\ }\bibfield
  {title} {\bibinfo {title} {Sequential establishment of stripe patterns in an
  expanding cell population},\ }\href {https://doi.org/10.1126/science.1209042}
  {\bibfield  {journal} {\bibinfo  {journal} {Science}\ }\textbf {\bibinfo
  {volume} {334}},\ \bibinfo {pages} {238} (\bibinfo {year}
  {2011})}\BibitemShut {NoStop}%
\bibitem [{\citenamefont {Cates}\ and\ \citenamefont
  {Tailleur}(2015)}]{cates2015motility}%
  \BibitemOpen
  \bibfield  {author} {\bibinfo {author} {\bibfnamefont {M.~E.}\ \bibnamefont
  {Cates}}\ and\ \bibinfo {author} {\bibfnamefont {J.}~\bibnamefont
  {Tailleur}},\ }\bibfield  {title} {\bibinfo {title} {Motility-induced phase
  separation},\ }\href
  {https://doi.org/https://doi.org/10.1146/annurev-conmatphys-031214-014710}
  {\bibfield  {journal} {\bibinfo  {journal} {Annu. Rev. Condens. Matter
  Phys.}\ }\textbf {\bibinfo {volume} {6}},\ \bibinfo {pages} {219} (\bibinfo
  {year} {2015})}\BibitemShut {NoStop}%
\bibitem [{\citenamefont {Curatolo}\ \emph {et~al.}(2020)\citenamefont
  {Curatolo}, \citenamefont {Zhou}, \citenamefont {Zhao}, \citenamefont {Liu},
  \citenamefont {Daerr}, \citenamefont {Tailleur},\ and\ \citenamefont
  {Huang}}]{curatolo2020cooperative}%
  \BibitemOpen
  \bibfield  {author} {\bibinfo {author} {\bibfnamefont {A.}~\bibnamefont
  {Curatolo}}, \bibinfo {author} {\bibfnamefont {N.}~\bibnamefont {Zhou}},
  \bibinfo {author} {\bibfnamefont {Y.}~\bibnamefont {Zhao}}, \bibinfo {author}
  {\bibfnamefont {C.}~\bibnamefont {Liu}}, \bibinfo {author} {\bibfnamefont
  {A.}~\bibnamefont {Daerr}}, \bibinfo {author} {\bibfnamefont
  {J.}~\bibnamefont {Tailleur}},\ and\ \bibinfo {author} {\bibfnamefont
  {J.}~\bibnamefont {Huang}},\ }\bibfield  {title} {\bibinfo {title}
  {Cooperative pattern formation in multi-component bacterial systems through
  reciprocal motility regulation},\ }\href
  {https://doi.org/https://doi.org/10.1038/s41567-020-0964-z} {\bibfield
  {journal} {\bibinfo  {journal} {Nat. Phys.}\ }\textbf {\bibinfo {volume}
  {16}},\ \bibinfo {pages} {1152} (\bibinfo {year} {2020})}\BibitemShut
  {NoStop}%
\bibitem [{\citenamefont {Zhao}\ \emph {et~al.}(2023)\citenamefont {Zhao},
  \citenamefont {Kosmrlj},\ and\ \citenamefont {Datta}}]{zhao2023chemotactic}%
  \BibitemOpen
  \bibfield  {author} {\bibinfo {author} {\bibfnamefont {H.}~\bibnamefont
  {Zhao}}, \bibinfo {author} {\bibfnamefont {A.}~\bibnamefont {Kosmrlj}},\ and\
  \bibinfo {author} {\bibfnamefont {S.~S.}\ \bibnamefont {Datta}},\ }\bibfield
  {title} {\bibinfo {title} {Chemotactic motility-induced phase separation},\
  }\href {https://doi.org/https://doi.org/10.1103/PhysRevLett.131.118301}
  {\bibfield  {journal} {\bibinfo  {journal} {Phys. Rev. Lett.}\ }\textbf
  {\bibinfo {volume} {131}},\ \bibinfo {pages} {118301} (\bibinfo {year}
  {2023})}\BibitemShut {NoStop}%
\bibitem [{\citenamefont {Fily}\ and\ \citenamefont
  {Marchetti}(2012)}]{fily2012athermal}%
  \BibitemOpen
  \bibfield  {author} {\bibinfo {author} {\bibfnamefont {Y.}~\bibnamefont
  {Fily}}\ and\ \bibinfo {author} {\bibfnamefont {M.~C.}\ \bibnamefont
  {Marchetti}},\ }\bibfield  {title} {\bibinfo {title} {Athermal phase
  separation of self-propelled particles with no alignment},\ }\href
  {https://doi.org/https://doi.org/10.1103/PhysRevLett.108.235702} {\bibfield
  {journal} {\bibinfo  {journal} {Phys. Rev. Lett.}\ }\textbf {\bibinfo
  {volume} {108}},\ \bibinfo {pages} {235702} (\bibinfo {year}
  {2012})}\BibitemShut {NoStop}%
\bibitem [{\citenamefont {Cates}\ \emph {et~al.}(2010)\citenamefont {Cates},
  \citenamefont {Marenduzzo}, \citenamefont {Pagonabarraga},\ and\
  \citenamefont {Tailleur}}]{cates2010arrested}%
  \BibitemOpen
  \bibfield  {author} {\bibinfo {author} {\bibfnamefont {M.~E.}\ \bibnamefont
  {Cates}}, \bibinfo {author} {\bibfnamefont {D.}~\bibnamefont {Marenduzzo}},
  \bibinfo {author} {\bibfnamefont {I.}~\bibnamefont {Pagonabarraga}},\ and\
  \bibinfo {author} {\bibfnamefont {J.}~\bibnamefont {Tailleur}},\ }\bibfield
  {title} {\bibinfo {title} {Arrested phase separation in reproducing bacteria
  creates a generic route to pattern formation},\ }\href
  {https://doi.org/https://doi.org/10.1073/pnas.1001994107} {\bibfield
  {journal} {\bibinfo  {journal} {Proc. Natl. Acad. Sci. U.S.A.}\ }\textbf
  {\bibinfo {volume} {107}},\ \bibinfo {pages} {11715} (\bibinfo {year}
  {2010})}\BibitemShut {NoStop}%
\bibitem [{\citenamefont {Dinelli}\ \emph {et~al.}(2023)\citenamefont
  {Dinelli}, \citenamefont {O’Byrne}, \citenamefont {Curatolo}, \citenamefont
  {Zhao}, \citenamefont {Sollich},\ and\ \citenamefont
  {Tailleur}}]{dinelli2023nonreciprocity}%
  \BibitemOpen
  \bibfield  {author} {\bibinfo {author} {\bibfnamefont {A.}~\bibnamefont
  {Dinelli}}, \bibinfo {author} {\bibfnamefont {J.}~\bibnamefont {O’Byrne}},
  \bibinfo {author} {\bibfnamefont {A.}~\bibnamefont {Curatolo}}, \bibinfo
  {author} {\bibfnamefont {Y.}~\bibnamefont {Zhao}}, \bibinfo {author}
  {\bibfnamefont {P.}~\bibnamefont {Sollich}},\ and\ \bibinfo {author}
  {\bibfnamefont {J.}~\bibnamefont {Tailleur}},\ }\bibfield  {title} {\bibinfo
  {title} {Non-reciprocity across scales in active mixtures},\ }\href
  {https://doi.org/https://doi.org/10.1038/s41467-023-42713-5} {\bibfield
  {journal} {\bibinfo  {journal} {Nat. Commun.}\ }\textbf {\bibinfo {volume}
  {14}},\ \bibinfo {pages} {7035} (\bibinfo {year} {2023})}\BibitemShut
  {NoStop}%
\bibitem [{\citenamefont {Duan}\ \emph {et~al.}(2023)\citenamefont {Duan},
  \citenamefont {Agudo-Canalejo}, \citenamefont {Golestanian},\ and\
  \citenamefont {Mahault}}]{duan2023dynamical}%
  \BibitemOpen
  \bibfield  {author} {\bibinfo {author} {\bibfnamefont {Y.}~\bibnamefont
  {Duan}}, \bibinfo {author} {\bibfnamefont {J.}~\bibnamefont
  {Agudo-Canalejo}}, \bibinfo {author} {\bibfnamefont {R.}~\bibnamefont
  {Golestanian}},\ and\ \bibinfo {author} {\bibfnamefont {B.}~\bibnamefont
  {Mahault}},\ }\bibfield  {title} {\bibinfo {title} {Dynamical pattern
  formation without self-attraction in quorum-sensing active matter: the
  interplay between nonreciprocity and motility},\ }\href
  {https://doi.org/https://doi.org/10.1103/PhysRevLett.131.148301} {\bibfield
  {journal} {\bibinfo  {journal} {Phys. Rev. Lett.}\ }\textbf {\bibinfo
  {volume} {131}},\ \bibinfo {pages} {148301} (\bibinfo {year}
  {2023})}\BibitemShut {NoStop}%
\bibitem [{\citenamefont {Pisegna}\ \emph {et~al.}(2024)\citenamefont
  {Pisegna}, \citenamefont {Saha},\ and\ \citenamefont
  {Golestanian}}]{pisegna2024emergent}%
  \BibitemOpen
  \bibfield  {author} {\bibinfo {author} {\bibfnamefont {G.}~\bibnamefont
  {Pisegna}}, \bibinfo {author} {\bibfnamefont {S.}~\bibnamefont {Saha}},\ and\
  \bibinfo {author} {\bibfnamefont {R.}~\bibnamefont {Golestanian}},\
  }\bibfield  {title} {\bibinfo {title} {Emergent polar order in nonpolar
  mixtures with nonreciprocal interactions},\ }\href
  {https://doi.org/https://doi.org/10.1073/pnas.2407705121} {\bibfield
  {journal} {\bibinfo  {journal} {Proc. Natl. Acad. Sci. U.S.A.}\ }\textbf
  {\bibinfo {volume} {121}},\ \bibinfo {pages} {e2407705121} (\bibinfo {year}
  {2024})}\BibitemShut {NoStop}%
\bibitem [{\citenamefont {Caprini}\ and\ \citenamefont {Marini
  Bettolo~Marconi}(2025)}]{caprini2025bubble}%
  \BibitemOpen
  \bibfield  {author} {\bibinfo {author} {\bibfnamefont {L.}~\bibnamefont
  {Caprini}}\ and\ \bibinfo {author} {\bibfnamefont {U.}~\bibnamefont {Marini
  Bettolo~Marconi}},\ }\bibfield  {title} {\bibinfo {title} {Bubble phase
  induced by odd interactions in chiral systems},\ }\href
  {https://doi.org/https://doi.org/10.1063/5.0262594} {\bibfield  {journal}
  {\bibinfo  {journal} {J. Chem. Phys.}\ }\textbf {\bibinfo {volume} {162}},\
  \bibinfo {pages} {161101} (\bibinfo {year} {2025})}\BibitemShut {NoStop}%
\bibitem [{\citenamefont {Fox}(1986{\natexlab{a}})}]{fox1986uniform}%
  \BibitemOpen
  \bibfield  {author} {\bibinfo {author} {\bibfnamefont {R.~F.}\ \bibnamefont
  {Fox}},\ }\bibfield  {title} {\bibinfo {title} {Uniform convergence to an
  effective fokker-planck equation for weakly colored noise},\ }\href
  {https://doi.org/https://doi.org/10.1103/PhysRevA.34.4525} {\bibfield
  {journal} {\bibinfo  {journal} {Phys. Rev. A}\ }\textbf {\bibinfo {volume}
  {34}},\ \bibinfo {pages} {4525} (\bibinfo {year}
  {1986}{\natexlab{a}})}\BibitemShut {NoStop}%
\bibitem [{\citenamefont {Fox}(1986{\natexlab{b}})}]{fox1986functional}%
  \BibitemOpen
  \bibfield  {author} {\bibinfo {author} {\bibfnamefont {R.~F.}\ \bibnamefont
  {Fox}},\ }\bibfield  {title} {\bibinfo {title} {Functional-calculus approach
  to stochastic differential equations},\ }\href
  {https://doi.org/https://doi.org/10.1103/PhysRevA.33.467} {\bibfield
  {journal} {\bibinfo  {journal} {Phys. Rev. A}\ }\textbf {\bibinfo {volume}
  {33}},\ \bibinfo {pages} {467} (\bibinfo {year}
  {1986}{\natexlab{b}})}\BibitemShut {NoStop}%
\bibitem [{\citenamefont {Faetti}\ \emph {et~al.}(1988)\citenamefont {Faetti},
  \citenamefont {Fronzoni}, \citenamefont {Grigolini},\ and\ \citenamefont
  {Mannella}}]{faetti1988projection}%
  \BibitemOpen
  \bibfield  {author} {\bibinfo {author} {\bibfnamefont {S.}~\bibnamefont
  {Faetti}}, \bibinfo {author} {\bibfnamefont {L.}~\bibnamefont {Fronzoni}},
  \bibinfo {author} {\bibfnamefont {P.}~\bibnamefont {Grigolini}},\ and\
  \bibinfo {author} {\bibfnamefont {R.}~\bibnamefont {Mannella}},\ }\bibfield
  {title} {\bibinfo {title} {The projection approach to the fokker-planck
  equation. i. colored gaussian noise},\ }\href
  {https://doi.org/https://doi.org/10.1007/BF01019735} {\bibfield  {journal}
  {\bibinfo  {journal} {J. Stat. Phys.}\ }\textbf {\bibinfo {volume} {52}},\
  \bibinfo {pages} {951} (\bibinfo {year} {1988})}\BibitemShut {NoStop}%
\bibitem [{\citenamefont {Schnitzer}(1993)}]{schnitzer1993theory}%
  \BibitemOpen
  \bibfield  {author} {\bibinfo {author} {\bibfnamefont {M.~J.}\ \bibnamefont
  {Schnitzer}},\ }\bibfield  {title} {\bibinfo {title} {Theory of continuum
  random walks and application to chemotaxis},\ }\href
  {https://doi.org/https://doi.org/10.1103/PhysRevE.48.2553} {\bibfield
  {journal} {\bibinfo  {journal} {Phys. Rev. E}\ }\textbf {\bibinfo {volume}
  {48}},\ \bibinfo {pages} {2553} (\bibinfo {year} {1993})}\BibitemShut
  {NoStop}%
\bibitem [{\citenamefont {Cates}\ and\ \citenamefont
  {Tailleur}(2013)}]{cates2013when}%
  \BibitemOpen
  \bibfield  {author} {\bibinfo {author} {\bibfnamefont {M.~E.}\ \bibnamefont
  {Cates}}\ and\ \bibinfo {author} {\bibfnamefont {J.}~\bibnamefont
  {Tailleur}},\ }\bibfield  {title} {\bibinfo {title} {When are active brownian
  particles and run-and-tumble particles equivalent? consequences for
  motility-induced phase separation},\ }\href
  {https://doi.org/10.1209/0295-5075/101/20010} {\bibfield  {journal} {\bibinfo
   {journal} {EPL}\ }\textbf {\bibinfo {volume} {101}},\ \bibinfo {pages}
  {20010} (\bibinfo {year} {2013})}\BibitemShut {NoStop}%
\bibitem [{\citenamefont {Wittmann}\ \emph
  {et~al.}(2017{\natexlab{a}})\citenamefont {Wittmann}, \citenamefont {Maggi},
  \citenamefont {Sharma}, \citenamefont {Scacchi}, \citenamefont {Brader},\
  and\ \citenamefont {Marconi}}]{wittmann2017effective}%
  \BibitemOpen
  \bibfield  {author} {\bibinfo {author} {\bibfnamefont {R.}~\bibnamefont
  {Wittmann}}, \bibinfo {author} {\bibfnamefont {C.}~\bibnamefont {Maggi}},
  \bibinfo {author} {\bibfnamefont {A.}~\bibnamefont {Sharma}}, \bibinfo
  {author} {\bibfnamefont {A.}~\bibnamefont {Scacchi}}, \bibinfo {author}
  {\bibfnamefont {J.~M.}\ \bibnamefont {Brader}},\ and\ \bibinfo {author}
  {\bibfnamefont {U.~M.~B.}\ \bibnamefont {Marconi}},\ }\bibfield  {title}
  {\bibinfo {title} {Effective equilibrium states in the colored-noise model
  for active matter i. pairwise forces in the fox and unified colored noise
  approximations},\ }\href
  {https://doi.org/https://doi.org/10.1088/1742-5468/aa8c1f} {\bibfield
  {journal} {\bibinfo  {journal} {J. Stat. Mech.}\ }\textbf {\bibinfo {volume}
  {2017}},\ \bibinfo {pages} {113207} (\bibinfo {year}
  {2017}{\natexlab{a}})}\BibitemShut {NoStop}%
\bibitem [{\citenamefont {O’Byrne}\ and\ \citenamefont
  {Tailleur}(2020)}]{o2020lamellar}%
  \BibitemOpen
  \bibfield  {author} {\bibinfo {author} {\bibfnamefont {J.}~\bibnamefont
  {O’Byrne}}\ and\ \bibinfo {author} {\bibfnamefont {J.}~\bibnamefont
  {Tailleur}},\ }\bibfield  {title} {\bibinfo {title} {Lamellar to micellar
  phases and beyond: when tactic active systems admit free energy
  functionals},\ }\href
  {https://doi.org/https://doi.org/10.1103/PhysRevLett.125.208003} {\bibfield
  {journal} {\bibinfo  {journal} {Phys. Rev. Lett.}\ }\textbf {\bibinfo
  {volume} {125}},\ \bibinfo {pages} {208003} (\bibinfo {year}
  {2020})}\BibitemShut {NoStop}%
\bibitem [{\citenamefont {Dinelli}\ \emph {et~al.}(2024)\citenamefont
  {Dinelli}, \citenamefont {O’Byrne},\ and\ \citenamefont
  {Tailleur}}]{dinelli2024fluctuating}%
  \BibitemOpen
  \bibfield  {author} {\bibinfo {author} {\bibfnamefont {A.}~\bibnamefont
  {Dinelli}}, \bibinfo {author} {\bibfnamefont {J.}~\bibnamefont {O’Byrne}},\
  and\ \bibinfo {author} {\bibfnamefont {J.}~\bibnamefont {Tailleur}},\
  }\bibfield  {title} {\bibinfo {title} {Fluctuating hydrodynamics of active
  particles interacting via taxis and quorum sensing: static and dynamics},\
  }\href {https://doi.org/10.1088/1751-8121/ad72bc} {\bibfield  {journal}
  {\bibinfo  {journal} {J. Phys. A: Math. Theor.}\ }\textbf {\bibinfo {volume}
  {57}},\ \bibinfo {pages} {395002} (\bibinfo {year} {2024})}\BibitemShut
  {NoStop}%
\bibitem [{\citenamefont {Pisegna}\ and\ \citenamefont
  {Saha}(2025)}]{pisegna2025spinning}%
  \BibitemOpen
  \bibfield  {author} {\bibinfo {author} {\bibfnamefont {G.}~\bibnamefont
  {Pisegna}}\ and\ \bibinfo {author} {\bibfnamefont {S.}~\bibnamefont {Saha}},\
  }\bibfield  {title} {\bibinfo {title} {Spinning mixtures: nonreciprocity
  transfers chirality across scales in scalar densities},\ }\href@noop {}
  {\bibfield  {journal} {\bibinfo  {journal} {arXiv preprint arXiv:2509.07630}\
  } (\bibinfo {year} {2025})}\BibitemShut {NoStop}%
\bibitem [{\citenamefont {Burekovi{\'c}}\ \emph {et~al.}(2026)\citenamefont
  {Burekovi{\'c}}, \citenamefont {De~Luca}, \citenamefont {Cates},\ and\
  \citenamefont {Nardini}}]{burekovic2026active}%
  \BibitemOpen
  \bibfield  {author} {\bibinfo {author} {\bibfnamefont {S.}~\bibnamefont
  {Burekovi{\'c}}}, \bibinfo {author} {\bibfnamefont {F.}~\bibnamefont
  {De~Luca}}, \bibinfo {author} {\bibfnamefont {M.~E.}\ \bibnamefont {Cates}},\
  and\ \bibinfo {author} {\bibfnamefont {C.}~\bibnamefont {Nardini}},\
  }\bibfield  {title} {\bibinfo {title} {Active cahn--hilliard theory for
  non-equilibrium phase separation: quantitative macroscopic predictions and a
  microscopic derivation},\ }\href@noop {} {\bibfield  {journal} {\bibinfo
  {journal} {arXiv preprint arXiv:2601.16539}\ } (\bibinfo {year}
  {2026})}\BibitemShut {NoStop}%
\bibitem [{\citenamefont {Berg}(2004)}]{berg2004coli}%
  \BibitemOpen
  \bibfield  {author} {\bibinfo {author} {\bibfnamefont {H.~C.}\ \bibnamefont
  {Berg}},\ }\href@noop {} {\emph {\bibinfo {title} {E. coli in Motion}}}\
  (\bibinfo  {publisher} {Springer},\ \bibinfo {year} {2004})\BibitemShut
  {NoStop}%
\bibitem [{\citenamefont {Kurzthaler}\ \emph {et~al.}(2024)\citenamefont
  {Kurzthaler}, \citenamefont {Zhao}, \citenamefont {Zhou}, \citenamefont
  {Schwarz-Linek}, \citenamefont {Devailly}, \citenamefont {Arlt},
  \citenamefont {Huang}, \citenamefont {Poon}, \citenamefont {Franosch},
  \citenamefont {Tailleur} \emph {et~al.}}]{kurzthaler2024characterization}%
  \BibitemOpen
  \bibfield  {author} {\bibinfo {author} {\bibfnamefont {C.}~\bibnamefont
  {Kurzthaler}}, \bibinfo {author} {\bibfnamefont {Y.}~\bibnamefont {Zhao}},
  \bibinfo {author} {\bibfnamefont {N.}~\bibnamefont {Zhou}}, \bibinfo {author}
  {\bibfnamefont {J.}~\bibnamefont {Schwarz-Linek}}, \bibinfo {author}
  {\bibfnamefont {C.}~\bibnamefont {Devailly}}, \bibinfo {author}
  {\bibfnamefont {J.}~\bibnamefont {Arlt}}, \bibinfo {author} {\bibfnamefont
  {J.-D.}\ \bibnamefont {Huang}}, \bibinfo {author} {\bibfnamefont {W.~C.}\
  \bibnamefont {Poon}}, \bibinfo {author} {\bibfnamefont {T.}~\bibnamefont
  {Franosch}}, \bibinfo {author} {\bibfnamefont {J.}~\bibnamefont {Tailleur}},
  \emph {et~al.},\ }\bibfield  {title} {\bibinfo {title} {Characterization and
  control of the run-and-tumble dynamics of escherichia coli},\ }\href
  {https://doi.org/https://doi.org/10.1103/PhysRevLett.132.038302} {\bibfield
  {journal} {\bibinfo  {journal} {Phys. Rev. Lett.}\ }\textbf {\bibinfo
  {volume} {132}},\ \bibinfo {pages} {038302} (\bibinfo {year}
  {2024})}\BibitemShut {NoStop}%
\bibitem [{\citenamefont {Golestanian}\ \emph {et~al.}(2007)\citenamefont
  {Golestanian}, \citenamefont {Liverpool},\ and\ \citenamefont
  {Ajdari}}]{golestanian2007designing}%
  \BibitemOpen
  \bibfield  {author} {\bibinfo {author} {\bibfnamefont {R.}~\bibnamefont
  {Golestanian}}, \bibinfo {author} {\bibfnamefont {T.}~\bibnamefont
  {Liverpool}},\ and\ \bibinfo {author} {\bibfnamefont {A.}~\bibnamefont
  {Ajdari}},\ }\bibfield  {title} {\bibinfo {title} {Designing phoretic
  micro-and nano-swimmers},\ }\href
  {https://doi.org/https://doi.org/10.1088/1367-2630/9/5/126} {\bibfield
  {journal} {\bibinfo  {journal} {New J. Phys.}\ }\textbf {\bibinfo {volume}
  {9}},\ \bibinfo {pages} {126} (\bibinfo {year} {2007})}\BibitemShut {NoStop}%
\bibitem [{\citenamefont {Theurkauff}\ \emph {et~al.}(2012)\citenamefont
  {Theurkauff}, \citenamefont {Cottin-Bizonne}, \citenamefont {Palacci},
  \citenamefont {Ybert},\ and\ \citenamefont
  {Bocquet}}]{theurkauff2012dynamic}%
  \BibitemOpen
  \bibfield  {author} {\bibinfo {author} {\bibfnamefont {I.}~\bibnamefont
  {Theurkauff}}, \bibinfo {author} {\bibfnamefont {C.}~\bibnamefont
  {Cottin-Bizonne}}, \bibinfo {author} {\bibfnamefont {J.}~\bibnamefont
  {Palacci}}, \bibinfo {author} {\bibfnamefont {C.}~\bibnamefont {Ybert}},\
  and\ \bibinfo {author} {\bibfnamefont {L.}~\bibnamefont {Bocquet}},\
  }\bibfield  {title} {\bibinfo {title} {Dynamic clustering in active colloidal
  suspensions with chemical signaling},\ }\href
  {https://doi.org/https://doi.org/10.1103/PhysRevLett.108.268303} {\bibfield
  {journal} {\bibinfo  {journal} {Phys. Rev. Lett.}\ }\textbf {\bibinfo
  {volume} {108}},\ \bibinfo {pages} {268303} (\bibinfo {year}
  {2012})}\BibitemShut {NoStop}%
\bibitem [{\citenamefont {Sep{\'u}lveda}\ \emph {et~al.}(2013)\citenamefont
  {Sep{\'u}lveda}, \citenamefont {Petitjean}, \citenamefont {Cochet},
  \citenamefont {Grasland-Mongrain}, \citenamefont {Silberzan},\ and\
  \citenamefont {Hakim}}]{sepulveda2013collective}%
  \BibitemOpen
  \bibfield  {author} {\bibinfo {author} {\bibfnamefont {N.}~\bibnamefont
  {Sep{\'u}lveda}}, \bibinfo {author} {\bibfnamefont {L.}~\bibnamefont
  {Petitjean}}, \bibinfo {author} {\bibfnamefont {O.}~\bibnamefont {Cochet}},
  \bibinfo {author} {\bibfnamefont {E.}~\bibnamefont {Grasland-Mongrain}},
  \bibinfo {author} {\bibfnamefont {P.}~\bibnamefont {Silberzan}},\ and\
  \bibinfo {author} {\bibfnamefont {V.}~\bibnamefont {Hakim}},\ }\bibfield
  {title} {\bibinfo {title} {Collective cell motion in an epithelial sheet can
  be quantitatively described by a stochastic interacting particle model},\
  }\href {https://doi.org/https://doi.org/10.1371/journal.pcbi.1002944}
  {\bibfield  {journal} {\bibinfo  {journal} {PLoS Comput. Biol.}\ }\textbf
  {\bibinfo {volume} {9}},\ \bibinfo {pages} {e1002944} (\bibinfo {year}
  {2013})}\BibitemShut {NoStop}%
\bibitem [{\citenamefont {Szamel}(2014)}]{szamel2014self}%
  \BibitemOpen
  \bibfield  {author} {\bibinfo {author} {\bibfnamefont {G.}~\bibnamefont
  {Szamel}},\ }\bibfield  {title} {\bibinfo {title} {Self-propelled particle in
  an external potential: Existence of an effective temperature},\ }\href
  {https://doi.org/https://doi.org/10.1103/PhysRevE.90.012111} {\bibfield
  {journal} {\bibinfo  {journal} {Phys. Rev. E}\ }\textbf {\bibinfo {volume}
  {90}},\ \bibinfo {pages} {012111} (\bibinfo {year} {2014})}\BibitemShut
  {NoStop}%
\bibitem [{\citenamefont {Wittmann}\ \emph
  {et~al.}(2017{\natexlab{b}})\citenamefont {Wittmann}, \citenamefont
  {Marconi}, \citenamefont {Maggi},\ and\ \citenamefont
  {Brader}}]{wittmann2017effective2}%
  \BibitemOpen
  \bibfield  {author} {\bibinfo {author} {\bibfnamefont {R.}~\bibnamefont
  {Wittmann}}, \bibinfo {author} {\bibfnamefont {U.~M.~B.}\ \bibnamefont
  {Marconi}}, \bibinfo {author} {\bibfnamefont {C.}~\bibnamefont {Maggi}},\
  and\ \bibinfo {author} {\bibfnamefont {J.~M.}\ \bibnamefont {Brader}},\
  }\bibfield  {title} {\bibinfo {title} {Effective equilibrium states in the
  colored-noise model for active matter ii. a unified framework for phase
  equilibria, structure and mechanical properties},\ }\href
  {https://doi.org/https://doi.org/10.1088/1742-5468/aa8c37} {\bibfield
  {journal} {\bibinfo  {journal} {J. Stat. Mech.}\ }\textbf {\bibinfo {volume}
  {2017}},\ \bibinfo {pages} {113208} (\bibinfo {year}
  {2017}{\natexlab{b}})}\BibitemShut {NoStop}%
\bibitem [{\citenamefont {Martin}\ \emph {et~al.}(2021)\citenamefont {Martin},
  \citenamefont {O'Byrne}, \citenamefont {Cates}, \citenamefont {Fodor},
  \citenamefont {Nardini}, \citenamefont {Tailleur},\ and\ \citenamefont
  {Van~Wijland}}]{martin2021statistical}%
  \BibitemOpen
  \bibfield  {author} {\bibinfo {author} {\bibfnamefont {D.}~\bibnamefont
  {Martin}}, \bibinfo {author} {\bibfnamefont {J.}~\bibnamefont {O'Byrne}},
  \bibinfo {author} {\bibfnamefont {M.~E.}\ \bibnamefont {Cates}}, \bibinfo
  {author} {\bibfnamefont {{\'E}.}~\bibnamefont {Fodor}}, \bibinfo {author}
  {\bibfnamefont {C.}~\bibnamefont {Nardini}}, \bibinfo {author} {\bibfnamefont
  {J.}~\bibnamefont {Tailleur}},\ and\ \bibinfo {author} {\bibfnamefont
  {F.}~\bibnamefont {Van~Wijland}},\ }\bibfield  {title} {\bibinfo {title}
  {Statistical mechanics of active ornstein-uhlenbeck particles},\ }\href
  {https://doi.org/https://doi.org/10.1103/PhysRevE.103.032607} {\bibfield
  {journal} {\bibinfo  {journal} {Phys. Rev. E}\ }\textbf {\bibinfo {volume}
  {103}},\ \bibinfo {pages} {032607} (\bibinfo {year} {2021})}\BibitemShut
  {NoStop}%
\bibitem [{\citenamefont {Solon}\ \emph {et~al.}(2015)\citenamefont {Solon},
  \citenamefont {Cates},\ and\ \citenamefont {Tailleur}}]{solon2015active}%
  \BibitemOpen
  \bibfield  {author} {\bibinfo {author} {\bibfnamefont {A.~P.}\ \bibnamefont
  {Solon}}, \bibinfo {author} {\bibfnamefont {M.~E.}\ \bibnamefont {Cates}},\
  and\ \bibinfo {author} {\bibfnamefont {J.}~\bibnamefont {Tailleur}},\
  }\bibfield  {title} {\bibinfo {title} {Active brownian particles and
  run-and-tumble particles: A comparative study},\ }\href
  {https://doi.org/https://doi.org/10.1140/epjst/e2015-02457-0} {\bibfield
  {journal} {\bibinfo  {journal} {Eur. Phys. J. Spec. Top.}\ }\textbf {\bibinfo
  {volume} {224}},\ \bibinfo {pages} {1231} (\bibinfo {year}
  {2015})}\BibitemShut {NoStop}%
\bibitem [{\citenamefont {Ghosh}\ and\ \citenamefont
  {Gov}(2014)}]{ghosh2014dynamics}%
  \BibitemOpen
  \bibfield  {author} {\bibinfo {author} {\bibfnamefont {A.}~\bibnamefont
  {Ghosh}}\ and\ \bibinfo {author} {\bibfnamefont {N.~S.}\ \bibnamefont
  {Gov}},\ }\bibfield  {title} {\bibinfo {title} {Dynamics of active
  semiflexible polymers},\ }\href {https://doi.org/10.1016/j.bpj.2014.07.034}
  {\bibfield  {journal} {\bibinfo  {journal} {Biophys. J.}\ }\textbf {\bibinfo
  {volume} {107}},\ \bibinfo {pages} {1065} (\bibinfo {year}
  {2014})}\BibitemShut {NoStop}%
\bibitem [{\citenamefont {Isele-Holder}\ \emph {et~al.}(2015)\citenamefont
  {Isele-Holder}, \citenamefont {Elgeti},\ and\ \citenamefont
  {Gompper}}]{isele2015self}%
  \BibitemOpen
  \bibfield  {author} {\bibinfo {author} {\bibfnamefont {R.~E.}\ \bibnamefont
  {Isele-Holder}}, \bibinfo {author} {\bibfnamefont {J.}~\bibnamefont
  {Elgeti}},\ and\ \bibinfo {author} {\bibfnamefont {G.}~\bibnamefont
  {Gompper}},\ }\bibfield  {title} {\bibinfo {title} {Self-propelled worm-like
  filaments: spontaneous spiral formation, structure, and dynamics},\ }\href
  {https://doi.org/https://doi.org/10.1039/C5SM01683E} {\bibfield  {journal}
  {\bibinfo  {journal} {Soft Matter}\ }\textbf {\bibinfo {volume} {11}},\
  \bibinfo {pages} {7181} (\bibinfo {year} {2015})}\BibitemShut {NoStop}%
\bibitem [{\citenamefont {Winkler}\ \emph {et~al.}(2017)\citenamefont
  {Winkler}, \citenamefont {Elgeti},\ and\ \citenamefont
  {Gompper}}]{winkler2017active}%
  \BibitemOpen
  \bibfield  {author} {\bibinfo {author} {\bibfnamefont {R.~G.}\ \bibnamefont
  {Winkler}}, \bibinfo {author} {\bibfnamefont {J.}~\bibnamefont {Elgeti}},\
  and\ \bibinfo {author} {\bibfnamefont {G.}~\bibnamefont {Gompper}},\
  }\bibfield  {title} {\bibinfo {title} {Active polymers—emergent
  conformational and dynamical properties: A brief review},\ }\href
  {https://doi.org/https://doi.org/10.7566/JPSJ.86.101014} {\bibfield
  {journal} {\bibinfo  {journal} {J. Phys. Soc. Jpn.}\ }\textbf {\bibinfo
  {volume} {86}},\ \bibinfo {pages} {101014} (\bibinfo {year}
  {2017})}\BibitemShut {NoStop}%
\bibitem [{\citenamefont {Bianco}\ \emph {et~al.}(2018)\citenamefont {Bianco},
  \citenamefont {Locatelli},\ and\ \citenamefont
  {Malgaretti}}]{bianco2018globulelike}%
  \BibitemOpen
  \bibfield  {author} {\bibinfo {author} {\bibfnamefont {V.}~\bibnamefont
  {Bianco}}, \bibinfo {author} {\bibfnamefont {E.}~\bibnamefont {Locatelli}},\
  and\ \bibinfo {author} {\bibfnamefont {P.}~\bibnamefont {Malgaretti}},\
  }\bibfield  {title} {\bibinfo {title} {Globulelike conformation and enhanced
  diffusion of active polymers},\ }\href
  {https://doi.org/https://doi.org/10.1103/PhysRevLett.121.217802} {\bibfield
  {journal} {\bibinfo  {journal} {Phys. Rev. Lett.}\ }\textbf {\bibinfo
  {volume} {121}},\ \bibinfo {pages} {217802} (\bibinfo {year}
  {2018})}\BibitemShut {NoStop}%
\bibitem [{\citenamefont {Winkler}\ and\ \citenamefont
  {Gompper}(2020)}]{winkler2020physics}%
  \BibitemOpen
  \bibfield  {author} {\bibinfo {author} {\bibfnamefont {R.~G.}\ \bibnamefont
  {Winkler}}\ and\ \bibinfo {author} {\bibfnamefont {G.}~\bibnamefont
  {Gompper}},\ }\bibfield  {title} {\bibinfo {title} {The physics of active
  polymers and filaments},\ }\href
  {https://doi.org/https://doi.org/10.1063/5.0011466} {\bibfield  {journal}
  {\bibinfo  {journal} {J. Chem. Phys.}\ }\textbf {\bibinfo {volume} {153}},\
  \bibinfo {pages} {040901} (\bibinfo {year} {2020})}\BibitemShut {NoStop}%
\bibitem [{\citenamefont {Pfreundt}\ \emph {et~al.}(2023)\citenamefont
  {Pfreundt}, \citenamefont {S{\l}omka}, \citenamefont {Schneider},
  \citenamefont {Sengupta}, \citenamefont {Carrara}, \citenamefont {Fernandez},
  \citenamefont {Ackermann},\ and\ \citenamefont
  {Stocker}}]{pfreundt2023controlled}%
  \BibitemOpen
  \bibfield  {author} {\bibinfo {author} {\bibfnamefont {U.}~\bibnamefont
  {Pfreundt}}, \bibinfo {author} {\bibfnamefont {J.}~\bibnamefont {S{\l}omka}},
  \bibinfo {author} {\bibfnamefont {G.}~\bibnamefont {Schneider}}, \bibinfo
  {author} {\bibfnamefont {A.}~\bibnamefont {Sengupta}}, \bibinfo {author}
  {\bibfnamefont {F.}~\bibnamefont {Carrara}}, \bibinfo {author} {\bibfnamefont
  {V.}~\bibnamefont {Fernandez}}, \bibinfo {author} {\bibfnamefont
  {M.}~\bibnamefont {Ackermann}},\ and\ \bibinfo {author} {\bibfnamefont
  {R.}~\bibnamefont {Stocker}},\ }\bibfield  {title} {\bibinfo {title}
  {Controlled motility in the cyanobacterium trichodesmium regulates aggregate
  architecture},\ }\href
  {https://doi.org/https://doi.org/10.1126/science.adf2753} {\bibfield
  {journal} {\bibinfo  {journal} {Science}\ }\textbf {\bibinfo {volume}
  {380}},\ \bibinfo {pages} {830} (\bibinfo {year} {2023})}\BibitemShut
  {NoStop}%
\bibitem [{\citenamefont {Dedenon}\ \emph {et~al.}(2026)\citenamefont
  {Dedenon}, \citenamefont {Blanch-Mercader}, \citenamefont {Kruse},\ and\
  \citenamefont {Elgeti}}]{dedenon2026importance}%
  \BibitemOpen
  \bibfield  {author} {\bibinfo {author} {\bibfnamefont {M.}~\bibnamefont
  {Dedenon}}, \bibinfo {author} {\bibfnamefont {C.}~\bibnamefont
  {Blanch-Mercader}}, \bibinfo {author} {\bibfnamefont {K.}~\bibnamefont
  {Kruse}},\ and\ \bibinfo {author} {\bibfnamefont {J.}~\bibnamefont
  {Elgeti}},\ }\bibfield  {title} {\bibinfo {title} {The importance of being
  discrete: fluctuations, defects, and density-orientation coupling in
  agent-based active nematics},\ }\href@noop {} {\bibfield  {journal} {\bibinfo
   {journal} {New Journal of Physics}\ }\textbf {\bibinfo {volume} {28}},\
  \bibinfo {pages} {024401} (\bibinfo {year} {2026})}\BibitemShut {NoStop}%
\bibitem [{\citenamefont {Hu}\ \emph {et~al.}(2025)\citenamefont {Hu},
  \citenamefont {Chen}, \citenamefont {Zhu}, \citenamefont {Lin}, \citenamefont
  {Nie},\ and\ \citenamefont {Xia}}]{hu2025cargo}%
  \BibitemOpen
  \bibfield  {author} {\bibinfo {author} {\bibfnamefont {X.}~\bibnamefont
  {Hu}}, \bibinfo {author} {\bibfnamefont {W.}~\bibnamefont {Chen}}, \bibinfo
  {author} {\bibfnamefont {Z.}~\bibnamefont {Zhu}}, \bibinfo {author}
  {\bibfnamefont {J.}~\bibnamefont {Lin}}, \bibinfo {author} {\bibfnamefont
  {D.}~\bibnamefont {Nie}},\ and\ \bibinfo {author} {\bibfnamefont
  {Y.}~\bibnamefont {Xia}},\ }\bibfield  {title} {\bibinfo {title} {Cargo
  carrying by a spring connected chiral micro-swimmer in a square channel},\
  }\href {https://doi.org/https://doi.org/10.1017/jfm.2025.405} {\bibfield
  {journal} {\bibinfo  {journal} {J. Fluid Mech.}\ }\textbf {\bibinfo {volume}
  {1011}},\ \bibinfo {pages} {A50} (\bibinfo {year} {2025})}\BibitemShut
  {NoStop}%
\bibitem [{\citenamefont {Vuijk}\ \emph {et~al.}(2021)\citenamefont {Vuijk},
  \citenamefont {Merlitz}, \citenamefont {Lang}, \citenamefont {Sharma},\ and\
  \citenamefont {Sommer}}]{vuijk2021chemotaxis}%
  \BibitemOpen
  \bibfield  {author} {\bibinfo {author} {\bibfnamefont {H.~D.}\ \bibnamefont
  {Vuijk}}, \bibinfo {author} {\bibfnamefont {H.}~\bibnamefont {Merlitz}},
  \bibinfo {author} {\bibfnamefont {M.}~\bibnamefont {Lang}}, \bibinfo {author}
  {\bibfnamefont {A.}~\bibnamefont {Sharma}},\ and\ \bibinfo {author}
  {\bibfnamefont {J.-U.}\ \bibnamefont {Sommer}},\ }\bibfield  {title}
  {\bibinfo {title} {Chemotaxis of cargo-carrying self-propelled particles},\
  }\href {https://doi.org/https://doi.org/10.1103/PhysRevLett.126.208102}
  {\bibfield  {journal} {\bibinfo  {journal} {Phys. Rev. Lett.}\ }\textbf
  {\bibinfo {volume} {126}},\ \bibinfo {pages} {208102} (\bibinfo {year}
  {2021})}\BibitemShut {NoStop}%
\bibitem [{\citenamefont {Muzzeddu}\ \emph {et~al.}(2023)\citenamefont
  {Muzzeddu}, \citenamefont {Rold{\'a}n}, \citenamefont {Gambassi},\ and\
  \citenamefont {Sharma}}]{muzzeddu2023taxis}%
  \BibitemOpen
  \bibfield  {author} {\bibinfo {author} {\bibfnamefont {P.~L.}\ \bibnamefont
  {Muzzeddu}}, \bibinfo {author} {\bibfnamefont {{\'E}.}~\bibnamefont
  {Rold{\'a}n}}, \bibinfo {author} {\bibfnamefont {A.}~\bibnamefont
  {Gambassi}},\ and\ \bibinfo {author} {\bibfnamefont {A.}~\bibnamefont
  {Sharma}},\ }\bibfield  {title} {\bibinfo {title} {Taxis of cargo-carrying
  microswimmers in traveling activity waves},\ }\href
  {https://doi.org/10.1209/0295-5075/acd8e9} {\bibfield  {journal} {\bibinfo
  {journal} {EPL}\ }\textbf {\bibinfo {volume} {142}},\ \bibinfo {pages}
  {67001} (\bibinfo {year} {2023})}\BibitemShut {NoStop}%
\bibitem [{\citenamefont {Muzzeddu}\ \emph {et~al.}(2022)\citenamefont
  {Muzzeddu}, \citenamefont {Vuijk}, \citenamefont {L{\"o}wen}, \citenamefont
  {Sommer},\ and\ \citenamefont {Sharma}}]{muzzeddu2022active}%
  \BibitemOpen
  \bibfield  {author} {\bibinfo {author} {\bibfnamefont {P.~L.}\ \bibnamefont
  {Muzzeddu}}, \bibinfo {author} {\bibfnamefont {H.~D.}\ \bibnamefont {Vuijk}},
  \bibinfo {author} {\bibfnamefont {H.}~\bibnamefont {L{\"o}wen}}, \bibinfo
  {author} {\bibfnamefont {J.-U.}\ \bibnamefont {Sommer}},\ and\ \bibinfo
  {author} {\bibfnamefont {A.}~\bibnamefont {Sharma}},\ }\bibfield  {title}
  {\bibinfo {title} {Active chiral molecules in activity gradients},\ }\href
  {https://doi.org/https://doi.org/10.1063/5.0109817} {\bibfield  {journal}
  {\bibinfo  {journal} {J. Chem. Phys.}\ }\textbf {\bibinfo {volume} {157}},\
  \bibinfo {pages} {134902} (\bibinfo {year} {2022})}\BibitemShut {NoStop}%
\bibitem [{\citenamefont {Caprini}\ \emph {et~al.}(2025)\citenamefont
  {Caprini}, \citenamefont {Abdoli}, \citenamefont {Marconi},\ and\
  \citenamefont {L{\"o}wen}}]{caprini2025spontaneous}%
  \BibitemOpen
  \bibfield  {author} {\bibinfo {author} {\bibfnamefont {L.}~\bibnamefont
  {Caprini}}, \bibinfo {author} {\bibfnamefont {I.}~\bibnamefont {Abdoli}},
  \bibinfo {author} {\bibfnamefont {U.~M.~B.}\ \bibnamefont {Marconi}},\ and\
  \bibinfo {author} {\bibfnamefont {H.}~\bibnamefont {L{\"o}wen}},\ }\bibfield
  {title} {\bibinfo {title} {Spontaneous self-wrapping in chiral active
  polymers},\ }\bibfield  {journal} {\bibinfo  {journal} {Newton}\ }\textbf
  {\bibinfo {volume} {1}},\ \href
  {https://doi.org/https://doi.org/10.1016/j.newton.2025.100253}
  {https://doi.org/10.1016/j.newton.2025.100253} (\bibinfo {year}
  {2025})\BibitemShut {NoStop}%
\bibitem [{\citenamefont {Valecha}\ \emph {et~al.}(2025)\citenamefont
  {Valecha}, \citenamefont {Vahid}, \citenamefont {Muzzeddu}, \citenamefont
  {Sommer},\ and\ \citenamefont {Sharma}}]{valecha2025active}%
  \BibitemOpen
  \bibfield  {author} {\bibinfo {author} {\bibfnamefont {B.}~\bibnamefont
  {Valecha}}, \bibinfo {author} {\bibfnamefont {H.}~\bibnamefont {Vahid}},
  \bibinfo {author} {\bibfnamefont {P.~L.}\ \bibnamefont {Muzzeddu}}, \bibinfo
  {author} {\bibfnamefont {J.-U.}\ \bibnamefont {Sommer}},\ and\ \bibinfo
  {author} {\bibfnamefont {A.}~\bibnamefont {Sharma}},\ }\bibfield  {title}
  {\bibinfo {title} {Active transport of cargo-carrying and interconnected
  chiral particles},\ }\href
  {https://doi.org/https://doi.org/10.1039/D5SM00170F} {\bibfield  {journal}
  {\bibinfo  {journal} {Soft Matter}\ }\textbf {\bibinfo {volume} {21}},\
  \bibinfo {pages} {3384} (\bibinfo {year} {2025})}\BibitemShut {NoStop}%
\bibitem [{\citenamefont {Dinelli}\ and\ \citenamefont
  {Muzzeddu}(2026)}]{dinelli2026PRL}%
  \BibitemOpen
  \bibfield  {author} {\bibinfo {author} {\bibfnamefont {A.}~\bibnamefont
  {Dinelli}}\ and\ \bibinfo {author} {\bibfnamefont {P.~L.}\ \bibnamefont
  {Muzzeddu}},\ }\bibfield  {title} {\bibinfo {title} {Unifying hydrodynamic
  theory for motility-regulated active matter: form single particles to
  interacting polymers},\ }\href@noop {} {\bibfield  {journal} {\bibinfo
  {journal} {Companion paper}\ } (\bibinfo {year} {2026})}\BibitemShut
  {NoStop}%
\bibitem [{\citenamefont {Pavliotis}\ and\ \citenamefont
  {Stuart}(2008)}]{pavliotis2008multiscale}%
  \BibitemOpen
  \bibfield  {author} {\bibinfo {author} {\bibfnamefont {G.}~\bibnamefont
  {Pavliotis}}\ and\ \bibinfo {author} {\bibfnamefont {A.}~\bibnamefont
  {Stuart}},\ }\href@noop {} {\emph {\bibinfo {title} {Multiscale methods:
  averaging and homogenization}}}\ (\bibinfo  {publisher} {Springer Science \&
  Business Media},\ \bibinfo {year} {2008})\BibitemShut {NoStop}%
\bibitem [{\citenamefont {Duan}\ \emph {et~al.}(2025)\citenamefont {Duan},
  \citenamefont {Agudo-Canalejo}, \citenamefont {Golestanian},\ and\
  \citenamefont {Mahault}}]{duan2025phase}%
  \BibitemOpen
  \bibfield  {author} {\bibinfo {author} {\bibfnamefont {Y.}~\bibnamefont
  {Duan}}, \bibinfo {author} {\bibfnamefont {J.}~\bibnamefont
  {Agudo-Canalejo}}, \bibinfo {author} {\bibfnamefont {R.}~\bibnamefont
  {Golestanian}},\ and\ \bibinfo {author} {\bibfnamefont {B.}~\bibnamefont
  {Mahault}},\ }\bibfield  {title} {\bibinfo {title} {Phase coexistence in
  nonreciprocal quorum-sensing active matter},\ }\href
  {https://doi.org/https://doi.org/10.1103/PhysRevResearch.7.013234} {\bibfield
   {journal} {\bibinfo  {journal} {Phys. Rev. Res.}\ }\textbf {\bibinfo
  {volume} {7}},\ \bibinfo {pages} {013234} (\bibinfo {year}
  {2025})}\BibitemShut {NoStop}%
\bibitem [{\citenamefont {O’Byrne}\ and\ \citenamefont
  {Cates}(2024)}]{o2024geometric}%
  \BibitemOpen
  \bibfield  {author} {\bibinfo {author} {\bibfnamefont {J.}~\bibnamefont
  {O’Byrne}}\ and\ \bibinfo {author} {\bibfnamefont {M.~E.}\ \bibnamefont
  {Cates}},\ }\bibfield  {title} {\bibinfo {title} {Geometric theory of
  (extended) time-reversal symmetries in stochastic processes: I. finite
  dimension},\ }\href {https://doi.org/10.1088/1742-5468/ad8f2b} {\bibfield
  {journal} {\bibinfo  {journal} {J. Stat. Mech.}\ }\textbf {\bibinfo {volume}
  {2024}},\ \bibinfo {pages} {113207} (\bibinfo {year} {2024})}\BibitemShut
  {NoStop}%
\bibitem [{\citenamefont {O’Byrne}\ and\ \citenamefont
  {Cates}(2025)}]{o2025geometric}%
  \BibitemOpen
  \bibfield  {author} {\bibinfo {author} {\bibfnamefont {J.}~\bibnamefont
  {O’Byrne}}\ and\ \bibinfo {author} {\bibfnamefont {M.}~\bibnamefont
  {Cates}},\ }\bibfield  {title} {\bibinfo {title} {Geometric theory of
  (extended) time-reversal symmetries in stochastic processes: Ii. field
  theory},\ }\href {https://doi.org/10.1088/1742-5468/add0a2} {\bibfield
  {journal} {\bibinfo  {journal} {J. Stat. Mech.}\ }\textbf {\bibinfo {volume}
  {2025}},\ \bibinfo {pages} {053204} (\bibinfo {year} {2025})}\BibitemShut
  {NoStop}%
\bibitem [{\citenamefont {Doi}\ and\ \citenamefont
  {Edwards}(1988)}]{doi1988theory}%
  \BibitemOpen
  \bibfield  {author} {\bibinfo {author} {\bibfnamefont {M.}~\bibnamefont
  {Doi}}\ and\ \bibinfo {author} {\bibfnamefont {S.~F.}\ \bibnamefont
  {Edwards}},\ }\href@noop {} {\emph {\bibinfo {title} {The theory of polymer
  dynamics}}},\ Vol.~\bibinfo {volume} {73}\ (\bibinfo  {publisher} {Oxford
  university press},\ \bibinfo {year} {1988})\BibitemShut {NoStop}%
\bibitem [{\citenamefont {Gardiner}\ \emph {et~al.}(2004)\citenamefont
  {Gardiner} \emph {et~al.}}]{gardiner2004handbook}%
  \BibitemOpen
  \bibfield  {author} {\bibinfo {author} {\bibfnamefont {C.~W.}\ \bibnamefont
  {Gardiner}} \emph {et~al.},\ }\href@noop {} {\emph {\bibinfo {title}
  {Handbook of stochastic methods}}},\ Vol.~\bibinfo {volume} {3}\ (\bibinfo
  {publisher} {springer Berlin},\ \bibinfo {year} {2004})\BibitemShut {NoStop}%
\bibitem [{\citenamefont {Green}(1954)}]{green1954markoff}%
  \BibitemOpen
  \bibfield  {author} {\bibinfo {author} {\bibfnamefont {M.~S.}\ \bibnamefont
  {Green}},\ }\bibfield  {title} {\bibinfo {title} {Markoff random processes
  and the statistical mechanics of time-dependent phenomena. ii. irreversible
  processes in fluids},\ }\href
  {https://doi.org/https://doi.org/10.1063/1.1740082} {\bibfield  {journal}
  {\bibinfo  {journal} {J. Chem. Phys.}\ }\textbf {\bibinfo {volume} {22}},\
  \bibinfo {pages} {398} (\bibinfo {year} {1954})}\BibitemShut {NoStop}%
\bibitem [{\citenamefont {Kubo}\ \emph {et~al.}(2012)\citenamefont {Kubo},
  \citenamefont {Toda},\ and\ \citenamefont
  {Hashitsume}}]{kubo2012statistical}%
  \BibitemOpen
  \bibfield  {author} {\bibinfo {author} {\bibfnamefont {R.}~\bibnamefont
  {Kubo}}, \bibinfo {author} {\bibfnamefont {M.}~\bibnamefont {Toda}},\ and\
  \bibinfo {author} {\bibfnamefont {N.}~\bibnamefont {Hashitsume}},\
  }\href@noop {} {\emph {\bibinfo {title} {Statistical physics II:
  nonequilibrium statistical mechanics}}},\ Vol.~\bibinfo {volume} {31}\
  (\bibinfo  {publisher} {Springer Science \& Business Media},\ \bibinfo {year}
  {2012})\BibitemShut {NoStop}%
\bibitem [{\citenamefont {Hargus}\ \emph {et~al.}(2021)\citenamefont {Hargus},
  \citenamefont {Epstein},\ and\ \citenamefont {Mandadapu}}]{hargus2021odd}%
  \BibitemOpen
  \bibfield  {author} {\bibinfo {author} {\bibfnamefont {C.}~\bibnamefont
  {Hargus}}, \bibinfo {author} {\bibfnamefont {J.~M.}\ \bibnamefont
  {Epstein}},\ and\ \bibinfo {author} {\bibfnamefont {K.~K.}\ \bibnamefont
  {Mandadapu}},\ }\bibfield  {title} {\bibinfo {title} {Odd diffusivity of
  chiral random motion},\ }\href
  {https://doi.org/https://doi.org/10.1103/PhysRevLett.127.178001} {\bibfield
  {journal} {\bibinfo  {journal} {Phys. Rev. Lett.}\ }\textbf {\bibinfo
  {volume} {127}},\ \bibinfo {pages} {178001} (\bibinfo {year}
  {2021})}\BibitemShut {NoStop}%
\bibitem [{\citenamefont {Scholz}\ \emph {et~al.}(2018)\citenamefont {Scholz},
  \citenamefont {Jahanshahi}, \citenamefont {Ldov},\ and\ \citenamefont
  {L{\"o}wen}}]{scholz2018inertial}%
  \BibitemOpen
  \bibfield  {author} {\bibinfo {author} {\bibfnamefont {C.}~\bibnamefont
  {Scholz}}, \bibinfo {author} {\bibfnamefont {S.}~\bibnamefont {Jahanshahi}},
  \bibinfo {author} {\bibfnamefont {A.}~\bibnamefont {Ldov}},\ and\ \bibinfo
  {author} {\bibfnamefont {H.}~\bibnamefont {L{\"o}wen}},\ }\bibfield  {title}
  {\bibinfo {title} {Inertial delay of self-propelled particles},\ }\href
  {https://doi.org/https://doi.org/10.1038/s41467-018-07596-x} {\bibfield
  {journal} {\bibinfo  {journal} {Nat. Commun.}\ }\textbf {\bibinfo {volume}
  {9}},\ \bibinfo {pages} {5156} (\bibinfo {year} {2018})}\BibitemShut
  {NoStop}%
\bibitem [{\citenamefont {L{\"o}wen}(2020)}]{lowen2020inertial}%
  \BibitemOpen
  \bibfield  {author} {\bibinfo {author} {\bibfnamefont {H.}~\bibnamefont
  {L{\"o}wen}},\ }\bibfield  {title} {\bibinfo {title} {Inertial effects of
  self-propelled particles: From active brownian to active langevin motion},\
  }\href {https://doi.org/https://doi.org/10.1063/1.5134455} {\bibfield
  {journal} {\bibinfo  {journal} {J. Chem. Phys.}\ }\textbf {\bibinfo {volume}
  {152}},\ \bibinfo {pages} {040901} (\bibinfo {year} {2020})}\BibitemShut
  {NoStop}%
\bibitem [{\citenamefont {Caprini}\ \emph {et~al.}(2022)\citenamefont
  {Caprini}, \citenamefont {Gupta},\ and\ \citenamefont
  {L{\"o}wen}}]{caprini2022role}%
  \BibitemOpen
  \bibfield  {author} {\bibinfo {author} {\bibfnamefont {L.}~\bibnamefont
  {Caprini}}, \bibinfo {author} {\bibfnamefont {R.~K.}\ \bibnamefont {Gupta}},\
  and\ \bibinfo {author} {\bibfnamefont {H.}~\bibnamefont {L{\"o}wen}},\
  }\bibfield  {title} {\bibinfo {title} {Role of rotational inertia for
  collective phenomena in active matter},\ }\href
  {https://doi.org/https://doi.org/10.1039/D2CP02940E} {\bibfield  {journal}
  {\bibinfo  {journal} {Phys. Chem. Chem. Phys.}\ }\textbf {\bibinfo {volume}
  {24}},\ \bibinfo {pages} {24910} (\bibinfo {year} {2022})}\BibitemShut
  {NoStop}%
\bibitem [{\citenamefont {Lisin}\ \emph {et~al.}(2022)\citenamefont {Lisin},
  \citenamefont {Vaulina}, \citenamefont {Lisina},\ and\ \citenamefont
  {Petrov}}]{lisin2022motion}%
  \BibitemOpen
  \bibfield  {author} {\bibinfo {author} {\bibfnamefont {E.}~\bibnamefont
  {Lisin}}, \bibinfo {author} {\bibfnamefont {O.}~\bibnamefont {Vaulina}},
  \bibinfo {author} {\bibfnamefont {I.}~\bibnamefont {Lisina}},\ and\ \bibinfo
  {author} {\bibfnamefont {O.}~\bibnamefont {Petrov}},\ }\bibfield  {title}
  {\bibinfo {title} {Motion of a self-propelled particle with rotational
  inertia},\ }\href {https://doi.org/https://doi.org/10.1039/D2CP01313D}
  {\bibfield  {journal} {\bibinfo  {journal} {Phys. Chem. Chem. Phys.}\
  }\textbf {\bibinfo {volume} {24}},\ \bibinfo {pages} {14150} (\bibinfo {year}
  {2022})}\BibitemShut {NoStop}%
\bibitem [{\citenamefont {Sprenger}\ \emph {et~al.}(2023)\citenamefont
  {Sprenger}, \citenamefont {Caprini}, \citenamefont {L{\"o}wen},\ and\
  \citenamefont {Wittmann}}]{sprenger2023dynamics}%
  \BibitemOpen
  \bibfield  {author} {\bibinfo {author} {\bibfnamefont {A.~R.}\ \bibnamefont
  {Sprenger}}, \bibinfo {author} {\bibfnamefont {L.}~\bibnamefont {Caprini}},
  \bibinfo {author} {\bibfnamefont {H.}~\bibnamefont {L{\"o}wen}},\ and\
  \bibinfo {author} {\bibfnamefont {R.}~\bibnamefont {Wittmann}},\ }\bibfield
  {title} {\bibinfo {title} {Dynamics of active particles with translational
  and rotational inertia},\ }\href {https://doi.org/10.1088/1361-648X/accd36}
  {\bibfield  {journal} {\bibinfo  {journal} {J. Phys. Condens. Matter}\
  }\textbf {\bibinfo {volume} {35}},\ \bibinfo {pages} {305101} (\bibinfo
  {year} {2023})}\BibitemShut {NoStop}%
\bibitem [{\citenamefont {Detcheverry}(2017)}]{detcheverry2017generalized}%
  \BibitemOpen
  \bibfield  {author} {\bibinfo {author} {\bibfnamefont {F.}~\bibnamefont
  {Detcheverry}},\ }\bibfield  {title} {\bibinfo {title} {Generalized
  run-and-turn motions: From bacteria to l{\'e}vy walks},\ }\href
  {https://doi.org/https://doi.org/10.1103/PhysRevE.96.012415} {\bibfield
  {journal} {\bibinfo  {journal} {Phys. Rev. E}\ }\textbf {\bibinfo {volume}
  {96}},\ \bibinfo {pages} {012415} (\bibinfo {year} {2017})}\BibitemShut
  {NoStop}%
\bibitem [{\citenamefont {Shaebani}\ \emph {et~al.}(2022)\citenamefont
  {Shaebani}, \citenamefont {Rieger},\ and\ \citenamefont
  {Sadjadi}}]{shaebani2022kinematics}%
  \BibitemOpen
  \bibfield  {author} {\bibinfo {author} {\bibfnamefont {M.~R.}\ \bibnamefont
  {Shaebani}}, \bibinfo {author} {\bibfnamefont {H.}~\bibnamefont {Rieger}},\
  and\ \bibinfo {author} {\bibfnamefont {Z.}~\bibnamefont {Sadjadi}},\
  }\bibfield  {title} {\bibinfo {title} {Kinematics of persistent random
  walkers with two distinct modes of motion},\ }\href
  {https://doi.org/https://doi.org/10.1103/PhysRevE.106.034105} {\bibfield
  {journal} {\bibinfo  {journal} {Phys. Rev. E}\ }\textbf {\bibinfo {volume}
  {106}},\ \bibinfo {pages} {034105} (\bibinfo {year} {2022})}\BibitemShut
  {NoStop}%
\bibitem [{\citenamefont {Datta}\ \emph {et~al.}(2024)\citenamefont {Datta},
  \citenamefont {Beta},\ and\ \citenamefont {Grossmann}}]{datta2024random}%
  \BibitemOpen
  \bibfield  {author} {\bibinfo {author} {\bibfnamefont {A.}~\bibnamefont
  {Datta}}, \bibinfo {author} {\bibfnamefont {C.}~\bibnamefont {Beta}},\ and\
  \bibinfo {author} {\bibfnamefont {R.}~\bibnamefont {Grossmann}},\ }\bibfield
  {title} {\bibinfo {title} {Random walks of intermittently self-propelled
  particles},\ }\href
  {https://doi.org/https://doi.org/10.1103/PhysRevResearch.6.043281} {\bibfield
   {journal} {\bibinfo  {journal} {Phys. Rev. Res.}\ }\textbf {\bibinfo
  {volume} {6}},\ \bibinfo {pages} {043281} (\bibinfo {year}
  {2024})}\BibitemShut {NoStop}%
\bibitem [{\citenamefont {Caraglio}\ \emph {et~al.}(2024)\citenamefont
  {Caraglio}, \citenamefont {Kaur}, \citenamefont {Fiderer}, \citenamefont
  {L{\'o}pez-Incera}, \citenamefont {Briegel}, \citenamefont {Franosch},\ and\
  \citenamefont {Mu{\~n}oz-Gil}}]{caraglio2024learning}%
  \BibitemOpen
  \bibfield  {author} {\bibinfo {author} {\bibfnamefont {M.}~\bibnamefont
  {Caraglio}}, \bibinfo {author} {\bibfnamefont {H.}~\bibnamefont {Kaur}},
  \bibinfo {author} {\bibfnamefont {L.~J.}\ \bibnamefont {Fiderer}}, \bibinfo
  {author} {\bibfnamefont {A.}~\bibnamefont {L{\'o}pez-Incera}}, \bibinfo
  {author} {\bibfnamefont {H.~J.}\ \bibnamefont {Briegel}}, \bibinfo {author}
  {\bibfnamefont {T.}~\bibnamefont {Franosch}},\ and\ \bibinfo {author}
  {\bibfnamefont {G.}~\bibnamefont {Mu{\~n}oz-Gil}},\ }\bibfield  {title}
  {\bibinfo {title} {Learning how to find targets in the micro-world: the case
  of intermittent active brownian particles},\ }\href
  {https://doi.org/10.1039/D3SM01680C} {\bibfield  {journal} {\bibinfo
  {journal} {Soft Matter}\ }\textbf {\bibinfo {volume} {20}},\ \bibinfo {pages}
  {2008} (\bibinfo {year} {2024})}\BibitemShut {NoStop}%
\bibitem [{\citenamefont {Santra}\ \emph {et~al.}(2024)\citenamefont {Santra},
  \citenamefont {Olsen},\ and\ \citenamefont {Gupta}}]{santra2024dynamics}%
  \BibitemOpen
  \bibfield  {author} {\bibinfo {author} {\bibfnamefont {I.}~\bibnamefont
  {Santra}}, \bibinfo {author} {\bibfnamefont {K.~S.}\ \bibnamefont {Olsen}},\
  and\ \bibinfo {author} {\bibfnamefont {D.}~\bibnamefont {Gupta}},\ }\bibfield
   {title} {\bibinfo {title} {Dynamics of switching processes: general results
  and applications in intermittent active motion},\ }\href
  {https://doi.org/https://doi.org/10.1039/D4SM01054J} {\bibfield  {journal}
  {\bibinfo  {journal} {Soft Matter}\ }\textbf {\bibinfo {volume} {20}},\
  \bibinfo {pages} {9360} (\bibinfo {year} {2024})}\BibitemShut {NoStop}%
\bibitem [{\citenamefont {Foster}\ and\ \citenamefont
  {Smyth}(1980)}]{foster1980light}%
  \BibitemOpen
  \bibfield  {author} {\bibinfo {author} {\bibfnamefont {K.~W.}\ \bibnamefont
  {Foster}}\ and\ \bibinfo {author} {\bibfnamefont {R.~D.}\ \bibnamefont
  {Smyth}},\ }\bibfield  {title} {\bibinfo {title} {Light antennas in
  phototactic algae},\ }\href
  {https://doi.org/https://doi.org/10.1128/mr.44.4.572-630.1980} {\bibfield
  {journal} {\bibinfo  {journal} {Microbiol. Rev.}\ }\textbf {\bibinfo {volume}
  {44}},\ \bibinfo {pages} {572} (\bibinfo {year} {1980})}\BibitemShut
  {NoStop}%
\bibitem [{\citenamefont {Ji}\ and\ \citenamefont
  {Huang}(2023)}]{ji2023durotaxis}%
  \BibitemOpen
  \bibfield  {author} {\bibinfo {author} {\bibfnamefont {C.}~\bibnamefont
  {Ji}}\ and\ \bibinfo {author} {\bibfnamefont {Y.}~\bibnamefont {Huang}},\
  }\bibfield  {title} {\bibinfo {title} {Durotaxis and negative durotaxis:
  where should cells go?},\ }\href
  {https://doi.org/https://doi.org/10.1038/s42003-023-05554-y} {\bibfield
  {journal} {\bibinfo  {journal} {Commun. Biol.}\ }\textbf {\bibinfo {volume}
  {6}},\ \bibinfo {pages} {1169} (\bibinfo {year} {2023})}\BibitemShut
  {NoStop}%
\bibitem [{\citenamefont {Howse}\ \emph {et~al.}(2007)\citenamefont {Howse},
  \citenamefont {Jones}, \citenamefont {Ryan}, \citenamefont {Gough},
  \citenamefont {Vafabakhsh},\ and\ \citenamefont
  {Golestanian}}]{howse2007self}%
  \BibitemOpen
  \bibfield  {author} {\bibinfo {author} {\bibfnamefont {J.~R.}\ \bibnamefont
  {Howse}}, \bibinfo {author} {\bibfnamefont {R.~A.}\ \bibnamefont {Jones}},
  \bibinfo {author} {\bibfnamefont {A.~J.}\ \bibnamefont {Ryan}}, \bibinfo
  {author} {\bibfnamefont {T.}~\bibnamefont {Gough}}, \bibinfo {author}
  {\bibfnamefont {R.}~\bibnamefont {Vafabakhsh}},\ and\ \bibinfo {author}
  {\bibfnamefont {R.}~\bibnamefont {Golestanian}},\ }\bibfield  {title}
  {\bibinfo {title} {Self-motile colloidal particles: from directed propulsion
  to random walk},\ }\href
  {https://doi.org/https://doi.org/10.1103/PhysRevLett.99.048102} {\bibfield
  {journal} {\bibinfo  {journal} {Phys. Rev. Lett.}\ }\textbf {\bibinfo
  {volume} {99}},\ \bibinfo {pages} {048102} (\bibinfo {year}
  {2007})}\BibitemShut {NoStop}%
\bibitem [{\citenamefont {Soto}\ and\ \citenamefont
  {Golestanian}(2014)}]{soto2014self}%
  \BibitemOpen
  \bibfield  {author} {\bibinfo {author} {\bibfnamefont {R.}~\bibnamefont
  {Soto}}\ and\ \bibinfo {author} {\bibfnamefont {R.}~\bibnamefont
  {Golestanian}},\ }\bibfield  {title} {\bibinfo {title} {Self-assembly of
  catalytically active colloidal molecules: tailoring activity through surface
  chemistry},\ }\href
  {https://doi.org/https://doi.org/10.1103/PhysRevLett.112.068301} {\bibfield
  {journal} {\bibinfo  {journal} {Phys. Rev. Lett.}\ }\textbf {\bibinfo
  {volume} {112}},\ \bibinfo {pages} {068301} (\bibinfo {year}
  {2014})}\BibitemShut {NoStop}%
\bibitem [{\citenamefont {Saha}\ \emph {et~al.}(2014)\citenamefont {Saha},
  \citenamefont {Golestanian},\ and\ \citenamefont
  {Ramaswamy}}]{saha2014clusters}%
  \BibitemOpen
  \bibfield  {author} {\bibinfo {author} {\bibfnamefont {S.}~\bibnamefont
  {Saha}}, \bibinfo {author} {\bibfnamefont {R.}~\bibnamefont {Golestanian}},\
  and\ \bibinfo {author} {\bibfnamefont {S.}~\bibnamefont {Ramaswamy}},\
  }\bibfield  {title} {\bibinfo {title} {Clusters, asters, and collective
  oscillations in chemotactic colloids},\ }\href
  {https://doi.org/https://doi.org/10.1103/PhysRevE.89.062316} {\bibfield
  {journal} {\bibinfo  {journal} {Phys. Rev. E}\ }\textbf {\bibinfo {volume}
  {89}},\ \bibinfo {pages} {062316} (\bibinfo {year} {2014})}\BibitemShut
  {NoStop}%
\bibitem [{\citenamefont {Wang}\ and\ \citenamefont
  {Gao}(2012)}]{wang2012nano}%
  \BibitemOpen
  \bibfield  {author} {\bibinfo {author} {\bibfnamefont {J.}~\bibnamefont
  {Wang}}\ and\ \bibinfo {author} {\bibfnamefont {W.}~\bibnamefont {Gao}},\
  }\bibfield  {title} {\bibinfo {title} {Nano/microscale motors: biomedical
  opportunities and challenges},\ }\href
  {https://doi.org/https://doi.org/10.1021/nn3028997} {\bibfield  {journal}
  {\bibinfo  {journal} {ACS nano}\ }\textbf {\bibinfo {volume} {6}},\ \bibinfo
  {pages} {5745} (\bibinfo {year} {2012})}\BibitemShut {NoStop}%
\bibitem [{\citenamefont {Jain}\ and\ \citenamefont
  {Thakur}(2022)}]{jain2022cargo}%
  \BibitemOpen
  \bibfield  {author} {\bibinfo {author} {\bibfnamefont {N.}~\bibnamefont
  {Jain}}\ and\ \bibinfo {author} {\bibfnamefont {S.}~\bibnamefont {Thakur}},\
  }\bibfield  {title} {\bibinfo {title} {Cargo transportation using an active
  polymer},\ }\href {https://doi.org/https://doi.org/10.1063/5.0119830}
  {\bibfield  {journal} {\bibinfo  {journal} {AIP Advances}\ }\textbf {\bibinfo
  {volume} {12}},\ \bibinfo {pages} {115211} (\bibinfo {year}
  {2022})}\BibitemShut {NoStop}%
\bibitem [{\citenamefont {O’Byrne}\ \emph {et~al.}(2022)\citenamefont
  {O’Byrne}, \citenamefont {Kafri}, \citenamefont {Tailleur},\ and\
  \citenamefont {van Wijland}}]{o2022time}%
  \BibitemOpen
  \bibfield  {author} {\bibinfo {author} {\bibfnamefont {J.}~\bibnamefont
  {O’Byrne}}, \bibinfo {author} {\bibfnamefont {Y.}~\bibnamefont {Kafri}},
  \bibinfo {author} {\bibfnamefont {J.}~\bibnamefont {Tailleur}},\ and\
  \bibinfo {author} {\bibfnamefont {F.}~\bibnamefont {van Wijland}},\
  }\bibfield  {title} {\bibinfo {title} {Time irreversibility in active matter,
  from micro to macro},\ }\href
  {https://doi.org/https://doi.org/10.1038/s42254-021-00406-2} {\bibfield
  {journal} {\bibinfo  {journal} {Nat. Rev. Phys.}\ }\textbf {\bibinfo {volume}
  {4}},\ \bibinfo {pages} {167} (\bibinfo {year} {2022})}\BibitemShut {NoStop}%
\bibitem [{\citenamefont {Arlt}\ \emph {et~al.}(2018)\citenamefont {Arlt},
  \citenamefont {Martinez}, \citenamefont {Dawson}, \citenamefont {Pilizota},\
  and\ \citenamefont {Poon}}]{arlt2018painting}%
  \BibitemOpen
  \bibfield  {author} {\bibinfo {author} {\bibfnamefont {J.}~\bibnamefont
  {Arlt}}, \bibinfo {author} {\bibfnamefont {V.~A.}\ \bibnamefont {Martinez}},
  \bibinfo {author} {\bibfnamefont {A.}~\bibnamefont {Dawson}}, \bibinfo
  {author} {\bibfnamefont {T.}~\bibnamefont {Pilizota}},\ and\ \bibinfo
  {author} {\bibfnamefont {W.~C.}\ \bibnamefont {Poon}},\ }\bibfield  {title}
  {\bibinfo {title} {Painting with light-powered bacteria},\ }\href
  {https://doi.org/https://doi.org/10.1038/s41467-018-03161-8} {\bibfield
  {journal} {\bibinfo  {journal} {Nat. Commun.}\ }\textbf {\bibinfo {volume}
  {9}},\ \bibinfo {pages} {768} (\bibinfo {year} {2018})}\BibitemShut {NoStop}%
\bibitem [{\citenamefont {Frangipane}\ \emph {et~al.}(2018)\citenamefont
  {Frangipane}, \citenamefont {Dell'Arciprete}, \citenamefont {Petracchini},
  \citenamefont {Maggi}, \citenamefont {Saglimbeni}, \citenamefont {Bianchi},
  \citenamefont {Vizsnyiczai}, \citenamefont {Bernardini},\ and\ \citenamefont
  {Di~Leonardo}}]{frangipane2018dynamic}%
  \BibitemOpen
  \bibfield  {author} {\bibinfo {author} {\bibfnamefont {G.}~\bibnamefont
  {Frangipane}}, \bibinfo {author} {\bibfnamefont {D.}~\bibnamefont
  {Dell'Arciprete}}, \bibinfo {author} {\bibfnamefont {S.}~\bibnamefont
  {Petracchini}}, \bibinfo {author} {\bibfnamefont {C.}~\bibnamefont {Maggi}},
  \bibinfo {author} {\bibfnamefont {F.}~\bibnamefont {Saglimbeni}}, \bibinfo
  {author} {\bibfnamefont {S.}~\bibnamefont {Bianchi}}, \bibinfo {author}
  {\bibfnamefont {G.}~\bibnamefont {Vizsnyiczai}}, \bibinfo {author}
  {\bibfnamefont {M.~L.}\ \bibnamefont {Bernardini}},\ and\ \bibinfo {author}
  {\bibfnamefont {R.}~\bibnamefont {Di~Leonardo}},\ }\bibfield  {title}
  {\bibinfo {title} {Dynamic density shaping of photokinetic e. coli},\ }\href
  {https://doi.org/https://doi.org/10.7554/eLife.36608} {\bibfield  {journal}
  {\bibinfo  {journal} {Elife}\ }\textbf {\bibinfo {volume} {7}},\ \bibinfo
  {pages} {e36608} (\bibinfo {year} {2018})}\BibitemShut {NoStop}%
\bibitem [{\citenamefont {Metzger}\ \emph {et~al.}(2024)\citenamefont
  {Metzger}, \citenamefont {Ro},\ and\ \citenamefont
  {Tailleur}}]{metzger2024revisiting}%
  \BibitemOpen
  \bibfield  {author} {\bibinfo {author} {\bibfnamefont {J.}~\bibnamefont
  {Metzger}}, \bibinfo {author} {\bibfnamefont {S.}~\bibnamefont {Ro}},\ and\
  \bibinfo {author} {\bibfnamefont {J.}~\bibnamefont {Tailleur}},\ }\bibfield
  {title} {\bibinfo {title} {Revisiting the ratchet principle: When hidden
  symmetries prevent steady currents},\ }\href@noop {} {\bibfield  {journal}
  {\bibinfo  {journal} {arXiv preprint arXiv:2412.07851}\ } (\bibinfo {year}
  {2024})}\BibitemShut {NoStop}%
\bibitem [{\citenamefont {Metzger}\ \emph {et~al.}(2025)\citenamefont
  {Metzger}, \citenamefont {Ro},\ and\ \citenamefont
  {Tailleur}}]{metzger2025exceptions}%
  \BibitemOpen
  \bibfield  {author} {\bibinfo {author} {\bibfnamefont {J.}~\bibnamefont
  {Metzger}}, \bibinfo {author} {\bibfnamefont {S.}~\bibnamefont {Ro}},\ and\
  \bibinfo {author} {\bibfnamefont {J.}~\bibnamefont {Tailleur}},\ }\bibfield
  {title} {\bibinfo {title} {Exceptions to the ratchet principle in active and
  passive stochastic dynamics},\ }\href@noop {} {\bibfield  {journal} {\bibinfo
   {journal} {arXiv preprint arXiv:2503.11902}\ } (\bibinfo {year}
  {2025})}\BibitemShut {NoStop}%
\bibitem [{\citenamefont {Muzzeddu}\ \emph {et~al.}(2024)\citenamefont
  {Muzzeddu}, \citenamefont {Gambassi}, \citenamefont {Sommer},\ and\
  \citenamefont {Sharma}}]{muzzeddu2024migration}%
  \BibitemOpen
  \bibfield  {author} {\bibinfo {author} {\bibfnamefont {P.~L.}\ \bibnamefont
  {Muzzeddu}}, \bibinfo {author} {\bibfnamefont {A.}~\bibnamefont {Gambassi}},
  \bibinfo {author} {\bibfnamefont {J.-U.}\ \bibnamefont {Sommer}},\ and\
  \bibinfo {author} {\bibfnamefont {A.}~\bibnamefont {Sharma}},\ }\bibfield
  {title} {\bibinfo {title} {Migration and separation of polymers in nonuniform
  active baths},\ }\href
  {https://doi.org/https://doi.org/10.1103/PhysRevLett.133.118102} {\bibfield
  {journal} {\bibinfo  {journal} {Phys. Rev. Lett.}\ }\textbf {\bibinfo
  {volume} {133}},\ \bibinfo {pages} {118102} (\bibinfo {year}
  {2024})}\BibitemShut {NoStop}%
\bibitem [{\citenamefont {Liebchen}\ and\ \citenamefont
  {Levis}(2022)}]{liebchen2022chiral}%
  \BibitemOpen
  \bibfield  {author} {\bibinfo {author} {\bibfnamefont {B.}~\bibnamefont
  {Liebchen}}\ and\ \bibinfo {author} {\bibfnamefont {D.}~\bibnamefont
  {Levis}},\ }\bibfield  {title} {\bibinfo {title} {Chiral active matter},\
  }\href {https://doi.org/10.1209/0295-5075/ac8f69} {\bibfield  {journal}
  {\bibinfo  {journal} {EPL}\ }\textbf {\bibinfo {volume} {139}},\ \bibinfo
  {pages} {67001} (\bibinfo {year} {2022})}\BibitemShut {NoStop}%
\bibitem [{\citenamefont {Pattanayak}\ \emph {et~al.}(2024)\citenamefont
  {Pattanayak}, \citenamefont {Shee}, \citenamefont {Chaudhuri},\ and\
  \citenamefont {Chaudhuri}}]{pattanayak2024impact}%
  \BibitemOpen
  \bibfield  {author} {\bibinfo {author} {\bibfnamefont {A.}~\bibnamefont
  {Pattanayak}}, \bibinfo {author} {\bibfnamefont {A.}~\bibnamefont {Shee}},
  \bibinfo {author} {\bibfnamefont {D.}~\bibnamefont {Chaudhuri}},\ and\
  \bibinfo {author} {\bibfnamefont {A.}~\bibnamefont {Chaudhuri}},\ }\bibfield
  {title} {\bibinfo {title} {Impact of torque on active brownian particle:
  exact moments in two and three dimensions},\ }\href
  {https://doi.org/10.1088/1367-2630/ad6a32} {\bibfield  {journal} {\bibinfo
  {journal} {New J. Phys.}\ }\textbf {\bibinfo {volume} {26}},\ \bibinfo
  {pages} {083024} (\bibinfo {year} {2024})}\BibitemShut {NoStop}%
\bibitem [{\citenamefont {Kalz}\ \emph {et~al.}(2024)\citenamefont {Kalz},
  \citenamefont {Sharma},\ and\ \citenamefont {Metzler}}]{kalz2024field}%
  \BibitemOpen
  \bibfield  {author} {\bibinfo {author} {\bibfnamefont {E.}~\bibnamefont
  {Kalz}}, \bibinfo {author} {\bibfnamefont {A.}~\bibnamefont {Sharma}},\ and\
  \bibinfo {author} {\bibfnamefont {R.}~\bibnamefont {Metzler}},\ }\bibfield
  {title} {\bibinfo {title} {Field theory of active chiral hard disks: a
  first-principles approach to steric interactions},\ }\href
  {https://doi.org/10.1088/1751-8121/ad5089} {\bibfield  {journal} {\bibinfo
  {journal} {J. Phys. A: Math. Theor.}\ }\textbf {\bibinfo {volume} {57}},\
  \bibinfo {pages} {265002} (\bibinfo {year} {2024})}\BibitemShut {NoStop}%
\bibitem [{\citenamefont {Massana-Cid}\ \emph {et~al.}(2021)\citenamefont
  {Massana-Cid}, \citenamefont {Levis}, \citenamefont {Hern{\'a}ndez},
  \citenamefont {Pagonabarraga},\ and\ \citenamefont
  {Tierno}}]{massana2021arrested}%
  \BibitemOpen
  \bibfield  {author} {\bibinfo {author} {\bibfnamefont {H.}~\bibnamefont
  {Massana-Cid}}, \bibinfo {author} {\bibfnamefont {D.}~\bibnamefont {Levis}},
  \bibinfo {author} {\bibfnamefont {R.~J.~H.}\ \bibnamefont {Hern{\'a}ndez}},
  \bibinfo {author} {\bibfnamefont {I.}~\bibnamefont {Pagonabarraga}},\ and\
  \bibinfo {author} {\bibfnamefont {P.}~\bibnamefont {Tierno}},\ }\bibfield
  {title} {\bibinfo {title} {Arrested phase separation in chiral fluids of
  colloidal spinners},\ }\href
  {https://doi.org/https://doi.org/10.1103/PhysRevResearch.3.L042021}
  {\bibfield  {journal} {\bibinfo  {journal} {Phys. Rev. Res.}\ }\textbf
  {\bibinfo {volume} {3}},\ \bibinfo {pages} {L042021} (\bibinfo {year}
  {2021})}\BibitemShut {NoStop}%
\bibitem [{\citenamefont {Caprini}\ \emph {et~al.}(2024)\citenamefont
  {Caprini}, \citenamefont {Liebchen},\ and\ \citenamefont
  {L{\"o}wen}}]{caprini2024self}%
  \BibitemOpen
  \bibfield  {author} {\bibinfo {author} {\bibfnamefont {L.}~\bibnamefont
  {Caprini}}, \bibinfo {author} {\bibfnamefont {B.}~\bibnamefont {Liebchen}},\
  and\ \bibinfo {author} {\bibfnamefont {H.}~\bibnamefont {L{\"o}wen}},\
  }\bibfield  {title} {\bibinfo {title} {Self-reverting vortices in chiral
  active matter},\ }\href
  {https://doi.org/https://doi.org/10.1038/s42005-024-01637-2} {\bibfield
  {journal} {\bibinfo  {journal} {Commun. Phys.}\ }\textbf {\bibinfo {volume}
  {7}},\ \bibinfo {pages} {153} (\bibinfo {year} {2024})}\BibitemShut {NoStop}%
\bibitem [{\citenamefont {Liebchen}\ and\ \citenamefont
  {Levis}(2017)}]{liebchen2017collective}%
  \BibitemOpen
  \bibfield  {author} {\bibinfo {author} {\bibfnamefont {B.}~\bibnamefont
  {Liebchen}}\ and\ \bibinfo {author} {\bibfnamefont {D.}~\bibnamefont
  {Levis}},\ }\bibfield  {title} {\bibinfo {title} {Collective behavior of
  chiral active matter: Pattern formation and enhanced flocking},\ }\href
  {https://doi.org/https://doi.org/10.1103/PhysRevLett.119.058002} {\bibfield
  {journal} {\bibinfo  {journal} {Phys. Rev. Lett.}\ }\textbf {\bibinfo
  {volume} {119}},\ \bibinfo {pages} {058002} (\bibinfo {year}
  {2017})}\BibitemShut {NoStop}%
\bibitem [{\citenamefont {Lauga}\ \emph {et~al.}(2006)\citenamefont {Lauga},
  \citenamefont {DiLuzio}, \citenamefont {Whitesides},\ and\ \citenamefont
  {Stone}}]{lauga2006swimming}%
  \BibitemOpen
  \bibfield  {author} {\bibinfo {author} {\bibfnamefont {E.}~\bibnamefont
  {Lauga}}, \bibinfo {author} {\bibfnamefont {W.~R.}\ \bibnamefont {DiLuzio}},
  \bibinfo {author} {\bibfnamefont {G.~M.}\ \bibnamefont {Whitesides}},\ and\
  \bibinfo {author} {\bibfnamefont {H.~A.}\ \bibnamefont {Stone}},\ }\bibfield
  {title} {\bibinfo {title} {Swimming in circles: motion of bacteria near solid
  boundaries},\ }\href
  {https://doi.org/https://doi.org/10.1529/biophysj.105.069401} {\bibfield
  {journal} {\bibinfo  {journal} {Biophys. J}\ }\textbf {\bibinfo {volume}
  {90}},\ \bibinfo {pages} {400} (\bibinfo {year} {2006})}\BibitemShut
  {NoStop}%
\bibitem [{\citenamefont {Friedrich}\ and\ \citenamefont
  {J{\"u}licher}(2007)}]{friedrich2007chemotaxis}%
  \BibitemOpen
  \bibfield  {author} {\bibinfo {author} {\bibfnamefont {B.~M.}\ \bibnamefont
  {Friedrich}}\ and\ \bibinfo {author} {\bibfnamefont {F.}~\bibnamefont
  {J{\"u}licher}},\ }\bibfield  {title} {\bibinfo {title} {Chemotaxis of sperm
  cells},\ }\href {https://doi.org/https://doi.org/10.1073/pnas.0703530104}
  {\bibfield  {journal} {\bibinfo  {journal} {Proc. Natl. Acad. Sci. U.S.A.}\
  }\textbf {\bibinfo {volume} {104}},\ \bibinfo {pages} {13256} (\bibinfo
  {year} {2007})}\BibitemShut {NoStop}%
\bibitem [{\citenamefont {Purcell}(2014)}]{purcell2014life}%
  \BibitemOpen
  \bibfield  {author} {\bibinfo {author} {\bibfnamefont {E.~M.}\ \bibnamefont
  {Purcell}},\ }\bibfield  {title} {\bibinfo {title} {Life at low reynolds
  number},\ }in\ \href@noop {} {\emph {\bibinfo {booktitle} {Physics and our
  world: reissue of the proceedings of a symposium in honor of Victor F
  Weisskopf}}}\ (\bibinfo {organization} {World Scientific},\ \bibinfo {year}
  {2014})\ pp.\ \bibinfo {pages} {47--67}\BibitemShut {NoStop}%
\bibitem [{\citenamefont {Lefranc}\ \emph {et~al.}(2025)\citenamefont
  {Lefranc}, \citenamefont {Dinelli}, \citenamefont {Fern{\'a}ndez-Rico},
  \citenamefont {Dullens}, \citenamefont {Tailleur},\ and\ \citenamefont
  {Bartolo}}]{lefranc2025synthetic}%
  \BibitemOpen
  \bibfield  {author} {\bibinfo {author} {\bibfnamefont {T.}~\bibnamefont
  {Lefranc}}, \bibinfo {author} {\bibfnamefont {A.}~\bibnamefont {Dinelli}},
  \bibinfo {author} {\bibfnamefont {C.}~\bibnamefont {Fern{\'a}ndez-Rico}},
  \bibinfo {author} {\bibfnamefont {R.~P.}\ \bibnamefont {Dullens}}, \bibinfo
  {author} {\bibfnamefont {J.}~\bibnamefont {Tailleur}},\ and\ \bibinfo
  {author} {\bibfnamefont {D.}~\bibnamefont {Bartolo}},\ }\bibfield  {title}
  {\bibinfo {title} {Synthetic quorum sensing and absorbing phase transitions
  in colloidal active matter},\ }\href
  {https://doi.org/https://doi.org/10.1103/8csn-71jk} {\bibfield  {journal}
  {\bibinfo  {journal} {Phys. Rev. X}\ }\textbf {\bibinfo {volume} {15}},\
  \bibinfo {pages} {031050} (\bibinfo {year} {2025})}\BibitemShut {NoStop}%
\bibitem [{\citenamefont {Van~Kampen}(1992)}]{vankampen1992stochastic}%
  \BibitemOpen
  \bibfield  {author} {\bibinfo {author} {\bibfnamefont {N.~G.}\ \bibnamefont
  {Van~Kampen}},\ }\href@noop {} {\emph {\bibinfo {title} {Stochastic processes
  in physics and chemistry}}},\ Vol.~\bibinfo {volume} {1}\ (\bibinfo
  {publisher} {Elsevier},\ \bibinfo {year} {1992})\BibitemShut {NoStop}%
\bibitem [{\citenamefont {Mandelbrot}\ and\ \citenamefont
  {Van~Ness}(1968)}]{mandelbrot1968fractional}%
  \BibitemOpen
  \bibfield  {author} {\bibinfo {author} {\bibfnamefont {B.~B.}\ \bibnamefont
  {Mandelbrot}}\ and\ \bibinfo {author} {\bibfnamefont {J.~W.}\ \bibnamefont
  {Van~Ness}},\ }\bibfield  {title} {\bibinfo {title} {Fractional brownian
  motions, fractional noises and applications},\ }\href
  {https://doi.org/https://doi.org/10.1137/1010093} {\bibfield  {journal}
  {\bibinfo  {journal} {SIAM review}\ }\textbf {\bibinfo {volume} {10}},\
  \bibinfo {pages} {422} (\bibinfo {year} {1968})}\BibitemShut {NoStop}%
\bibitem [{\citenamefont {Gomez-Solano}\ and\ \citenamefont
  {Sevilla}(2020)}]{gomez2020active}%
  \BibitemOpen
  \bibfield  {author} {\bibinfo {author} {\bibfnamefont {J.~R.}\ \bibnamefont
  {Gomez-Solano}}\ and\ \bibinfo {author} {\bibfnamefont {F.~J.}\ \bibnamefont
  {Sevilla}},\ }\bibfield  {title} {\bibinfo {title} {Active particles with
  fractional rotational brownian motion},\ }\href
  {https://doi.org/10.1088/1742-5468/ab8553} {\bibfield  {journal} {\bibinfo
  {journal} {J. Stat. Mech.: Theory Exp.}\ }\textbf {\bibinfo {volume}
  {2020}}\bibinfo  {number} { (6)},\ \bibinfo {pages} {063213}}\BibitemShut
  {NoStop}%
\bibitem [{\citenamefont {Ravichandir}\ \emph {et~al.}(2025)\citenamefont
  {Ravichandir}, \citenamefont {Valecha}, \citenamefont {Muzzeddu},
  \citenamefont {Sommer},\ and\ \citenamefont
  {Sharma}}]{ravichandir2025transport}%
  \BibitemOpen
\bibfield  {number} {  }\bibfield  {author} {\bibinfo {author} {\bibfnamefont
  {S.}~\bibnamefont {Ravichandir}}, \bibinfo {author} {\bibfnamefont
  {B.}~\bibnamefont {Valecha}}, \bibinfo {author} {\bibfnamefont {P.~L.}\
  \bibnamefont {Muzzeddu}}, \bibinfo {author} {\bibfnamefont {J.-U.}\
  \bibnamefont {Sommer}},\ and\ \bibinfo {author} {\bibfnamefont
  {A.}~\bibnamefont {Sharma}},\ }\bibfield  {title} {\bibinfo {title}
  {Transport of partially active polymers in chemical gradients},\ }\href
  {https://doi.org/https://doi.org/10.1039/D4SM01357C} {\bibfield  {journal}
  {\bibinfo  {journal} {Soft Matter}\ }\textbf {\bibinfo {volume} {21}},\
  \bibinfo {pages} {1835} (\bibinfo {year} {2025})}\BibitemShut {NoStop}%
\bibitem [{\citenamefont {Epstein}\ and\ \citenamefont
  {Mandadapu}(2020)}]{epstein2020time}%
  \BibitemOpen
  \bibfield  {author} {\bibinfo {author} {\bibfnamefont {J.~M.}\ \bibnamefont
  {Epstein}}\ and\ \bibinfo {author} {\bibfnamefont {K.~K.}\ \bibnamefont
  {Mandadapu}},\ }\bibfield  {title} {\bibinfo {title} {Time-reversal symmetry
  breaking in two-dimensional nonequilibrium viscous fluids},\ }\href
  {https://doi.org/https://doi.org/10.1103/PhysRevE.101.052614} {\bibfield
  {journal} {\bibinfo  {journal} {Phys. Rev. E}\ }\textbf {\bibinfo {volume}
  {101}},\ \bibinfo {pages} {052614} (\bibinfo {year} {2020})}\BibitemShut
  {NoStop}%
\bibitem [{\citenamefont {Hargus}\ \emph {et~al.}(2020)\citenamefont {Hargus},
  \citenamefont {Klymko}, \citenamefont {Epstein},\ and\ \citenamefont
  {Mandadapu}}]{hargus2020time}%
  \BibitemOpen
  \bibfield  {author} {\bibinfo {author} {\bibfnamefont {C.}~\bibnamefont
  {Hargus}}, \bibinfo {author} {\bibfnamefont {K.}~\bibnamefont {Klymko}},
  \bibinfo {author} {\bibfnamefont {J.~M.}\ \bibnamefont {Epstein}},\ and\
  \bibinfo {author} {\bibfnamefont {K.~K.}\ \bibnamefont {Mandadapu}},\
  }\bibfield  {title} {\bibinfo {title} {Time reversal symmetry breaking and
  odd viscosity in active fluids: Green--kubo and nemd results},\ }\href
  {https://doi.org/https://doi.org/10.1063/5.0006441} {\bibfield  {journal}
  {\bibinfo  {journal} {J. Chem. Phys.}\ }\textbf {\bibinfo {volume} {152}},\
  \bibinfo {pages} {201102} (\bibinfo {year} {2020})}\BibitemShut {NoStop}%
\bibitem [{\citenamefont {Pohl}\ and\ \citenamefont
  {Stark}(2014)}]{pohl2014dynamic}%
  \BibitemOpen
  \bibfield  {author} {\bibinfo {author} {\bibfnamefont {O.}~\bibnamefont
  {Pohl}}\ and\ \bibinfo {author} {\bibfnamefont {H.}~\bibnamefont {Stark}},\
  }\bibfield  {title} {\bibinfo {title} {Dynamic clustering and chemotactic
  collapse of self-phoretic active particles},\ }\href
  {https://doi.org/https://doi.org/10.1103/PhysRevLett.112.238303} {\bibfield
  {journal} {\bibinfo  {journal} {Phys. Rev. Lett.}\ }\textbf {\bibinfo
  {volume} {112}},\ \bibinfo {pages} {238303} (\bibinfo {year}
  {2014})}\BibitemShut {NoStop}%
\bibitem [{\citenamefont {Sire}\ and\ \citenamefont
  {Chavanis}(2004)}]{sire2004postcollapse}%
  \BibitemOpen
  \bibfield  {author} {\bibinfo {author} {\bibfnamefont {C.}~\bibnamefont
  {Sire}}\ and\ \bibinfo {author} {\bibfnamefont {P.-H.}\ \bibnamefont
  {Chavanis}},\ }\bibfield  {title} {\bibinfo {title} {Postcollapse dynamics of
  self-gravitating brownian particles and bacterial populations},\ }\href
  {https://doi.org/https://doi.org/10.1103/PhysRevE.69.066109} {\bibfield
  {journal} {\bibinfo  {journal} {Phys. Rev. E}\ }\textbf {\bibinfo {volume}
  {69}},\ \bibinfo {pages} {066109} (\bibinfo {year} {2004})}\BibitemShut
  {NoStop}%
\bibitem [{\citenamefont {Solon}\ \emph {et~al.}(2018)\citenamefont {Solon},
  \citenamefont {Stenhammar}, \citenamefont {Cates}, \citenamefont {Kafri},\
  and\ \citenamefont {Tailleur}}]{solon2018generalized}%
  \BibitemOpen
  \bibfield  {author} {\bibinfo {author} {\bibfnamefont {A.~P.}\ \bibnamefont
  {Solon}}, \bibinfo {author} {\bibfnamefont {J.}~\bibnamefont {Stenhammar}},
  \bibinfo {author} {\bibfnamefont {M.~E.}\ \bibnamefont {Cates}}, \bibinfo
  {author} {\bibfnamefont {Y.}~\bibnamefont {Kafri}},\ and\ \bibinfo {author}
  {\bibfnamefont {J.}~\bibnamefont {Tailleur}},\ }\bibfield  {title} {\bibinfo
  {title} {Generalized thermodynamics of phase equilibria in scalar active
  matter},\ }\href {https://doi.org/https://doi.org/10.1103/PhysRevE.97.020602}
  {\bibfield  {journal} {\bibinfo  {journal} {Phys. Rev. E}\ }\textbf {\bibinfo
  {volume} {97}},\ \bibinfo {pages} {020602} (\bibinfo {year}
  {2018})}\BibitemShut {NoStop}%
\bibitem [{\citenamefont {O'Byrne}\ \emph {et~al.}(2023)\citenamefont
  {O'Byrne}, \citenamefont {Solon}, \citenamefont {Tailleur},\ and\
  \citenamefont {Zhao}}]{obyrne2023introduction}%
  \BibitemOpen
  \bibfield  {author} {\bibinfo {author} {\bibfnamefont {J.}~\bibnamefont
  {O'Byrne}}, \bibinfo {author} {\bibfnamefont {A.}~\bibnamefont {Solon}},
  \bibinfo {author} {\bibfnamefont {J.}~\bibnamefont {Tailleur}},\ and\
  \bibinfo {author} {\bibfnamefont {Y.}~\bibnamefont {Zhao}},\ }\href@noop {}
  {\emph {\bibinfo {title} {An introduction to motility-induced phase
  separation}}}\ (\bibinfo  {publisher} {The Royal Society of Chemistry},\
  \bibinfo {year} {2023})\BibitemShut {NoStop}%
\bibitem [{\citenamefont {Dean}(1996)}]{dean1996langevin}%
  \BibitemOpen
  \bibfield  {author} {\bibinfo {author} {\bibfnamefont {D.~S.}\ \bibnamefont
  {Dean}},\ }\bibfield  {title} {\bibinfo {title} {Langevin equation for the
  density of a system of interacting langevin processes},\ }\href
  {https://doi.org/10.1088/0305-4470/29/24/001} {\bibfield  {journal} {\bibinfo
   {journal} {J. Phys. A}\ }\textbf {\bibinfo {volume} {29}},\ \bibinfo {pages}
  {L613} (\bibinfo {year} {1996})}\BibitemShut {NoStop}%
\bibitem [{\citenamefont {Kuroda}\ and\ \citenamefont
  {Miyazaki}(2023)}]{kuroda2023microscopic}%
  \BibitemOpen
  \bibfield  {author} {\bibinfo {author} {\bibfnamefont {Y.}~\bibnamefont
  {Kuroda}}\ and\ \bibinfo {author} {\bibfnamefont {K.}~\bibnamefont
  {Miyazaki}},\ }\bibfield  {title} {\bibinfo {title} {Microscopic theory for
  hyperuniformity in two-dimensional chiral active fluid},\ }\href
  {https://doi.org/10.1088/1742-5468/ad0639} {\bibfield  {journal} {\bibinfo
  {journal} {J. Stat. Mech.}\ }\textbf {\bibinfo {volume} {2023}},\ \bibinfo
  {pages} {103203} (\bibinfo {year} {2023})}\BibitemShut {NoStop}%
\bibitem [{\citenamefont {Illien}(2025)}]{illien2025dean}%
  \BibitemOpen
  \bibfield  {author} {\bibinfo {author} {\bibfnamefont {P.}~\bibnamefont
  {Illien}},\ }\bibfield  {title} {\bibinfo {title} {The dean--kawasaki
  equation and stochastic density functional theory},\ }\href
  {https://doi.org/10.1088/1361-6633/adee2e} {\bibfield  {journal} {\bibinfo
  {journal} {Rep. Prog. Phys.}\ }\textbf {\bibinfo {volume} {88}},\ \bibinfo
  {pages} {086601} (\bibinfo {year} {2025})}\BibitemShut {NoStop}%
\bibitem [{\citenamefont {Fodor}\ \emph {et~al.}(2016)\citenamefont {Fodor},
  \citenamefont {Nardini}, \citenamefont {Cates}, \citenamefont {Tailleur},
  \citenamefont {Visco},\ and\ \citenamefont {Van~Wijland}}]{fodor2016far}%
  \BibitemOpen
  \bibfield  {author} {\bibinfo {author} {\bibfnamefont {{\'E}.}~\bibnamefont
  {Fodor}}, \bibinfo {author} {\bibfnamefont {C.}~\bibnamefont {Nardini}},
  \bibinfo {author} {\bibfnamefont {M.~E.}\ \bibnamefont {Cates}}, \bibinfo
  {author} {\bibfnamefont {J.}~\bibnamefont {Tailleur}}, \bibinfo {author}
  {\bibfnamefont {P.}~\bibnamefont {Visco}},\ and\ \bibinfo {author}
  {\bibfnamefont {F.}~\bibnamefont {Van~Wijland}},\ }\bibfield  {title}
  {\bibinfo {title} {How far from equilibrium is active matter?},\ }\href
  {https://doi.org/https://doi.org/10.1103/PhysRevLett.117.038103} {\bibfield
  {journal} {\bibinfo  {journal} {Phys. Rev. Lett.}\ }\textbf {\bibinfo
  {volume} {117}},\ \bibinfo {pages} {038103} (\bibinfo {year}
  {2016})}\BibitemShut {NoStop}%
\bibitem [{\citenamefont {Chaikin}\ \emph {et~al.}(1995)\citenamefont
  {Chaikin}, \citenamefont {Lubensky},\ and\ \citenamefont
  {Witten}}]{chaikin1995principles}%
  \BibitemOpen
  \bibfield  {author} {\bibinfo {author} {\bibfnamefont {P.~M.}\ \bibnamefont
  {Chaikin}}, \bibinfo {author} {\bibfnamefont {T.~C.}\ \bibnamefont
  {Lubensky}},\ and\ \bibinfo {author} {\bibfnamefont {T.~A.}\ \bibnamefont
  {Witten}},\ }\href@noop {} {\emph {\bibinfo {title} {Principles of condensed
  matter physics}}},\ Vol.~\bibinfo {volume} {10}\ (\bibinfo  {publisher}
  {Cambridge university press Cambridge},\ \bibinfo {year} {1995})\BibitemShut
  {NoStop}%
\bibitem [{\citenamefont {Dinelli}\ \emph {et~al.}(2025)\citenamefont
  {Dinelli}, \citenamefont {Altieri},\ and\ \citenamefont
  {Tailleur}}]{dinelli2025random}%
  \BibitemOpen
  \bibfield  {author} {\bibinfo {author} {\bibfnamefont {A.}~\bibnamefont
  {Dinelli}}, \bibinfo {author} {\bibfnamefont {A.}~\bibnamefont {Altieri}},\
  and\ \bibinfo {author} {\bibfnamefont {J.}~\bibnamefont {Tailleur}},\
  }\bibfield  {title} {\bibinfo {title} {Random motility regulation drives the
  fragmentation of microbial ecosystems},\ }\href@noop {} {\bibfield  {journal}
  {\bibinfo  {journal} {arXiv preprint arXiv:2503.12692}\ } (\bibinfo {year}
  {2025})}\BibitemShut {NoStop}%
\bibitem [{\citenamefont {Brauns}\ and\ \citenamefont
  {Marchetti}(2024)}]{brauns2024nonreciprocal}%
  \BibitemOpen
  \bibfield  {author} {\bibinfo {author} {\bibfnamefont {F.}~\bibnamefont
  {Brauns}}\ and\ \bibinfo {author} {\bibfnamefont {M.~C.}\ \bibnamefont
  {Marchetti}},\ }\bibfield  {title} {\bibinfo {title} {Nonreciprocal pattern
  formation of conserved fields},\ }\href
  {https://doi.org/https://doi.org/10.1103/PhysRevX.14.021014} {\bibfield
  {journal} {\bibinfo  {journal} {Phys. Rev. X}\ }\textbf {\bibinfo {volume}
  {14}},\ \bibinfo {pages} {021014} (\bibinfo {year} {2024})}\BibitemShut
  {NoStop}%
\bibitem [{\citenamefont {Greve}\ \emph {et~al.}(2025)\citenamefont {Greve},
  \citenamefont {Lovato}, \citenamefont {Frohoff-H{\"u}lsmann},\ and\
  \citenamefont {Thiele}}]{greve2025coexistence}%
  \BibitemOpen
  \bibfield  {author} {\bibinfo {author} {\bibfnamefont {D.}~\bibnamefont
  {Greve}}, \bibinfo {author} {\bibfnamefont {G.}~\bibnamefont {Lovato}},
  \bibinfo {author} {\bibfnamefont {T.}~\bibnamefont {Frohoff-H{\"u}lsmann}},\
  and\ \bibinfo {author} {\bibfnamefont {U.}~\bibnamefont {Thiele}},\
  }\bibfield  {title} {\bibinfo {title} {Coexistence of uniform and oscillatory
  states resulting from nonreciprocity and conservation laws},\ }\href
  {https://doi.org/https://doi.org/10.1103/PhysRevLett.134.018303} {\bibfield
  {journal} {\bibinfo  {journal} {Phys. Rev. Lett.}\ }\textbf {\bibinfo
  {volume} {134}},\ \bibinfo {pages} {018303} (\bibinfo {year}
  {2025})}\BibitemShut {NoStop}%
\bibitem [{\citenamefont {Omar}\ \emph {et~al.}(2023)\citenamefont {Omar},
  \citenamefont {Row}, \citenamefont {Mallory},\ and\ \citenamefont
  {Brady}}]{omar2023mechanical}%
  \BibitemOpen
  \bibfield  {author} {\bibinfo {author} {\bibfnamefont {A.~K.}\ \bibnamefont
  {Omar}}, \bibinfo {author} {\bibfnamefont {H.}~\bibnamefont {Row}}, \bibinfo
  {author} {\bibfnamefont {S.~A.}\ \bibnamefont {Mallory}},\ and\ \bibinfo
  {author} {\bibfnamefont {J.~F.}\ \bibnamefont {Brady}},\ }\bibfield  {title}
  {\bibinfo {title} {Mechanical theory of nonequilibrium coexistence and
  motility-induced phase separation},\ }\href
  {https://doi.org/https://doi.org/10.1073/pnas.2219900120} {\bibfield
  {journal} {\bibinfo  {journal} {Proc. Natl. Acad. Sci. U.S.A.}\ }\textbf
  {\bibinfo {volume} {120}},\ \bibinfo {pages} {e2219900120} (\bibinfo {year}
  {2023})}\BibitemShut {NoStop}%
\bibitem [{\citenamefont {Nguyen}\ \emph {et~al.}(2025)\citenamefont {Nguyen},
  \citenamefont {Dinelli}, \citenamefont {Spera},\ and\ \citenamefont
  {Tailleur}}]{nguyen2025contact}%
  \BibitemOpen
  \bibfield  {author} {\bibinfo {author} {\bibfnamefont {Q.~M.}\ \bibnamefont
  {Nguyen}}, \bibinfo {author} {\bibfnamefont {A.}~\bibnamefont {Dinelli}},
  \bibinfo {author} {\bibfnamefont {G.}~\bibnamefont {Spera}},\ and\ \bibinfo
  {author} {\bibfnamefont {J.}~\bibnamefont {Tailleur}},\ }\bibfield  {title}
  {\bibinfo {title} {Contact forces in motility-regulated active matter},\
  }\href@noop {} {\bibfield  {journal} {\bibinfo  {journal} {arXiv preprint
  arXiv:2507.08964}\ } (\bibinfo {year} {2025})}\BibitemShut {NoStop}%
\bibitem [{\citenamefont {Abaurrea~Velasco}\ \emph {et~al.}(2018)\citenamefont
  {Abaurrea~Velasco}, \citenamefont {Abkenar}, \citenamefont {Gompper},\ and\
  \citenamefont {Auth}}]{abaurrea2018collective}%
  \BibitemOpen
  \bibfield  {author} {\bibinfo {author} {\bibfnamefont {C.}~\bibnamefont
  {Abaurrea~Velasco}}, \bibinfo {author} {\bibfnamefont {M.}~\bibnamefont
  {Abkenar}}, \bibinfo {author} {\bibfnamefont {G.}~\bibnamefont {Gompper}},\
  and\ \bibinfo {author} {\bibfnamefont {T.}~\bibnamefont {Auth}},\ }\bibfield
  {title} {\bibinfo {title} {Collective behavior of self-propelled rods with
  quorum sensing},\ }\href
  {https://doi.org/https://doi.org/10.1103/PhysRevE.98.022605} {\bibfield
  {journal} {\bibinfo  {journal} {Phys. Rev. E}\ }\textbf {\bibinfo {volume}
  {98}},\ \bibinfo {pages} {022605} (\bibinfo {year} {2018})}\BibitemShut
  {NoStop}%
\bibitem [{\citenamefont {Bertin}\ \emph {et~al.}(2006)\citenamefont {Bertin},
  \citenamefont {Droz},\ and\ \citenamefont
  {Gr{\'e}goire}}]{bertin2006boltzmann}%
  \BibitemOpen
  \bibfield  {author} {\bibinfo {author} {\bibfnamefont {E.}~\bibnamefont
  {Bertin}}, \bibinfo {author} {\bibfnamefont {M.}~\bibnamefont {Droz}},\ and\
  \bibinfo {author} {\bibfnamefont {G.}~\bibnamefont {Gr{\'e}goire}},\
  }\bibfield  {title} {\bibinfo {title} {Boltzmann and hydrodynamic description
  for self-propelled particles},\ }\href
  {https://doi.org/https://doi.org/10.1103/PhysRevE.74.022101} {\bibfield
  {journal} {\bibinfo  {journal} {Phys. Rev. E}\ }\textbf {\bibinfo {volume}
  {74}},\ \bibinfo {pages} {022101} (\bibinfo {year} {2006})}\BibitemShut
  {NoStop}%
\bibitem [{\citenamefont {Saintillan}\ and\ \citenamefont
  {Shelley}(2008)}]{saintillan2008instabilities}%
  \BibitemOpen
  \bibfield  {author} {\bibinfo {author} {\bibfnamefont {D.}~\bibnamefont
  {Saintillan}}\ and\ \bibinfo {author} {\bibfnamefont {M.~J.}\ \bibnamefont
  {Shelley}},\ }\bibfield  {title} {\bibinfo {title} {Instabilities and pattern
  formation in active particle suspensions: Kinetic theory and continuum
  simulations},\ }\href
  {https://doi.org/https://doi.org/10.1103/PhysRevLett.100.178103} {\bibfield
  {journal} {\bibinfo  {journal} {Phys. Rev. Lett.}\ }\textbf {\bibinfo
  {volume} {100}},\ \bibinfo {pages} {178103} (\bibinfo {year}
  {2008})}\BibitemShut {NoStop}%
\bibitem [{\citenamefont {Subramanian}\ and\ \citenamefont
  {Koch}(2009)}]{subramanian2009critical}%
  \BibitemOpen
  \bibfield  {author} {\bibinfo {author} {\bibfnamefont {G.}~\bibnamefont
  {Subramanian}}\ and\ \bibinfo {author} {\bibfnamefont {D.~L.}\ \bibnamefont
  {Koch}},\ }\bibfield  {title} {\bibinfo {title} {Critical bacterial
  concentration for the onset of collective swimming},\ }\href
  {https://doi.org/https://doi.org/10.1017/S002211200900706X} {\bibfield
  {journal} {\bibinfo  {journal} {J. Fluid Mech.}\ }\textbf {\bibinfo {volume}
  {632}},\ \bibinfo {pages} {359} (\bibinfo {year} {2009})}\BibitemShut
  {NoStop}%
\bibitem [{\citenamefont {Ihle}(2011)}]{ihle2011kinetic}%
  \BibitemOpen
  \bibfield  {author} {\bibinfo {author} {\bibfnamefont {T.}~\bibnamefont
  {Ihle}},\ }\bibfield  {title} {\bibinfo {title} {Kinetic theory of flocking:
  Derivation of hydrodynamic equations},\ }\href
  {https://doi.org/https://doi.org/10.1103/PhysRevE.83.030901} {\bibfield
  {journal} {\bibinfo  {journal} {Phys. Rev. E}\ }\textbf {\bibinfo {volume}
  {83}},\ \bibinfo {pages} {030901} (\bibinfo {year} {2011})}\BibitemShut
  {NoStop}%
\bibitem [{\citenamefont {Solon}\ and\ \citenamefont
  {Tailleur}(2013)}]{solon2013revisiting}%
  \BibitemOpen
  \bibfield  {author} {\bibinfo {author} {\bibfnamefont {A.~P.}\ \bibnamefont
  {Solon}}\ and\ \bibinfo {author} {\bibfnamefont {J.}~\bibnamefont
  {Tailleur}},\ }\bibfield  {title} {\bibinfo {title} {Revisiting the flocking
  transition using active spins},\ }\href
  {https://doi.org/https://doi.org/10.1103/PhysRevLett.111.078101} {\bibfield
  {journal} {\bibinfo  {journal} {Phys. Rev. Lett.}\ }\textbf {\bibinfo
  {volume} {111}},\ \bibinfo {pages} {078101} (\bibinfo {year}
  {2013})}\BibitemShut {NoStop}%
\bibitem [{\citenamefont {Peshkov}\ \emph {et~al.}(2014)\citenamefont
  {Peshkov}, \citenamefont {Bertin}, \citenamefont {Ginelli},\ and\
  \citenamefont {Chat{\'e}}}]{peshkov2014boltzmann}%
  \BibitemOpen
  \bibfield  {author} {\bibinfo {author} {\bibfnamefont {A.}~\bibnamefont
  {Peshkov}}, \bibinfo {author} {\bibfnamefont {E.}~\bibnamefont {Bertin}},
  \bibinfo {author} {\bibfnamefont {F.}~\bibnamefont {Ginelli}},\ and\ \bibinfo
  {author} {\bibfnamefont {H.}~\bibnamefont {Chat{\'e}}},\ }\bibfield  {title}
  {\bibinfo {title} {Boltzmann-ginzburg-landau approach for continuous
  descriptions of generic vicsek-like models},\ }\href
  {https://doi.org/https://doi.org/10.1140/epjst/e2014-02193-y} {\bibfield
  {journal} {\bibinfo  {journal} {Eur. Phys. J. Spec. Top.}\ }\textbf {\bibinfo
  {volume} {223}},\ \bibinfo {pages} {1315} (\bibinfo {year}
  {2014})}\BibitemShut {NoStop}%
\bibitem [{\citenamefont {Weady}\ \emph {et~al.}(2022)\citenamefont {Weady},
  \citenamefont {Stein},\ and\ \citenamefont
  {Shelley}}]{weady2022thermodynamically}%
  \BibitemOpen
  \bibfield  {author} {\bibinfo {author} {\bibfnamefont {S.}~\bibnamefont
  {Weady}}, \bibinfo {author} {\bibfnamefont {D.~B.}\ \bibnamefont {Stein}},\
  and\ \bibinfo {author} {\bibfnamefont {M.~J.}\ \bibnamefont {Shelley}},\
  }\bibfield  {title} {\bibinfo {title} {Thermodynamically consistent
  coarse-graining of polar active fluids},\ }\href
  {https://doi.org/https://doi.org/10.1103/PhysRevFluids.7.063301} {\bibfield
  {journal} {\bibinfo  {journal} {Phys. Rev. Fluids}\ }\textbf {\bibinfo
  {volume} {7}},\ \bibinfo {pages} {063301} (\bibinfo {year}
  {2022})}\BibitemShut {NoStop}%
\end{thebibliography}%

\end{document}